# On the Role of Triadic Substructures in Complex Networks
*Marco Winkler*

**Ph.D. dissertation**

# On the Role of Triadic Substructures in Complex Networks

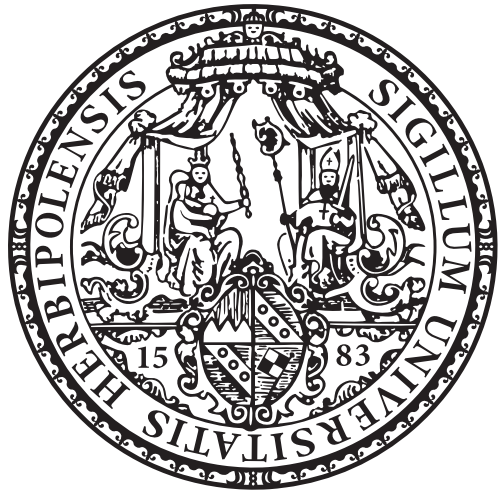

**Dissertation**
zur Erlangung des naturwissenschaftlichen Doktorgrades
der Julius-Maximilians-Universität Würzburg

vorgelegt von
**Marco Winkler**
aus Frankfurt am Main

Würzburg 2015

Marco Winkler
University of Würzburg
Institute for Theoretical Physics
Am Hubland
97074 Würzburg
Germany
marco.a.winkler@gmail.com
`www.mwinkler.eu`

*Today the network of relationships linking the human race to itself and to the rest of the biosphere is so complex that all aspects affect all others to an extraordinary degree. Someone should be studying the whole system, however crudely that has to be done, because no gluing together of partial studies of a complex nonlinear system can give a good idea of the behavior of the whole.*
Murray Gell-Mann, 1997

# Abstract

In the course of the growth of the Internet and due to increasing availability of data, over the last two decades, the field of network science has established itself as an own area of research. With quantitative scientists from computer science, mathematics, and physics working on datasets from biology, economics, sociology, political sciences, and many others, network science serves as a paradigm for interdisciplinary research.

One of the major goals in network science is to unravel the relationship between topological graph *structure* and a network's *function*. As evidence suggests, systems from the same fields, i.e. with similar function, tend to exhibit similar structure. However, it is still vague whether a similar graph structure automatically implies likewise function. This dissertation aims at helping to bridge this gap, while particularly focusing on the role of triadic structures.

After a general introduction to the main concepts of network science, existing work devoted to the relevance of triadic substructures is reviewed. A major challenge in modeling triadic structure is the fact that not all three-node subgraphs can be specified independently of each other, as pairs of nodes may participate in multiple of those triadic subgraphs.

In order to overcome this obstacle, we suggest a novel class of generative network models based on so called *Steiner triple systems*. The latter are partitions of a graph's vertices into pair-disjoint triples (Steiner triples). Thus, the configurations on Steiner triples can be specified independently of each other without overdetermining the network's link structure.

Subsequently, we investigate the most basic realization of this new class of models. We call it the *triadic random graph model (TRGM)*. The TRGM is parametrized by a probability distribution over all possible triadic subgraph patterns. In order to generate a network instantiation of the model, for all Steiner triples in the system, a pattern is drawn from the distribution and adjusted randomly on the Steiner triple. We calculate the degree distribution of the TRGM analytically and find it to be similar to a Poissonian distribution. Furthermore, it is shown that TRGMs possess non-trivial triadic structure. We discover inevitable correlations in the abundance of certain triadic subgraph patterns which should be taken into account when attributing functional relevance to particular *motifs* – patterns which occur significantly more frequently than expected at random. Beyond, the strong impact of the probability distributions on the Steiner triples on the occurrence of triadic subgraphs over the whole network is demonstrated. This interdependence allows us to design ensembles of networks with predefined triadic substructure. Hence, TRGMs help to overcome the lack of generative models needed for assessing the relevance of triadic structure.

We further investigate whether motifs occur homogeneously or heterogeneously distributed over a graph. Therefore, we study triadic subgraph structures in each node's neighborhood individually. In order to quantitatively measure structure from an individual node's perspective, we introduce an algorithm for *node-specific pattern mining* for both directed unsigned, and undirected signed networks. Analyzing real-world datasets, we find that there are networks in which motifs are distributed highly heterogeneously, bound to the proximity of only very few nodes. Moreover, we observe indication for the potential sensitivity of biological systems to a targeted removal of these critical vertices. In addition, we study whole graphs with respect to the homogeneity and homophily of their node-specific triadic structure. The former describes the similarity of subgraph distributions in the neighborhoods of individual vertices. The latter quantifies whether connected vertices are structurally more similar than non-connected ones. We discover these features to be characteristic for the networks' origins. More-over, clustering the vertices of graphs regarding their triadic structure, we investigate structural groups in the neural network of C. elegans, the international airport-connection network, and the global network of diplomatic sentiments between countries. For the latter we find evidence for the instability of triangles considered *socially unbalanced* according to sociological theories.

Finally, we utilize our TRGM to explore ensembles of networks with similar triadic substructure in terms of the evolution of dynamical processes acting on their nodes. Focusing on oscillators, coupled along the graphs' edges, we observe that certain triad motifs impose a clear signature on the systems' dynamics, even when embedded in a larger network structure.

# Zusammenfassung

Im Zuge des Wachstums des Internets und der Verfügbarkeit nie da gewesener Datenmengen, hat sich, während der letzten beiden Jahrzehnte, die Netzwerkwissenschaft zu einer eigenständigen Forschungsrichtung entwickelt. Mit Wissenschaftlern aus quantitativen Feldern wie der Informatik, Mathematik und Physik, die Datensätze aus Biologie, den Wirtschaftswissenschaften, Soziologie, Politikwissenschaft und vielen weiteren Anwendungsgebieten untersuchen, stellt die Netzwerkwissenschaft ein Paradebeispiel interdisziplinärer Forschung dar.

Eines der grundlegenden Ziele der Netzwerkwissenschaft ist es, den Zusammenhang zwischen der topologischen *Struktur* und der *Funktion* von Netzwerken herauszufinden. Es gibt zahlreiche Hinweise, dass Netz-werke aus den gleichen Bereichen, d.h. Systeme mit ähnlicher Funktion, auch ähnliche Graphstrukturen aufweisen. Es ist allerdings nach wie vor unklar, ob eine ähnliche Graphstruktur generell zu gleicher Funktionsweise führt. Es ist das Ziel der vorliegenden Dissertation, zur Klärung dieser Frage beizutragen. Das Hauptaugenmerk wird hierbei auf der Rolle von Dreiecksstrukturen liegen.

Nach einer allgemeinen Einführung der wichtigsten Grundlagen der Theorie komplexer Netzwerke, wird eine Übersicht über existierende Arbeiten zur Bedeutung von Dreiecksstrukturen gegeben. Eine der größten Herausforderungen bei der Modellierung triadischer Strukturen ist die Tatsache, dass nicht alle Dreiecksbeziehungen in einem Graphen unabhängig voneinander bestimmt werden können, da zwei Knoten an mehreren solcher Dreiecksbeziehungen beteiligt sein können.

Um dieses Problem zu lösen, führen wir, basierend auf sogenannten *Steiner-Tripel-Systemen*, eine neue Klasse generativer Netzwerkmodelle ein. Steiner-Tripel-Systeme sind Zerlegungen der Knoten eines

Graphen in paarfremde Tripel (Steiner-Tripel). Daher können die Konfigurationen auf Steiner-Tripeln unabhängig voneinander gewählt werden, ohne dass dies zu einer Überbestimmung der Netzwerkstruktur führen würde.

Anschließend untersuchen wir die grundlegendste Realisierung dieser neuen Klasse von Netzwerkmodellen, die wir das triadische Zufallsgraph-Modell (engl. *triadic random graph model, TRGM*) nennen. TRGMs werden durch eine Wahrscheinlichkeitsverteilung über alle möglichen Dreiecksstrukturen parametrisiert. Um ein konkretes Netzwerk zu erzeugen wird für jedes Steiner-Tripel eine Dreiecksstruktur gemäß der Wahrscheinlichkeitsverteilung gezogen und zufällig auf dem Tripel orientiert. Wir berechnen die Knotengradverteilung des TRGM analytisch und finden heraus, dass diese einer Poissonverteilung ähnelt. Des Weiteren wird gezeigt, dass TRGMs nichttriviale Dreiecksstrukturen aufweisen. Außerdem finden wir unvermeidliche Korrelationen im Auftreten bestimmter Subgraphen, derer man sich bewusst sein sollte. Insbesondere wenn es darum geht, die Bedeutung sogenannter *Motive* (Strukturen, die signifikant häufiger als zufällig erwartet auftreten) zu beurteilen. Darüber hinaus wird der starke Einfluss der Wahrscheinlichkeitsverteilung auf den Steiner-Tripeln, auf die generelle Dreiecksstruktur der erzeugten Netzwerke gezeigt. Diese Abhängigkeit ermöglicht es, Netzwerkensembles mit vorgegebener Dreiecksstruktur zu konzipieren. Daher helfen TRGMs dabei, den bestehenden Mangel an generativen Netzwerkmodellen, zur Beurteilung der Bedeutung triadischer Strukturen in Graphen, zu beheben.

Es wird ferner untersucht, wie homogen Motive räumlich über Graphstrukturen verteilt sind. Zu diesem Zweck untersuchen wir das Auftreten von Dreiecksstrukturen in der Umgebung jedes Knotens separat. Um die Struktur individueller Knoten quantitativ erfassen zu können, führen wir einen Algorithmus zur *knotenspezifischen Musterauswertung (node-specific pattern mining)* ein, der sowohl auf gerichtete, als auch auf Graphen mit positiven und negativen Kanten angewendet werden kann. Bei der Analyse realer Datensätze beobachten wir, dass Motive in einigen Netzen hochgradig heterogen verteilt, und auf die Umge-bung einiger, weniger Knoten beschränkt sind. Darüber hinaus finden wir Hinweise auf die mögliche Fehleranfälligkeit biologischer Systeme auf ein gezieltes Entfernen ebendieser Knoten. Des Weiteren studieren wir ganze Graphen bezüglich der Homogenität und Homophilie ihrer knotenspezifischen Dreiecksmuster. Erstere beschreibt die Ähnlichkeit der lokalen Dreiecksstrukturen zwischen verschiedenen Knoten. Letztere gibt an, ob sich verbundene Knoten begzüglich ihrer Dreiecksstruktur ähnlicher sind, als nicht verbundene Knoten. Wir stellen fest, dass diese Eigenschaften charakteristisch für die Herkunft der jeweiligen Netzwerke sind. Darüber hinaus gruppieren wir die Knoten verschiedener Systeme bezüglich der Ähnlichkeit ihrer lokalen Dreiecksstrukturen. Hierzu untersuchen wir das neuronale Netz von C. elegans, das internationale Flugverbindungsnetzwerk, sowie das Netzwerk internationaler Beziehungen zwischen Staaten. In Letzterem finden wir Hinweise darauf, dass Dreieckskonfigurationen, die nach soziologischen Theorien als *unbalanciert* gelten, besonders instabil sind.

Schließlich verwenden wir unser TRGM, um Netzwerkensembles mit ähnlicher Dreiecksstruktur bezüglich der Eigenschaften dynamischer Prozesse, die auf ihren Knoten ablaufen, zu untersuchen. Wir konzentrieren uns auf Oszillatoren, die entlang der Kanten der Graphen miteinander gekoppelt sind. Hierbei beobachten wir, dass bestimmte Dreiecksmotive charakteristische Merkmale im dynamischen Verhalten der Systeme hinterlassen. Dies ist auch der Fall, wenn die Motive in eine größere Netzwerkstruktur eingebettet sind.

# Acknowledgements

First and foremost, I would like to express my sincere gratitude to my supervisor Wolfgang Kinzel who offered me a position in his group and hence gave me the chance to work on this dissertation. Furthermore, he enabled me to present my results at conferences in Dresden, Berlin, Regensburg, Copenhagen, Berkeley, Boston, and Shenzhen.

I would also like to thank Jörg Reichardt with whom I collaborated closely during the first two years of my research for providing his expertise in the fields of network science and machine learning.

I particularly want to thank my collaborator Otti D'Huys for sharing her knowledge on the simulation of dynamical processes (and her Belgian chocolate) with me. Moreover, I am most grateful for proofreading this manuscript.

Let me also thank Björn Trauzettel and Jens Pflaum for agreeing to participate in the examination board as well as Haye Hinrichsen, Evimaria Terzi, Navid Dianti, Nima Dehmamy, and Nagendra Kumar Panduranga for helpful scientific discussions and suggestions ranging from statistical physics to data mining and political sciences.

In addition, I want to thank my present and former co-workers in the Computational Physics group in Würzburg – Georg Reents, Elka Korutcheva, Christian Drobnik, Johannes Falk, Sven Heiligenthal, Peter Janotta, Thomas Jüngling, Laura Lauerbach, David Luposchainsky, Lisa Morgan, Peter Reisenauer, Manuel Schottdorf, Jaegon Um, Christoph Wick, Steffen Zeeb, and Robert Ziener – for the pleasant working atmosphere. Especially I want to thank Nelly Meyer for her competent and kind assistance in all organizational and bureaucracy-related aspects.

Last, but not least, I want to express my gratefulness to my wife

Katharina, my parents Marion and Peter, as well as to my whole family and friends for their constant endorsement and support.

Finally, I want to thank the German National Merit Foundation (Studienstiftung des deutschen Volkes) and the German Academic Exchange Service (DAAD) for their support during my physics studies, as well as for affording a one year stay as a visiting student at Boston University. I thank the W.E. Heraeus-Stiftung for supporting the participation in the Spring Meetings 2011-2014 of the German Physical Society (DPG) in Dresden, Berlin, and Regensburg.

<div style="text-align: right">Marco Winkler</div>

<div style="text-align: right">Würzburg, June 2015</div>

# Contents

*Contents*

# 1. Introduction

The astonishing complexity of many systems does not arise from the intricacy of their constituents. It rather results from their interactions with each other. Comparatively primitive neurons may form brains as powerful as those of humans, thanks to the way they are synaptically wired. Proteins control whole organisms by regulating their gene-expression levels. Connected via optical fiber or telephone cable, single computers form the Internet, allowing to exchange information between individuals around the world within seconds. On the Internet, by linking to each other, individual websites with rather specific content establish the *world-wide web (www)* where basically any information can be retrieved from. Simple airplane routes, which typically operate between two airports, build the world's air-traffic system, allowing people to get from any major city in the world to any other within less than two days, even without the need and ability to charter an executive jet.

A particularly successful way of analyzing such systems as a whole has been the investigation of the network structure they form. As an example, Fig. 1.1 shows one of the most widely known networks. It is a snapshot of the Internet as it looked like in the year 2003. Already with the naked eye and without any quantitative analysis it becomes clear that there is some structure present in the system. The belief that the structure of a network is closely related to its function is at the bottom of complex-networks research. In other words, it is assumed that connections did not form solely by chance, but were, e.g. in an evolutionary process, selected for the system to work the way it is supposed to. To be able to precisely test which aspects of networks have emerged purely randomly and which of them may have established for

*1. Introduction*

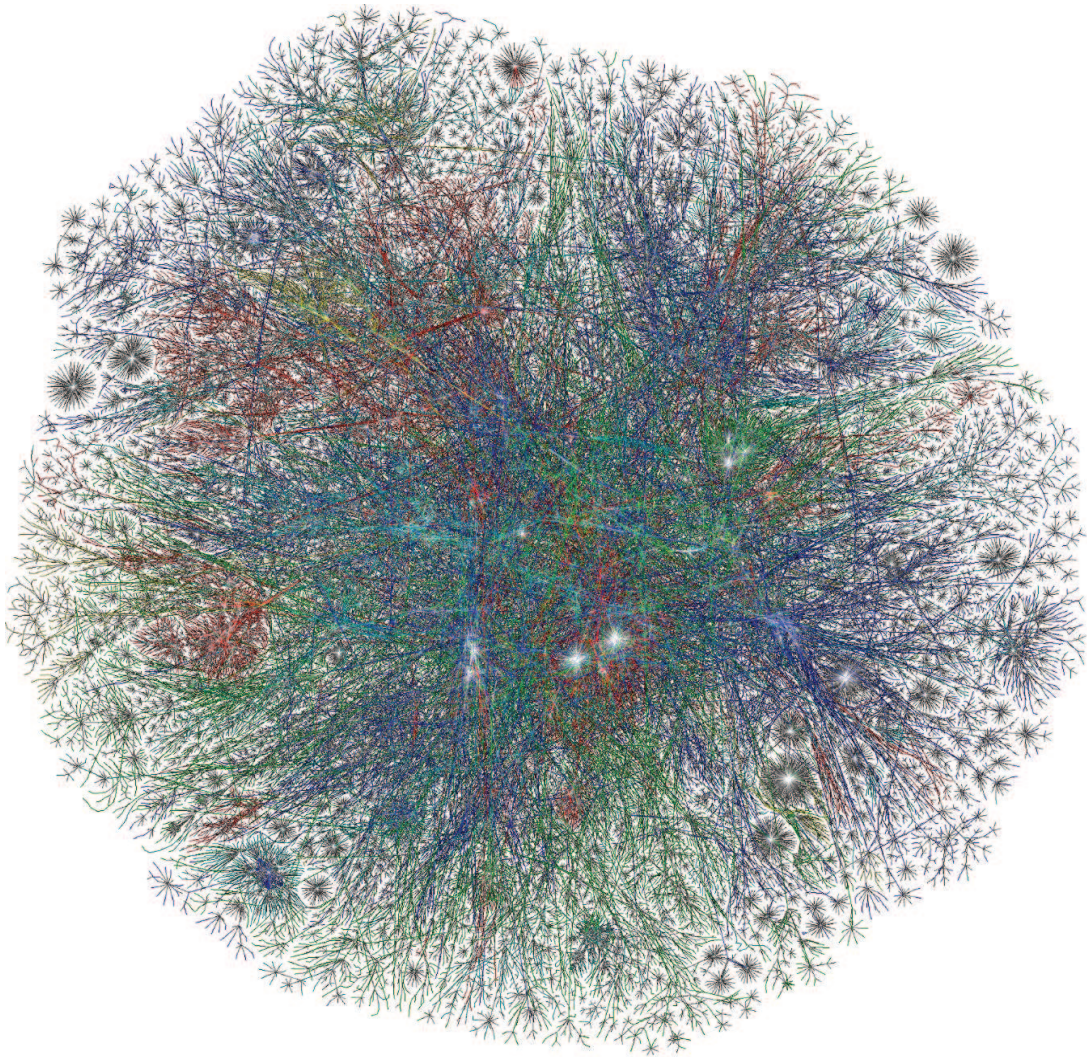

**Figure 1.1.:** Map of the Internet in 2003. Shown are groups of computers. Connections represent data packages exchanged between them. Positions in the map are not related to the physical location of the computers. Colors indicate Class A allocation of IP space to different registrars in the world: Asia Pacific - red, Europe/Middle East/Central Asia/Africa - green, North America - blue, Latin American and Caribbean - yellow, RFC1918 IP Addresses - cyan. Adapted from *The Opte Project / Barrett Lyon* (`www.opte.org`) under CC BY-NC 4.0 (`creativecommons.org`). Printed with permission.

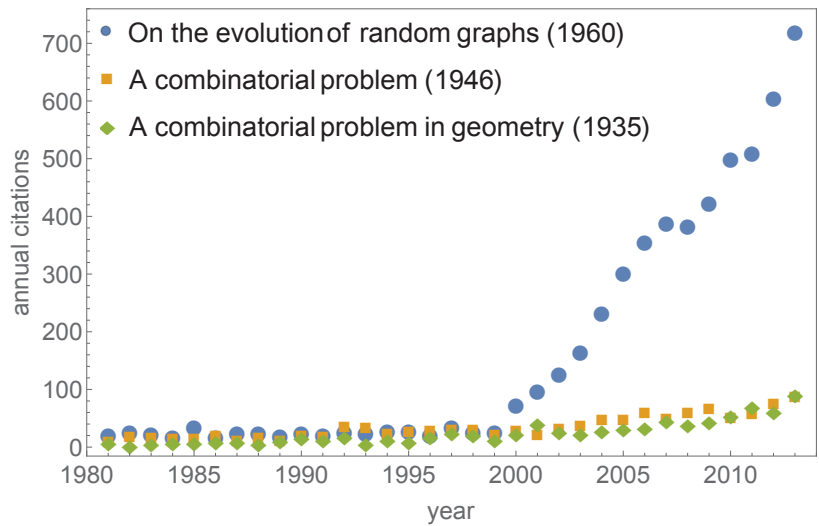

**Figure 1.2.:** Annual citations of the most cited article of Paul Erdös: P.E. and A. Rényi, Evolution 5, 17 (1960); for comparison we show P.E.'s two highest cited publications not related to random graphs: N. G. de Bruijn and P.E., Koninklijke Nederlandse Akademie v. Wetenschappen 49.49, 758-764 (1946); P.E. and G. Szekeres, Compositio Mathematica 2, 463-470 (1935). Source: Google Scholar, Oct 26, 2014.

some reason, one needs to be aware of the characteristics of systems where all connections have formed by chance with equal probability. Such systems have been studied intensively by the Hungarian mathematicians Paul Erdös (1913-1996) and Alfréd Rényi (1921-1970) in the middle of the 20th century.

In order to distinguish meaningful network structure from the random graphs investigated by Erdös and Rényi, many publications in network science refer to their seminal work *On the evolution of random graphs* [42] and – like this dissertation – cite the article. Hence, its number of citations is a good measure for the scientific interest in the area. Fig. 1.2 shows the annual number of citations between 1980 and 2013. Until the year 1999 the paper was cited between 20 and 40 times a year. Since then, the number of citations literally exploded, marking the genesis of network science as an own scientific field, though to date

*1. Introduction*

it is still in its developing process. Due to the adolescence of the field, today mostly physicists, mathematicians, and computer scientists work in complex-networks research. This is because they generally possess the quantitative skills to work with the abstract concept of networks. However, in 2013 Northeastern University in Boston, MA, launched the first Ph.D. program in network science as a self-contained subject.

The rise of complex-networks research comes along with the growth of the world-wide web. In 1999, Larry Page and Sergei Brin – the first network scientists to become billionaires – published the famous PageRank algorithm [114] that allows to rank websites based on their relevance in the web structure. About at the same time they launched their search engine Google for which they utilized PageRank. Over the last decade, the *www* has provided for a multitude of datasets that has been unprecedented. The advancing digitalization of all aspects of everyday life will most likely further promote the relevance of network science.

However, one of the best-known experiments in network science, which supported the small-world hypothesis, was conducted already back in 1967 by Stanley Milgram[1] [98]. His results are often summarized under the phrase *six degrees of separation*. The idea of his reserach was to figure out how 'distant' two randomly selected people are (on average) in the network of social relationships. In his experiment, two individuals are considered to be connected if they know each other on a first-name basis. So the distance between two acquaintances, say Alice and Bob, will be one. If Bob also knows Charlie who does not know Alice, Alice and Charlie have a distance of two steps and so forth. Milgram sent out letters to people living in the Midwest of America with the name of a stockbroker from Boston who should eventually receive the letter. In case they knew the target, they would send it directly to him. Otherwise, they should choose the most promising person among their acquaintances. The latter was then asked to continue with the process. Milgram found out that – among the letters that reached the target – the average chain length was only six, resulting in the conclusion that we live in a rather 'small' world. In 2002, Kleinfeld pointed out that Milgram's original experiment had several methodological shortcomings [80]. However, in 2008 Leskovec and Horvitz could show the small-world property on a large dataset of instant messenger users [85]. They found a mean distance of 6.6 steps in a network of around 240 million users. In 2011, Ugander, Karrer et al. from Facebook reported that 92% of the 721 million active users of the social network were connected within five degrees of separation with an average distance of 4.7 steps [139].

---

[1]Milgram (1933-1984) was an American psychologist and social scientist. The small-world experiment shall not be confused with his well-known psychological experiment about obedience and conscience, coined *Milgram experiment* [97].

Of course, one should ask now whether six or seven steps are a small number or not. For a brief estimation one can assume that every person knows around 100 people. One can guess that there are around $100^2$ people two steps away and $100^6 = 10^{12}$ people six steps away, which is much more than the $7 \times 10^9$ people living on Earth at present. However, it is the fact that people do not become acquaintances of random people that makes the small-world result interesting. Our society is highly structured in groups on various levels. There are families, school classes, sports teams etc. where anyone is acquainted with anyone else, such that this rough estimation will not work. It is further very unlikely to become friends with a person living on the opposite hemisphere of the planet; a few centuries ago this was even impossible: the subnetwork of the Aborigines in Australia or the Aztecs, Inkas, and Mayas in America was disconnected from the people living in Europe, Africa, and Asia. That the average distance is still so low is therefore indeed a remarkable result. For more details on the small-world phenomenon see e.g. [148].

Among mathematicians, a popular way to illustrate the small-world property is to evaluate Erdös numbers. If you have published a paper with Paul Erdös, you get assigned an Erdös number of one. If you have published with a co-author of Erdös you have Erdös number two and so forth. Fig. 1.3 shows the author's personal network of collaborators, including their shortest paths to Paul Erdös. Although none of the author's co-authors is a mathematician by training, with values of three

## 1. Introduction

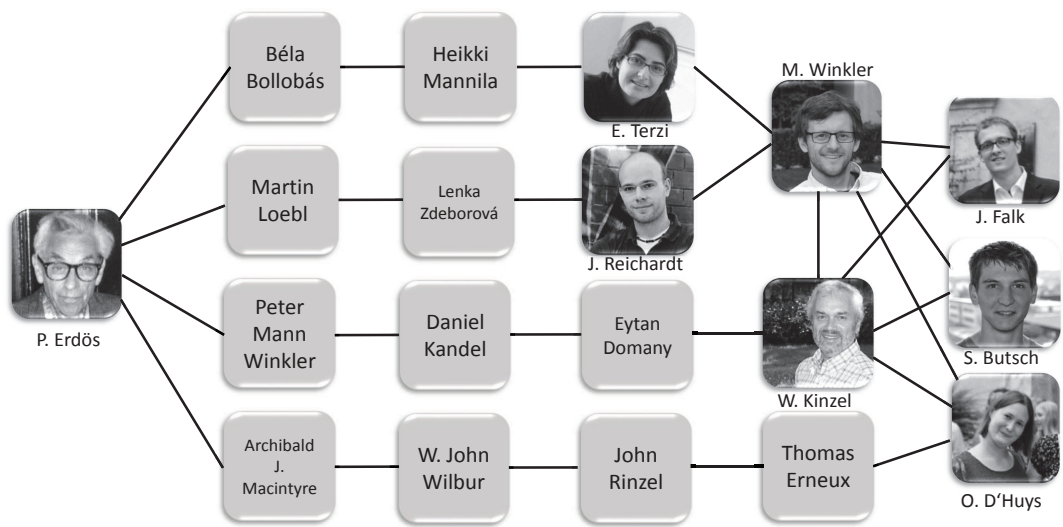

**Figure 1.3.:** Illustration of the small-world phenomenon. Although the scientific-collaboration network is huge, most scientist are only a few steps away from Paul Erdös. Picture of P.E. by Topsy Kretts under CC BY 3.0 (`creativecommons.org`).

to five the Erdös numbers are quite small. Of course, the choice of Erdös as the reference is completely arbitrary and a low Erdös number does not make any statement about the scientific impact of a person. Yet, it still serves as a neat example for the small-world pheonomenon.

Besides the average shortest-path length, sophisticated methodologies for network-analysis have been developed over the last decades; local measures that describe the properties of single units, global measures characterizing whole systems, clustering and community-detection algorithms to detect group structures, and *generative models* that allow to produce synthetic networks mimicing real systems in terms of certain characteristics. These methods allow to detect potentially functionally relevant structures and to better understand structural effects on the evolution of dynamical processes on networks, e.g. by means of computer simulations. Such dynamical processes range from neural activity to the propagation of rumors in social networks, the spreading of infectious diseases such as HIV, SARS, Ebola, or simply the flu between humans, or the evolution of traffic on transportation networks. In the following chapter, we will review complex-networks theory to the extent necessary for the remainder of this dissertation.

Chapter 3 will be particularly devoted to existing results on the role of triadic substructures in complex networks, the main focus of this work. We will present empirical evidence for their relevance and introduce the methodology commonly used to detect patterns that appear significantly more often than expected at random, so called *motifs*. It was hypothesized that they may have developed due to evolutionary advantages and might therefore also be functionally relevant [5, 6, 102, 129]. To test this conjecture, sound generative models are needed which successfully reproduce observed substructural features. To our best knowledge such models do not exist yet. We will discuss existing concepts as well as their shortcomings.

In Chapter 4, we will suggest a novel, very general framework to model such substructures using triadic entities as the building blocks of networks. The most basic realization of this new class of models, we call it the *triadic random graph model* (TRGM), will be investigated in detail in Chapter 5. In Chapter 6, we will extend the exististing motif-detection procedure to *node-specific pattern mining* (NoSPaM). The commonly used algorithm evaluates the abundance of subgraphs over the whole system. In contrast, our methodology considers the local neighborhood of every unit in the system separately and thus provides for a higher resolution. We will further present results of applying our novel tool to a variety of real-world datasets. The influence of triadic substructures on dynamical processes will be investigated in Chapter 7. For this purpose, we will use the triadic random graphs defined in Chapter 5 to examine how different substructures affect those processes.

# 2. Complex-Networks Theory

This chapter will focus on terminology, methodology, and tools in the context of complex network science that will be relevant in the following chapters of this work. For a broader introduction to complex-network science see e.g. the textbooks by Barabási [16] or Newman [108], or the review article by Boccaletti et al. [23].

## 2.1. Terminology

In this section the general terminology and definitions used in the remainder of this dissertation will be introduced.

### 2.1.1. Basic Entities and Types of Networks

*Nodes* are the fundamental units of networks. Their pairwise interactions are realized by *links*. Table 2.1 gives some examples for representations of these entities in real-world systems. Formally, there is a distinction between a network and its abstract representation, called the *graph*. A *graph* $\mathcal{G}(V, E)$ is described as a set of *vertices* (representing the nodes), $V$, and a set of *edges* (representing the links), $E$. However, in the scientific literature – including this dissertation – the terms network/graph, node/vertex, and link/edge are used interchangeably. Edges, and thus graphs, can be *directed* or *undirected*, *weighted*, or *signed*.

In directed networks links can be *unidirectional* or *bidirectional*. Unidirectional links have a source node and a target node. Consider e.g. Fig. 2.1(a) where website $A$ has a hyperlink to website $B$. Then $A$

## 2. Complex-Networks Theory

| Realization | Nodes | Links |
|---|---|---|
| social network | individuals | relationships between individuals |
| brain | neurons | synapses |
| airport connectivity | airports | flights |
| road map | cities | highways |
| subway system | stations | rails, subway lines |
| www | websites | hyperlinks |
| Internet | computers | telephone cables, optical fiber |
| publication citations | articles | citations between articles |
| author citations | authors | citations between authors (via articles) |
| co-authorship | authors | papers written together |
| electricity grid | power plants, consumers | high-voltage transmission lines |

**Table 2.1.:** Examples of the fundamental entities of a network in various realizations.

is the source and $B$ is the target node of a unidirectional link. If $A$ also has a hyperlink to page $C$ while $C$ also links to $A$, this is called a *bidirectional* link. In directed networks it is common to illustrate them with two arrowheads; in undirected ones the edges are drawn with no arrowheads whatsoever.

For some applications it is useful to consider weighted edges. The weight of an edge reflects its relevance or capacity, depending on the context. Suppose the toy network in Fig. 2.1(b) models neurons and their synaptic wiring. Then the state of neuron $C$ is much more determined by the state of neuron $B$ rather than $A$ or $D$.

In a social context it may be further relevant what the sentiment of a relationship is like. Fig. 2.1(c) shows the situation where both $A$ and $D$, as well as $B$ and $C$, feel sympathy for each other while $A$ and $B$, and $C$ and $D$ express mutual antagony. This is modelled by signed edges. Neither $A$ and $C$, nor $B$ and $D$ interact at all.

Moreover, it is possible to distinguish between different types of links. Consider e.g. Fig. 2.1(d) where the nodes could e.g. correspond to cities in a traffic network. The solid lines could indicate roads, the dashed lines rails, and the dotted line a flight connection. The system can be thought of as a composition of separate layers. Such networks are referred to as *multiplex networks* or *interconnected networks* whose intensive investigation has started only recently. In this example, there exists a path from any node in the graph to any other node. However, to get from node $A$ to node $D$ it is neccessary to switch layers.

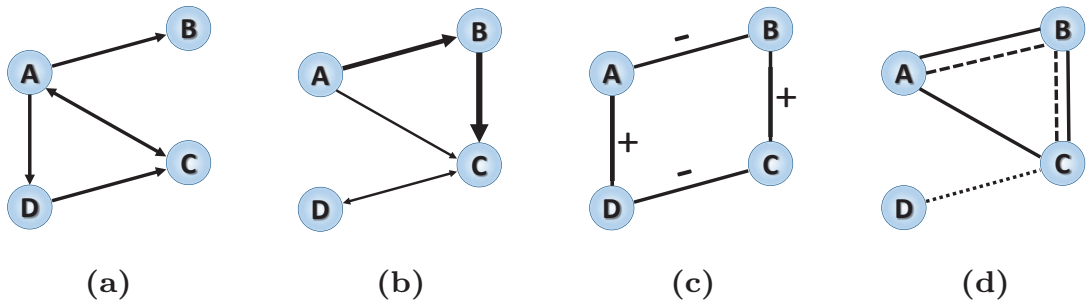

**Figure 2.1.:** Illustration of **(a)** an unweighted directed network, **(b)** a weighted directed network, **(c)** an unweighted undirected signed network, and **(d)** an unweighted undirected multiplex network.

The way in which systems are modelled as networks strongly depends on the goals, the available data, and the analytic tools that shall be applied. Usually, the knowledge about complex systems is incomplete. E.g. measured weights may vary in time and hence be unreliable. In such cases it can be useful to only model the link structure without accounting for weights. In addition, sometimes considering too many details may even prevent oneself from detecting existing structure on a higher level.

### 2.1.2. Network Substructures

The main focus of this work is on the role of substructures in complex networks. Their terminology shall be introduced now.

- A *dyad* is a set of two nodes. Hence, edges or links describe

## 2. Complex-Networks Theory

dyadic relationships. In a network with $N = |V|$ nodes there are $D = \binom{N}{2} = \frac{N(N-1)}{2}$ distinct dyads.

- Accordingly, a *triad* is a set of three nodes of which there are $T = \binom{N}{3} = \frac{N(N-1)(N-2)}{6}$ distinct ones.

- A *triple* is a *3-tuple* of nodes, i.e. an ordered list of three nodes.

- A *triangle* denotes three mutually interconnected nodes.

- A *subgraph* is a part of a network which considers only a subset of all nodes, including their mutual connections.

- A *subgraph configuration* is a specification of the connections in a subgraph, while accounting for node identities; e.g. dyad configuration $A \to B$ is distinct from dyad configuration $A \leftarrow B$.

- *Subgraph patterns* (or just *subgraphs*) are sets of nodes including their relations without accounting for node identities, i.e. isomorphic subgraph configurations are mapped to the same subgraph pattern; e.g. dyad pattern $A \to B$ is the same as dyad pattern $A \leftarrow B$. Apart from node permutations, there are 16 distinct triad patterns in directed unweighted networks as shown in Fig. 2.2. To illustrate the difference between configuration and pattern again, consider Fig. 2.3. Here, the three *configurations* corresponding to *pattern* number 4 are displayed.

- In *connected triadic subgraphs* all three participating vertices are part of at least one edge. In Fig. 2.2 this applies to patterns 4 through 16.

- A(n) *(anti)motif* is a subgraph pattern which is significantly over- (under-) represented in a network, as compared to some random null model. Motif analysis in networks will be discussed in detail in section 3.3.

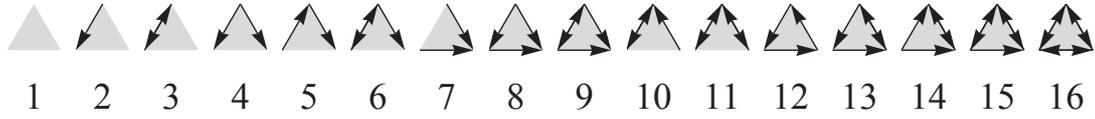

1 2 3 4 5 6 7 8 9 10 11 12 13 14 15 16

**Figure 2.2.:** All 16 possible non-isomorphic triadic subgraphs (subgraph patterns) in directed unweighted networks.

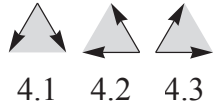

4.1 4.2 4.3

**Figure 2.3.:** The three isomorphic configurations belonging to pattern 4. Contrary to Fig. 2.2, here node identities matter.

### 2.1.3. Group Structure and Bipartite Networks

Many networks show community or group structure on a mesoscopic network level. The term *mesoscopic* refers to properties that are neither *local*, i.e. related to single (or very few) nodes, nor *global*, i.e. related to the system as a whole.

Group structure is for example ubiquitous in social systems on various scales. People can be grouped according to the continent they live on, their nationality, the region in a country or the city they live in. In this context, individuals are typically more likely to be linked to other members of the same group than to members of another group. To be linked can e.g. mean to be related or to know each other on a first-name basis. The detection of group structure is a powerful tool to complete only fragmentarily-known network structures, i.e. to predict as yet unobserved links [29, 76, 122].

There is another important type of network structure called *bipartite network*. In bipartite graphs, there are two groups of vertices of different types which exclusively connect to members of the other group. This is e.g. the case for product-consumer networks in which both products and customers are represented by nodes (see Fig. 2.4(a)). If a client buys an item, there is a link between their corresponding nodes. Bipartite graphs can be projected to unipartite graphs by connecting two vertices if they have a common neighbor in the other group. Fig. 2.4(b) and Fig. 2.4(c) show projections of the customer and the product graphs. Within the projections it is then again possible to find group structure. Concerning the example, individuals A and B may belong to the group of musicians, while C and D are parents of a baby. Amps and guitars are typical products for musicians, while parents of a new-born child are typical customers for baby food and diaper. Smart phones are popular among a wider range of people in our society.

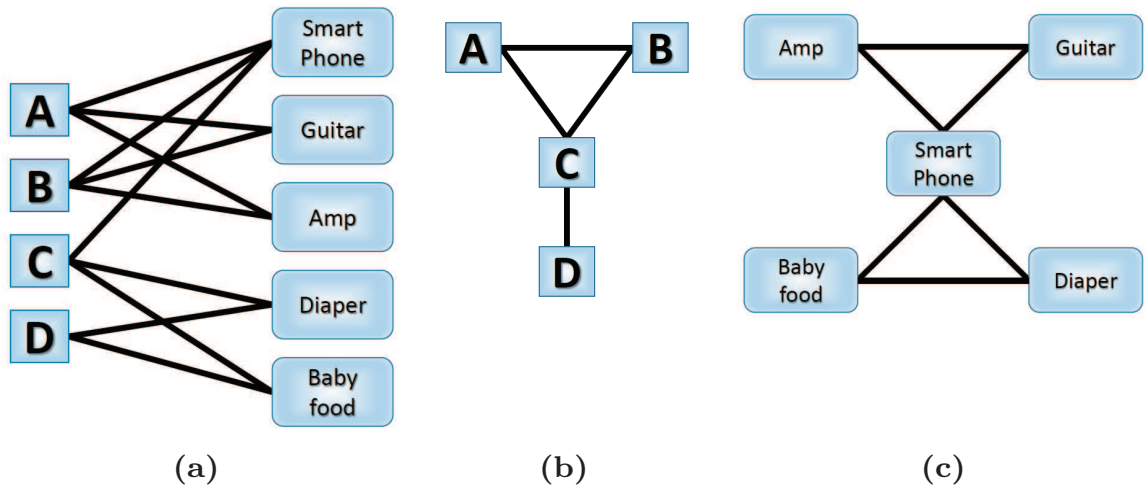

**Figure 2.4.:** **(a)** Bipartite network of consumers and products. **(b)** Projection of the consumer network. **(c)** Projection of the products network.

## 2.1.4. Adjacency Matrix and Coupling Matrix

Linear algebra is the language to describe the graph structure of complex networks mathematically. A network can be described by the so called *adjacency matrix* $\boldsymbol{A}$. It is:

$$A_{ij} = \begin{cases} 1 & \text{if there is a link from node } i \text{ to node } j \\ 0 & \text{otherwise} \end{cases} \quad (2.1)$$

For example, the adjacency matrix of the network shown in Fig. 2.1(a) is

$$\boldsymbol{A} = \begin{pmatrix} 0 & 1 & 1 & 1 \\ 0 & 0 & 0 & 0 \\ 1 & 0 & 0 & 0 \\ 0 & 0 & 1 & 0 \end{pmatrix}. \quad (2.2)$$

The adjacency matrix of undirected networks, $\boldsymbol{A}^{\text{undir}}$, is symmetric: $A_{ij}^{\text{undir}} = A_{ji}^{\text{undir}}$. Entries on the diagonal describe self interactions. If those are present in a system, this is usually specifically mentioned.

Weighted networks are modelled by a likewise weighted matrix,

$$G_{ij} = A_{ij} \, w_{ij} \quad (2.3)$$

where $w_{ij} \in \mathbb{R}$ indicates the strength of a link from node $i$ to node $j$. Negative weights apply to *signed networks*. We will term $\boldsymbol{G}$ the *coupling matrix* of a graph. In the literature, the matrix including the weights is sometimes referred to as the *weighted adjacency matrix*. For the sake of clarity and consistency, throughout this work, the adjacency matrix comprises the information about the presence and absence of links in the graph, while the weight is considered separately or included in a coupling matrix. As an example, the coupling matrix of Fig. 2.1(b) is given by

$$\boldsymbol{A} = \begin{pmatrix} 0 & 1 & 0.5 & 0 \\ 0 & 0 & 2 & 0 \\ 0 & 0 & 0 & 0.5 \\ 0 & 0 & 0.5 & 0 \end{pmatrix}. \quad (2.4)$$

## 2.2. Network Measures

In order to characterize and compare networks with each other, many analysis tools have been established. There are measures characterizing nodes, edges, or the whole network. In this section we will summarize those that will be relevant in the forthcoming chapters.

### 2.2.1. Node Degree

The most fundamental node measure is the *degree*. It denotes the number of adjacent edges to the node. In directed networks it is distinguished between *in* and *out degrees*. Both the in and out degree of a node $i$, $k_i^{in}$ and $k_i^{out}$, can be expressed in terms of the (unweighted) adjacency matrix $\boldsymbol{A}$:

$$
\begin{aligned}
k_i^{in} &= \sum_{j=1}^{|V|} A_{ji} \\
k_i^{out} &= \sum_{j=1}^{|V|} A_{ij}.
\end{aligned}
\tag{2.5}
$$

In undirected networks it is

$$k_i \equiv k_i^{in} = k_i^{out}. \tag{2.6}$$

### 2.2.2. Density of a Network

The *density*, $p$, of a graph $\mathcal{G}(V, E)$ indicates the ratio of all possible links which are actually realized. For undirected networks it is

$$p = \frac{|E|}{\binom{|V|}{2}} = \frac{2|E|}{|V|(|V|-1)}. \tag{2.7}$$

For directed networks it is

$$p = \frac{|E|}{|V|(|V|-1)} \tag{2.8}$$

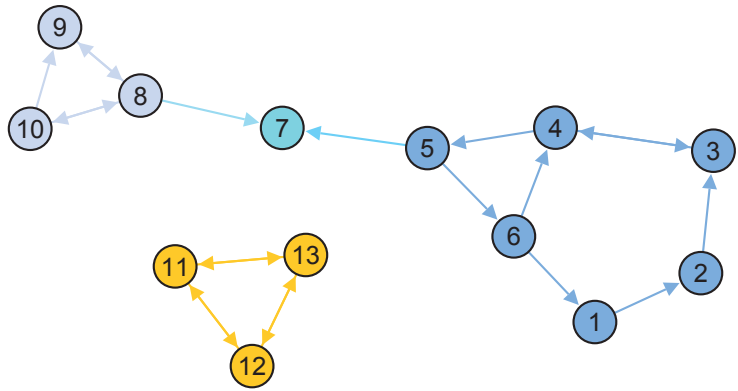

**Figure 2.5.:** Network with two weakly connected components (vertices 1-10 and 11-13) and four strongly connected components (vertices 1-6, 7, 8-10, and 11-13).

since every dyad may hold edges in either direction.

### 2.2.3. Connected Components

In undirected networks, *connected components* describe subsets of the graph for which there is a path between any pair of nodes. Subsets in which all pairs of nodes are directly linked are called *cliques*. In directed networks one can distinguish between *strongly* and *weakly connected components*.

The number and size of connected components is important for all kinds of dynamical processes, e.g. the propagation of information or diseases in social networks.

*Strongly connected components* of a directed graph are defined as a subset of nodes, $\mathcal{S} \subseteq \mathcal{G}$ such that for every pair of nodes $n_i, n_j \in \mathcal{S}$ there is a directed path both, from $n_i$ to $n_j$, and from $n_j$ to $n_i$. Fig. 2.5 illustrates the concept of connected components. There are four strongly connected components. The single node number 7 has no outgoing edges. It thus comprises a connected component with only one element on its own.

A special case of a strongly connected component is the *complete*

*2. Complex-Networks Theory*

*graph* (also referred to as the fully-connected or completely-connected network). In the complete graph, there is a bidirectional link between every pair of nodes.

For *weakly-connected components*, $\mathcal{W}$, the directionality of edges is being ignored. It is then sufficient if there is an undirected path connecting any pair of nodes $n_i, n_j \in \mathcal{W}$. Hence, in Fig. 2.5 there are two weakly connected components.

### 2.2.4. Clustering Coefficient

*Clustering coefficients* are important measures to characterize a network. They indicate to which extent the neighbors of a given vertex are also connected with each other, e.g., in a social context, how likely it is that two of your acquaintances also know each other.

Clustering coefficients were originally introduced for *undirected unweighted* networks [149]. Generalizations have been made for weighted [18, 127], directed [43], signed [31], and multiplex networks [36]. However, those generalizations are not unique. In Fig. 2.2, e.g. either of patterns 8, 9, and 12-16 may be considered as triangles. Therefore, when comparing clustering coefficients among directed and/or weighted networks one should be aware of the respective methodology applied.

In the following definitions we will stick with *undirected unweighted* networks.

**Local Clustering Coefficient**

The local *clustering coefficient* of a node $i$, as suggested by Watts and Strogatz in 1998 [149], is the fraction of pairs of node $i$'s neighbors that are also connected. It can be obtained from the adjacency matrix $\boldsymbol{A}$ by

$$C_i = \frac{\frac{1}{2}\sum_{j,k=1}^{|V|} A_{ij} A_{jk} A_{ki}}{\binom{k_i}{2}} = \frac{\left(\boldsymbol{A}^3\right)_{ii}}{k_i\left(k_i-1\right)}. \tag{2.9}$$

The factor $\frac{1}{2}$ in the numerator arises due to double counting of triangles. For example, suppose there is a triangle between nodes 1, 2, and 3, then both $A_{12}A_{23}A_{31}$ and $A_{13}A_{32}A_{21}$ contribute to the summation.

**Average Clustering Coefficient**

To evaluate the abundance of closed triads over whole graphs, Watts and Strogatz suggested to compute the average of the local clustering coefficients in (2.9) over all nodes in the network.

$$\langle C \rangle = \frac{1}{|V|} \sum_{i=1}^{|V|} C_i \qquad (2.10)$$

**Global Clustering Coefficient**

Another possible definition to measure the relative occurrence of closed triads is the global clustering coefficient. It is defined as the total ratio of nodes that share a common neighbor and are also connected.

$$C_{\text{global}} = \frac{\frac{1}{2} \sum_{i,j,k=1}^{|V|} A_{ij} A_{jk} A_{ki}}{\sum_{i=1}^{|V|} \frac{1}{2} k_i (k_i - 1)} = \frac{\text{Trace}\left(\boldsymbol{A}^3\right)}{\sum_{i=1}^{|V|} k_i (k_i - 1)} \qquad (2.11)$$

The numerator counts all triangles in the system three times (one time for evey participating node). The denominator counts all pairs of neighbors for every node $i$. The *average* clustering coefficient averages $C_i$ over all nodes, while the *global* clustering coefficient averages over the numerator and the denominator of the $C_i$ separately.

In the scientific literature, all of the mentioned clustering coefficients appear. Hence, one should pay attention to the exact definition being used. All of the above clustering coefficients have in common that $C = 0$ corresponds to a network without any triangles, e.g. a tree or a square lattice. Furthermore, $C = 1$ indicates a graph which is entirely composed of cliques.

## 2. Complex-Networks Theory

### 2.2.5. Shortest Paths

The distances between pairs of nodes in a graph are described by *shortest paths*. In a graph $\mathcal{G}(V, E)$, a *path* is a series of nodes $n_i \in V$, $\mathcal{P} = (n_1, n_2, ..., n_{k-1}, n_k)$, such that there exists a link between all consecutive nodes, i.e. $(n_i, n_{i+1}) \in E$.

The number of paths of length 2 from node $i$ to node $j$, can be expressed in terms of the adjacency matrix $\boldsymbol{A}$:

$$N_{ij}^{(2)} = \sum_k A_{ik} A_{kj} = (\boldsymbol{A}^2)_{ij} \qquad (2.12)$$

This expression can be generalized to paths of arbitrary length $l$:

$$N_{ij}^{(l)} = (\boldsymbol{A}^l)_{ij} \qquad (2.13)$$

For $l = 3$, the diagonal elements $(\boldsymbol{A}^3)_{ii}$ can be used to efficiently estimate the number of closed triangles in large graphs by computing the largest eigenvalues of $\boldsymbol{A}$ [136, 138].

The length of the shortest path, or *geodesic distance*, between two nodes $i$ and $j$ is denoted as $d(i, j)$. It shall be emphasized that in directed graphs this distance is not necessarily symmetric, i.e. $d(i, j) \neq d(i, j)$. The average shortest path length of a graph is then given by

$$l_\mathcal{G} = \frac{1}{|V|\,(|V| - 1)} \sum_{i \neq j} d(i, j). \qquad (2.14)$$

It shall not be confused with the *diameter* of a graph, which is the maximum of the geodesic distances over all pairs of vertices.

### 2.2.6. Betweenness Centrality

The *betweenness centrality* (or just *betweenness*) of a node $i$ measures its structural relevance, in particular for spreading processes. Introduced by Freeman [46] and independently by Anthonisse [13] in the 1970s, it indicates the weighted number of shortest paths in a graph that traverse node $i$.

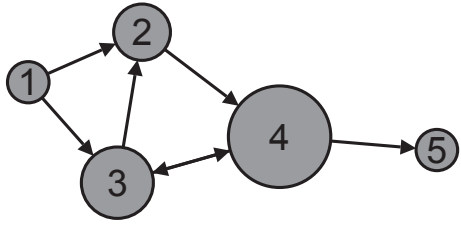

**Figure 2.6.:** Example network with vertex sizes being scaled with respect to the corresponding betweenness centralities.

The number of distinct geodesics, i.e. shortest paths, between a source node $s$ and a target node $t$ be denoted as $g_{st}$. The number of those geodesics between $s$ and $t$ that pass through node $i$ be denoted as $g_{st}^i$. Then the betweenness of node $i$ is defined as

$$b_i = \sum_{s \neq i} \sum_{t \neq i \wedge t \neq s} \frac{g_{st}^i}{g_{st}}. \qquad (2.15)$$

By convention, it is $\frac{g_{st}^i}{g_{st}} = 0$ if $g_{st} = 0$, i.e. there is no path from $s$ to $t$.

Fig. 2.6 illustrates the concept of betweenness centrality. Since node 1 has no incoming edges and node 5 has no outgoing ones, their betweenness is zero. Node two is part of both a shortest path from node 1 to 4 and from 1 to 5. However, there are alternative geodesics via node 3. Therefore, the betweenness of node 2 is $2 \times 0.5 = 1$. In addition to the geodesics between 1 and 4, and 1 and 5, node three is also part of the geodesic from 4 to 2 and its betweenness is 2. Finally, node 4 is part of all shortest paths from 1 to 5, 2 to 3, 2 to 5, and 3 to 5. Hence, it is $b_4 = 4$.

## 2.3. Network Models

To assess characteristic structural features of real network representations, it is important to compare them to the expectations of suitable *null models*. Null models match the system under investigation with respect to some specifically modelled characteristics $\{a_k\}$. Apart from these properties they are random.

A particularly important term when dealing with probabilistic models is the network *ensemble*. Network models provide for rules to construct graphs and typically involve several parameters. For fixed parameter values, an ensemble is the universe of all network instances that can be generated by the model, weighted with their probability of generation. When dealing with the properties of ensembles it is important to be clear that these properties are *expectation values* for the ensemble and do not necessarily match the properties of every single instance.

Appropriate null models allow to unravel the impact of properties $\{a_k\}$ on other properties $\{b_k\}$. Suppose, e.g., a graph $\mathcal{G}_{real}$ with a certain density $a_1 = p_{real}$ shall be examined. Its clustering coefficient is found to be $b_1 = C_{real}$. To unravel whether the observed clustering coefficient differs at all from complete randomness, a first step will be to evaluate the expected clustering coefficient in an *ensemble* of systems with the same density as $\mathcal{G}_{real}$. Except for the expected density, the ensemble shall be absolutely random. The class of models that fixes only the density is named after the two Hungarian mathematicians Erdös and Rényi [41, 42] and will be discussed in Section 2.3.1. We will further introduce the *configuration model* which allows for modelling specific degree sequences such as scale-free degree distributions which follow power laws.

### 2.3.1. Erdös-Rényi Model

The ensemble of undirected[1] random graphs only parametrized by their average link density was studied intensively by Erdös and Rényi in the 1950s and 1960s [41, 42]. It is among the most commonly used benchmarks to detect structural features in networks. Properties that differ significantly from the expectations for the corresponding Erdös-Rényi (ER) model suggest that there may be some (potentially functional) reason that led to their emergence. Conversely, everything that looks like a typical ER graph has most likely ocurred by sole coincidence. Therefore, it is of fundamental importance for complex-networks analysis to be familiar with the typical structural aspects of ER random graphs.

ER graph ensembles with $|V| = N$ nodes and each possible link being present with probability $p$ are typically denoted as $\mathcal{G}(N, p)$. Formally, the probability distribution for observing the network with adjacency matrix $\boldsymbol{A}$, conditioned on the parameter $p$, is then given by

$$\mathcal{P}(\boldsymbol{D} = \boldsymbol{A}|p) = \prod_{i=1}^{N-1} \prod_{j=i+1}^{N} \mathcal{P}(D_{ij} = D_{ji} = A_{ij} = A_{ji}|p)$$
$$= \prod_{i=1}^{N-1} \prod_{j=i+1}^{N} p^{A_{ij}} (1-p)^{1-A_{ij}} \quad (2.16)$$

where $\boldsymbol{D}$ denotes the *random variables* of the distribution and $\boldsymbol{A} \in \{0,1\}^N \times \{0,1\}^N$ their *values*. Because of the dyadic independence of the link-formation process the distribution factorizes.

The distribution in Eq. (2.16) covers all possible $2^{\binom{N}{2}}$ configurations. The probability of observing a network with $|E| = M$ edges follows the binomial distribution

$$\mathcal{P}(M) = \binom{\binom{N}{2}}{M} p^M (1-p)^{\binom{N}{2} - M} \quad (2.17)$$

---

[1] Although the classical ER model is undirected, it can easily be generalized to directed networks by setting any existing link to be unidirectional with probability $p_u$ and bidirectional with probability $p_b = 1 - p_u$. For unidirectional links between two vertices $V_1$ and $V_2$ the two configurations $V_1 \to V_2$ and $V_2 \to V_1$ shall then be equally likely.

*2. Complex-Networks Theory*

with expectation value

$$\langle M \rangle = \binom{N}{2} p = \frac{N(N-1)}{2} p \propto N^2 p \qquad (2.18)$$

and standard deviation

$$\sigma_M = \sqrt{\frac{N(N-1)}{2} p (1-p)} \propto N p. \qquad (2.19)$$

In the limit $N \to \infty$ the probability distribution has a sharp peak around $\langle M \rangle$. Hence, for large systems a definition of the ER model in terms of $\mathcal{G}\left(N, M \approx \frac{N(N-1)}{2} p\right)$, i.e. with a fixed number of edges, is equivalent to $\mathcal{G}(N, p)$. Of course, the number of edges, $M$, has to be specified as a positive integer value.

Since every node has $(N-1)$ potential neighbors to link to, the expected degree of each node is

$$\langle k \rangle = (N-1) p. \qquad (2.20)$$

The degrees of ER graphs are Poisson distributed. To derive this result consider that for a node to have degree $k = \kappa$ it must link to $\kappa$ nodes and **not** link to the remaining $N - 1 - \kappa$ nodes. Since there are $\binom{N-1}{\kappa}$ possibilities to realize a degree of $\kappa$ it is

$$\begin{aligned} \mathcal{P}(k = \kappa) &= \binom{N-1}{\kappa} p^\kappa (1-p)^{N-1-\kappa} \\ &= \frac{(N-1)!}{\kappa!\,(N-1-\kappa)!} \left(\frac{\langle k \rangle}{N-1}\right)^\kappa \left(1 - \frac{\langle k \rangle}{N-1}\right)^{N-1-\kappa}. \end{aligned} \qquad (2.21)$$

In the second row we inserted Eq. (2.20) to eliminate $p$. For large,

sparse networks, i.e. $N \to \infty$ and $\langle k \rangle < \infty$, we can write

$$\lim_{N \to \infty} \mathcal{P}(k = \kappa) = \lim_{N \to \infty} \underbrace{\frac{(N-1) \times (N-2) \times ... \times (N-\kappa)}{(N-1)^{\kappa}}}_{\to 1} \frac{\langle k \rangle^{\kappa}}{\kappa!}$$

$$\times \underbrace{\left(1 - \frac{\langle k \rangle}{N-1}\right)^{N-1}}_{\to e^{-\langle k \rangle}} \underbrace{\left(1 - \frac{\langle k \rangle}{N-1}\right)^{-\kappa}}_{\to 1}$$

$$= \frac{\langle k \rangle^{\kappa}}{\kappa!} e^{-\langle k \rangle} \tag{2.22}$$

i.e., the node degrees are Poisson distributed.

The calculation of the local clustering coefficients is straightforward. The probability that two neighbors of a particular node are also connected is $p$. Thus, it is $C_i^{\text{ER}} = p$ for all $i$ and therefore also $\langle C^{\text{ER}} \rangle = p$. Due to the homogeneity of the system, $C^{\text{global}}$ is also $p$.

$$C^{\text{ER}} = p = \frac{\langle k \rangle}{N-1} \tag{2.23}$$

This implies that large ER systems with finite average degree have vanishing clustering coefficients. A comparison of this clustering coefficient to those found in real datasets can be found on page 48.

### 2.3.2. Configuration Model

We have seen in Eq. (2.22) that ER graphs have a Poissonian degree distribution. However, generally this is not the case for real-world networks. Fig. 2.7 shows a variety of degree distributions from diverse fields. For comparison, we have also included the expected degree sequence for a graph with the same density as the displayed systems. It can be observed that the distributions are considerably broader than

*2. Complex-Networks Theory*

the Poissonians, while both low and high degrees occur with a comparatively high probability. In particular the *heavy tails*[2] of the distributions characterize many real systems, indicating that there are some nodes with exceptionally high degree. Some of them are even very well described by power laws, as can be seen from the double-logarithmic scale in Fig. 2.7.

In order to detect features of a network that are not determined by the degree distribution, it is important to have null models which are random except for their fixed degree sequences. Any properties of the network that deviate from the expectation for these null models cannot be sufficiently explained in terms of the node degrees alone. The *configuration model* defines such a null model for undirected graphs [103, 104].

In fact, the configuration model fixes the degree $k_i$ of every node $i$. Therefore, $i$ gets assigned $k_i$ 'stubs' or 'half edges'. An instance of the ensemble is then generated by matching the available stubs, i.e. two stubs are picked uniformly at random and connected until none of them are left. Since $\sum_i k_i = 2M$, the number of stubs has to be even.

It shall be noticed that the configuration model does neither prevent the formation of self edges, nor the creation of multiple edges between a pair of nodes. Of course, one could forbid their generation in the matching process. However, such networks would no longer be uniformly drawn from the set of all possible matches, invalidating many analytically known results for the configuration model [108, 111].

For all pairs of nodes $i$ and $j$ the probability of observing an edge connecting the two is

$$p_{ij} \approx \frac{k_i \, k_j}{2M}. \qquad (2.24)$$

The approximation becomes exact in the limit of large $M$ and given $k_i$ and $k_j$ [108]. It shall be mentioned that $p_{ij}$ does not depend on the presence of other edges in the system. The configuration model therefore falls into the class of *dyadic-independence models*.

---

[2] A distribution is called *heavy tailed* if their probabilities decay slower than exponentially.

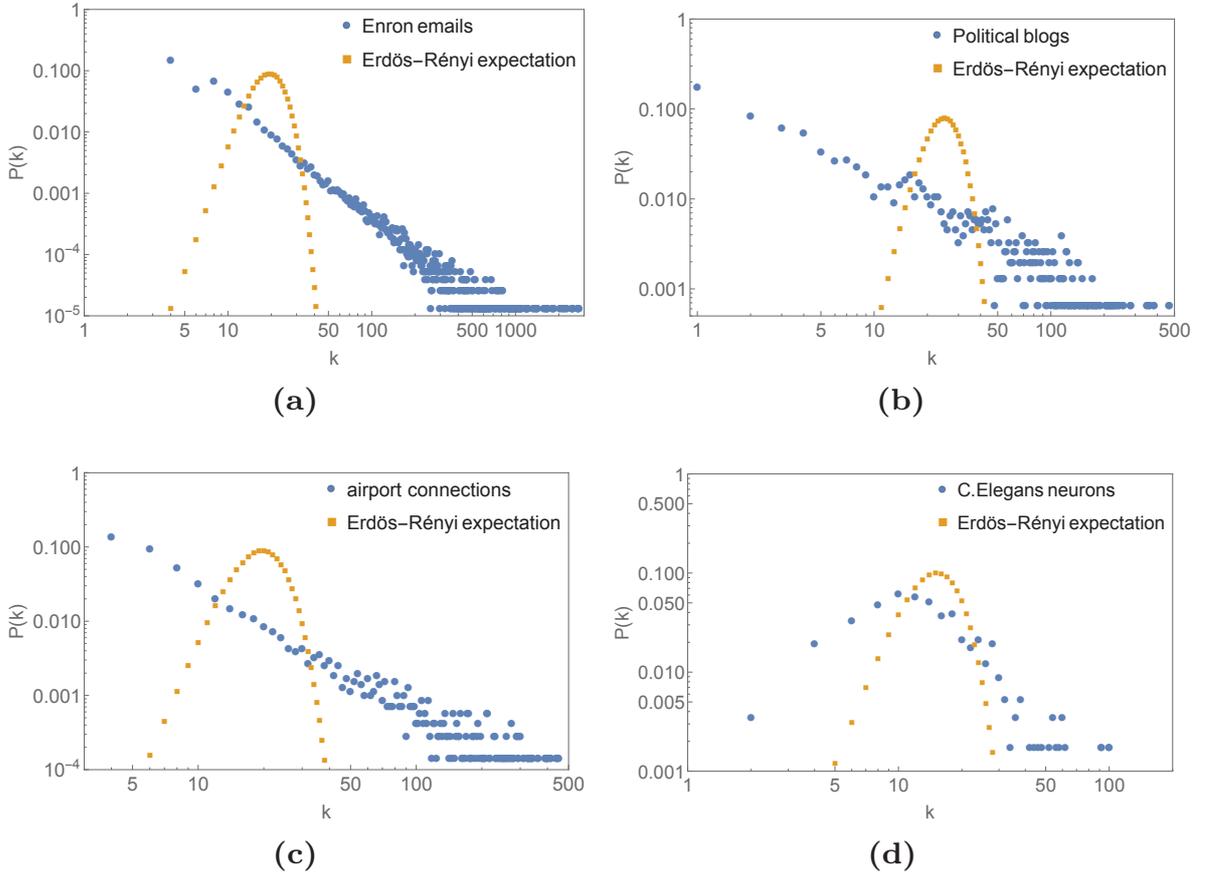

**Figure 2.7.:** Degree distributions observed in real-world networks compared to the expectation of the ER graph with the same link density as the examined system. Notice the double-logarithmic scale. In real systems, the degree distribution is typically much broader. **(a)** Email-communication network from Enron [88]. **(b)** Hyperlinks between political blogs [1, 109]. **(c)** Network of non-stop plane routes between airports [117]. **(e)** Neural network of *C. elegans* [109, 149, 150]. For details on the datasets, see page 167.

The clustering coefficient of the configuration model is (see e.g. [23, 108])

$$C_{\text{glob}}^{\text{config}} = \frac{1}{N} \frac{\left(\langle k^2 \rangle - \langle k \rangle\right)^2}{\langle k \rangle^3}. \qquad (2.25)$$

Like the ER model $\mathcal{G}(N, M)$ fixes the number of edges, the configuration model fixes the degree of each node. In analogy to the ER model $\mathcal{G}(N, p)$, one can define a model with expected degrees, where each connection between nodes $i$ and $j$ be present with probability $p_{ij} = c_i c_j / (2M)$. Furthermore, the configuration model can be generalized to directed graphs [38, 111]. In this case, every node gets assigned both 'in- and out-stubs' separately. During the matching process the two different types are connected with each other.

## 2.4. Exponential Random Graph Models

Both the Erdös-Rényi model $\mathcal{G}(N, p)$ and the model with expected node degrees are part of a broad class of models known as *exponential random graphs*, especially in the social-science literature sometimes also referred to as p* (p star) models [22, 45, 64, 125, 147]. These allow us to generate ensembles of networks with specified structural characteristics. This is useful, for example to test the impact of the latter on the evolution of dynamical processes. In Chapter 4 and 5 we will use the concept of exponential random graph models (ERGMs) to introduce novel network models based on triadic subgraphs.

ERGMs describe ensembles of graphs with preassigned expectation values for some network statistics $\{s_k\}$. These statistics may contain any kind of graph measures, e.g. the density, the degree-distribution, or the clustering coefficient. The probability for the random variable of adjacency matrices, $\boldsymbol{D} \in \{0,1\}^N \times \{0,1\}^N$, to assume a value $\boldsymbol{A}$ be denoted as $\mathcal{P}(\boldsymbol{D} = \boldsymbol{A})$. The distribution $\mathcal{P}(\boldsymbol{D})$ is normalized,

$$\sum_{\{\boldsymbol{D}\}} \mathcal{P}(\boldsymbol{D}) = 1, \qquad (2.26)$$

where the summation runs over all possible $N \times N$ matrices with entries 0 or 1. We want to fix the expectation value for all modeled statistics $s_k$ over the graph ensemble to the value of a given network with adjacency matrix $\boldsymbol{A}$. These constraints to the probability distribution can then be expressed via

$$\langle s_k \rangle = \sum_{\{\boldsymbol{D}\}} \mathcal{P}(\boldsymbol{D})\, s_k(\boldsymbol{D}) \stackrel{!}{=} s_k(\boldsymbol{A}). \tag{2.27}$$

Except for the chosen measures, the ensemble of generated networks shall be maximally random. This can be guaranteed by maximizing the entropy

$$S = -\sum_{\{\boldsymbol{D}\}} \mathcal{P}(\boldsymbol{D}) \ln \mathcal{P}(\boldsymbol{D}) \tag{2.28}$$

under consideration of Eq. (2.26) and the $\boldsymbol{A}$-dependent constraints of Eq. (2.27). In order to account for the constraints, we use the *Lagrange multipliers* $\alpha$ and $\{\theta_k\}$ to define the modified entropy

$$\widetilde{S} = -\sum_{\{\boldsymbol{D}\}} \mathcal{P}(\boldsymbol{D}) \ln \mathcal{P}(\boldsymbol{D}) + \alpha \left( \sum_{\{\boldsymbol{D}\}} \mathcal{P}(\boldsymbol{D}) - 1 \right)$$
$$+ \sum_k \theta_k \left( \sum_{\{\boldsymbol{D}\}} \mathcal{P}(\boldsymbol{D})\, s_k(\boldsymbol{D}) - s_k(\boldsymbol{A}) \right). \tag{2.29}$$

If the conditions in Eqs. (2.26) and (2.27) are fulfilled, it is $\widetilde{S} = S$. Maximizing $\widetilde{S}$ with respect to $\mathcal{P}(\boldsymbol{D})$ then yields

$$\begin{aligned} 0 &\stackrel{!}{=} \frac{\partial \widetilde{S}}{\partial \mathcal{P}(\boldsymbol{D})} \\ &= -\ln \mathcal{P}(\boldsymbol{D}) - 1 + \alpha + \sum_k \theta_k\, s_k(\boldsymbol{D}). \end{aligned} \tag{2.30}$$

## 2. Complex-Networks Theory

The derivatives of the entropy in terms of $\alpha$ and $\{\theta_k\}$ require the constraints of Eqs. (2.26) and (2.27), respectively, to be fulfilled. From Eq. (2.30) we find the entropy to be maximized for

$$\mathcal{P}\left(\boldsymbol{D}|\vec{\theta}\left[\vec{s}(\boldsymbol{A})\right]\right) = \frac{1}{Z} \exp\left(\vec{\theta}\left[\vec{s}(\boldsymbol{A})\right] \cdot \vec{s}(\boldsymbol{D})\right) \tag{2.31}$$

where we substituted the parameter $\alpha$ via $Z \equiv e^{1-\alpha}$. To fix the Lagrange multipliers $\theta_k$, one needs to insert Eq. (2.31) in Eq. (2.27) and solve for the $\theta_k$. Therefore, the $\theta_k$ depend on the statistics of the given adjacency matrix, $\vec{s}(\boldsymbol{A})$. From Eq. (2.26) we see that $Z$ is the partition function over all $2^{N(N-1)}$ possible adjacency matrices,

$$Z = \sum_{\boldsymbol{D}} e^{\vec{\theta}[\vec{s}(\boldsymbol{A})] \cdot \vec{s}(\boldsymbol{D})} = \prod_{i \neq j} \sum_{D_{ij}=0}^{1} e^{\vec{\theta}[\vec{s}(\boldsymbol{A})] \cdot \vec{s}(\boldsymbol{D})}. \tag{2.32}$$

Since ERGMs define Boltzmann-distributed ensembles over all networks with a given size $N$, they are referred to as *exponential* random graphs.

From a statistical-physics point of view the scalar product of $\vec{\theta}[\vec{s}(\boldsymbol{A})]$ and $\vec{s}(\boldsymbol{D})$ can be interpreted as the negative of a Hamiltonian

$$\mathcal{H} = -\vec{\theta}[\vec{s}(\boldsymbol{A})] \cdot \vec{s}(\boldsymbol{D}) \tag{2.33}$$

acting on the binary variables $D_{ij} \in \{0, 1\}$ of the adjacency matrices. This is equivalent to a Hamiltonian of likewise binary spin variables that can be either upward or downward oriented. The $\theta_k$ can then be interpreted as external fields which control their conjugate generalized forces $s_k$, like e.g. the magnetic field, $\vec{B}$, controls the magnetization $\vec{M}$. Furthermore, the $\theta_k$ can describe interdependencies between the occurrences of edges, like, for instance, the elements of a coupling matrix $\boldsymbol{J}$ model the interaction between spins. Suppose, e.g., that $s_1(\boldsymbol{D}) = D_{ij}D_{lm}$, then $\theta_1$ controls the probability for the simultaneous appearance of the link between nodes $i$ and $j$ and the link between nodes $l$ and $m$.

Defining further the *free energy* as

$$F = -\ln Z \qquad (2.34)$$

and inserting Eq. (2.31) in Eq. (2.28), with $E = \langle \mathcal{H} \rangle$, we obtain the relation

$$F = E - S. \qquad (2.35)$$

This formalism is in perfect analogy with the theory of statistical mechanics at constant temperature set to $T = 1$. We are then able to evaluate expectation values for all graph statistics, $s_k$, of the distribution for $\mathcal{P}\left(\boldsymbol{D}|\vec{\theta}\right)$ from the partition function and the respective free energy,

$$\langle s_k \rangle = -\frac{\partial F}{\partial \theta_k}. \qquad (2.36)$$

Conversely, the equation will facilitate to fix the parameters $\theta_k$ in Eq. (2.31) such that the $\langle s_k \rangle$ match their desired values $s_k(\boldsymbol{A})$. If this is the case, the adjacency matrix $\boldsymbol{A}$ is a typical instantiation of the ensemble defined by $\mathcal{H}$.

If the Hamiltonian, i.e. the graph statistics, contains only sums of entries of the adjacency matrix, e.g. the total number of edges or the degrees of nodes, the probability distribution in Eq. (2.31) factorizes

$$\mathcal{P}\left(\boldsymbol{D}|\vec{\theta}\right) = \prod_{i \neq j} \mathcal{P}\left(D_{ij}|\vec{\theta}\right). \qquad (2.37)$$

In this case, the probability for observing an edge is independent from the presence or absence of other edges. This can be seen in analogy to non-interacting spins in a magnetic field.

Models obeying Eq. (2.37), like ER graphs or the configuration model, for instance, are therefore called *dyadic independence models*. Both the inference of the parameters and eventually sampling from the learned probability distribution are tremendously facilitated for models showing dyadic independence. Instead of having to sample from the $2^{N(N-1)}$

*2. Complex-Networks Theory*

possible directed adjacency matrices, e.g., one only needs to sample $N(N-1)$ binary variables indicating whether the corresponding links are present or not. It shall be emphasized that – even for dyadic-independence models – in general the probability of creating an edge between $i$ and $j$ can differ from the one of creating an edge between $k$ and $l$, $\mathcal{P}\left(D_{ij}|\vec{\theta}\right) \neq \mathcal{P}\left(D_{kl}|\vec{\theta}\right)$. Eq. (2.37) only states that the events of creating edges are independent of each other.

### 2.4.1. Specified Density

The process of parameter inference shall be illustrated with the most simple ERGM, the Erdös-Rényi graph. The only statistics specified by the ER model is the density, $p$, of the graph or, equivalently, the expected number of undirected edges $\langle M \rangle = \frac{1}{2} N (N-1) p$. The Hamiltonian is thus

$$\mathcal{H}^{\text{ER}} = -\theta\, M(\boldsymbol{D}) = -\theta \sum_{i<j} D_{ij}. \qquad (2.38)$$

Because of the dyadic independence, the partition function can be simplified as follows:

$$\begin{aligned} Z^{\text{ER}} &= \prod_{i<j} \sum_{D_{ij}=0}^{1} e^{\theta\, D_{ij}} = \prod_{i<j} \left(1 + e^{\theta}\right) \\ &= \left(1 + e^{\theta}\right)^{\frac{N(N-1)}{2}}. \end{aligned} \qquad (2.39)$$

From the partial derivative of the free energy

$$F^{\text{ER}} = -\ln Z^{\text{ER}} = -\frac{N(N-1)}{2} \ln\left(1 + e^{\theta}\right) \qquad (2.40)$$

we obtain the relation between the density and the parameter $\theta$,

$$\begin{aligned}\langle M\rangle &= \frac{N(N-1)}{2}p = -\frac{\partial}{\partial \theta}F^{\mathrm{ER}} \\ &= \frac{N(N-1)}{2}\frac{e^{\theta}}{1+e^{\theta}} \\ \Leftrightarrow \theta &= \ln\frac{p}{1-p}.\end{aligned} \quad (2.41)$$

### 2.4.2. Specified Node Degrees

The Hamiltonian of the ERGM that specifies the expected degree of every single node in an undirected graph is given by

$$\begin{aligned}\mathcal{H} &= -\sum_{i}\theta_i\, k_i\,(\boldsymbol{D}) = -\sum_{i,j}\theta_i\, D_{ij} \\ &= -\sum_{i<j}(\theta_i+\theta_j)\, D_{ij}\end{aligned} \quad (2.42)$$

and yields the partition function [108]

$$\begin{aligned}Z &= \prod_{i<j}\sum_{D_{ij}=0}^{1} e^{(\theta_i+\theta_j)\, D_{ij}} \\ &= \prod_{i<j}\left(1+e^{\theta_i+\theta_j}\right).\end{aligned} \quad (2.43)$$

From the partial derivative of the free energy with respect to $\theta_m$,

$$\begin{aligned}\frac{\partial}{\partial \theta_m}F &= \frac{\partial}{\partial \theta_m}\sum_{i<j}\ln\left(1+e^{\theta_i+\theta_j}\right) \\ &= \sum_{i}\frac{e^{\theta_i+\theta_m}}{1+e^{\theta_i+\theta_m}} = \langle k_m\rangle,\end{aligned} \quad (2.44)$$

we obtain the set of equations to relate the desired expected degrees of nodes, $k_m$, to their parameters $\theta_m$ via $\langle k_m\rangle \stackrel{!}{=} k_m$.

## 2. Complex-Networks Theory

The expectations for the degree of single nodes depend on all other parameters of the model. Thus, Eq. (2.44) cannot be solved for $\theta_m$ explicitly. This problem can be overcome by iteratively updating the parameters. The goal is to achieve $k_m = \langle k_m \rangle$ for all $m$. This can be rewritten as

$$1 = \frac{k_m}{\langle k_m \rangle} = \frac{k_m}{\sum_i \frac{e^{\theta_i + \theta_m}}{1 + e^{\theta_i + \theta_m}}}. \tag{2.45}$$

With $\vec{\theta}^t$ being the vector of parameters at iteration step $t$ we can write

$$\exp\left(\theta_m^{t+1}\right) = \frac{k_m}{\sum_i \frac{e^{\theta_i^t + \theta_m^t}}{1 + e^{\theta_i^t + \theta_m^t}}} \exp\left(\theta_k^t\right). \tag{2.46}$$

If the expected degree is smaller (larger) than the desired one the corresponding parameter is increased (decreased). At convergence, all parameters $\{\theta_m\}$ have been learned properly.

Fig 2.8(a) gives an example of an undirected adjacency matrix, $\boldsymbol{A}$, that is used to learn the degree-fixing ERGM, i.e. the statistics $k_m(\boldsymbol{A})$ are used to infer the parameters of the model. Since the model is dyadically independent, instead of having to sample from a distribution in the space of the $2^{\frac{7 \cdot 6}{2}} \approx 2.1 \times 10^6$ possible adjacency matrices, we can simply draw every edge $e_{ij}$ $(i < j)$ with the learned probability specified at the corresponding position in the matrix of Fig 2.8(b).

Moreover, exponential random graph models are widely used to predict hitherto unknown links [29, 92, 122]. An ERGM whose parameters have been learned from an adjacency matrix, $\boldsymbol{A}$, defines an ensemble of graphs for which $\boldsymbol{A}$ is a typical instance. The underlying assumption for using ERGMs for link-prediction tasks is that other matrices, $\boldsymbol{D}$, that are sampled from the ensemble with high probability are also in reality more likely to be observed than others. For dyadically independent ERGMs this even transfers to single links. From the links that are not present in the training matrix in Fig. 2.8(a), with 32% probability, the one between nodes 4 and 5 would be the best guess from the ERGM that learned the degrees of $\boldsymbol{A}$. In contrast the existence of a link between nodes 6 and 7 would be considered rather unlikely (2%).

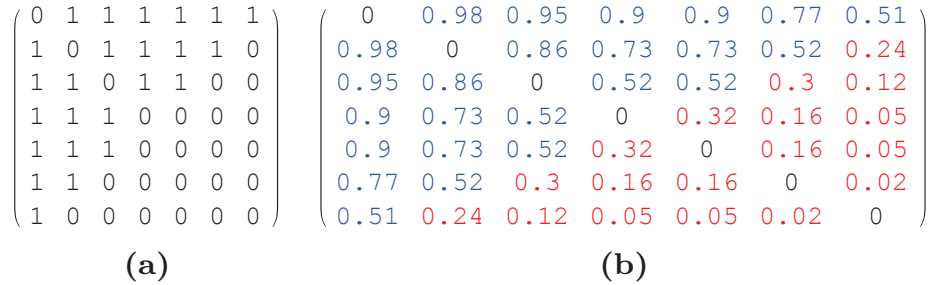

**Figure 2.8.:** **(a)** Observed adjacency matrix $\boldsymbol{A}$. **(b)** Probability matrix of a dyadic ERGM fixing the expected node degrees.

In addition to missing links, many real datasets contain links which in fact are not present. ERGMs have also proved beneficial in estimating the certainty of a detected link to actually be in place. In terms of the learned ERGM in Fig. 2.8, the most reliable link is the one between nodes 1 and 2, while the one between nodes 1 and 7 does only exist with a probability of 51%.

Models for link prediction are eminently helpful when testing the presence of links is very intricate [2], e.g. in many biochemical systems such as gene-transcription networks. In these cases ERGMs can provide for lists with the most promising candidates. Other prominent applications are the suggestions for friends in online social networks and product recommendations in customer-product networks.

## 2.5. Dynamics on Networks

Network function is often closely related to dynamical processes taking place *on* their graph structure. Road planners aim to design streets such that traffic jams are minimized. Medical scientists try to understand how synchronization of neural activity affects diseases such as Parkinson's, epilepsy, schizophrenia, or Alzheimer's [52, 137, 140] and

*2. Complex-Networks Theory*

how to manipulate it. Epidemiologists aim to develop vaccination and quarantine strategies that avoid pandemic spreading of both infectious diseases or computer viruses [53, 144].

### 2.5.1. Random Walks and Markov Chains

Among the most basic dynamical processes on networks are random walks: An agent starts at some node and follows the graph's links randomly. On the *www* the random walker can be thought of as a surfer that clicks on hyperlinks randomly.

For an agent at vertex $i$ the transition rate to vertex $j$ be denoted as the entry $T_{ij}$ of the transition matrix $\boldsymbol{T}$ with the normalization

$$\sum_{j=1}^{N} T_{ij} = 1 \quad , \quad T_{ij} \in [0,1]. \tag{2.47}$$

Matrices that obey Eq. (2.47) are called *row stochastic matrices*. Given the probabilities $\{p_j^t\}$ of finding the random walker at nodes $j$, the probability of finding him at node $i$ at time $t+1$ is

$$p_i^{t+1} = \sum_{j=1}^{N} T_{ji}\, p_j^t. \tag{2.48}$$

Such processes, where the transition rates do not depend on the path the random walker has taken so far, are called *Markov chains*.

Starting from an initial distribution, $\vec{p}^{\,0}$, the probabilities at time $t$ are then obtained by applying the transposed of the transition matrix $t$ times to the former,

$$\vec{p}^{\,t} = \left(\boldsymbol{T}^T\right)^t \vec{p}^{\,0}. \tag{2.49}$$

Since the $\vec{p}^{\,t}$ represent probability distributions, the normalization,

$$\sum_{i=1}^{N} p_i^t = 1, \tag{2.50}$$

2.5. Dynamics on Networksmust hold at all times. With Eqs. (2.48) and (2.47)

$$\sum_{i=1}^{N} p_i^{t+1} = \sum_{i=1}^{N}\sum_{j=1}^{N} T_{ji}\, p_j^t = \sum_{j=1}^{N} p_j^t \sum_{i=1}^{N} T_{ji} = \sum_{j=1}^{N} p_j^t \qquad (2.51)$$

we see that this is guaranteed for all times $t > 0$ if $\vec{p}^{\,0}$ is normalized.

Of particular relevance are *steady states* of the random walker, i.e. the *stationary distributions*, $\vec{p}^{\,\mathrm{stat}}$, for which

$$\vec{p}^{\,\mathrm{stat}} = \boldsymbol{T}^T \vec{p}^{\,\mathrm{stat}}. \qquad (2.52)$$

In these steady states, the probability of the random walker to be at a certain node is no longer time dependent and can be considered as a measure for the latter's relevance in the network. However, such steady states do not exist for all networks. One possible issue can e.g. be missing paths between pairs of nodes. If, for example, there are 'sink' nodes with no outgoing links, the random walker will, once it has entered such a vertex, stay there for all times. Similarly, 'source' nodes with no incoming edges may never be visited by a random walker at all times $t > 0$.

**PageRank**

The paradigm of a random walker on the web graph is at the bottom of the *PageRank* algorithm [114] that was initially used by the search engine Google to compare the relevance of websites. The walker can be thought of as an agent that randomly follows hyperlinks between the sites. Their importance can then be interpreted as the steady-state probability of being visited by the agent. The transition matrix is defined as

$$T_{ij}^{\mathrm{PR}} = \frac{A_{ij}}{k_i^{\mathrm{out}}}, \qquad (2.53)$$

implying that every of the $k_i^{\mathrm{out}}$ outbound links of node $i$ is taken with equal probability.

## 2. Complex-Networks Theory

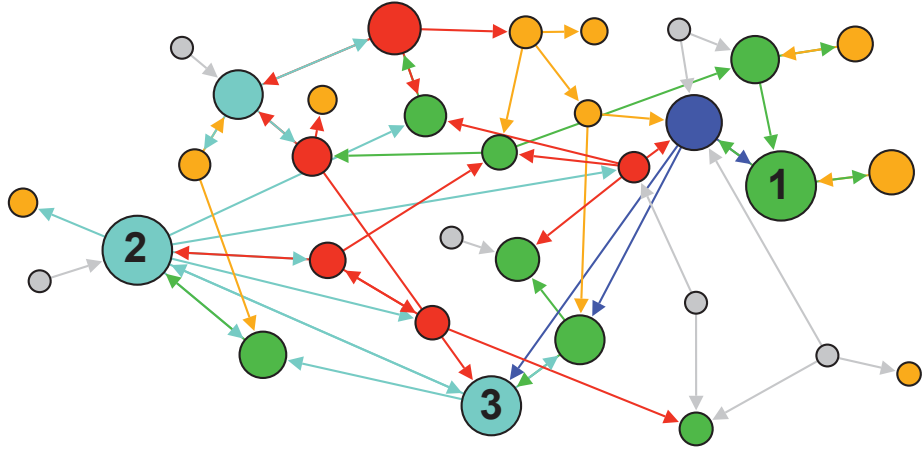

**Figure 2.9.:** Illustration of the PageRank for $d = 0.85$. Nodes with the same in-degree have the same color. Sizes are chosen proportional to the nodes' PageRank. Nodes with the three largest PageRanks are labeled 1, 2, and 3. It is the link structure of the network that makes them more likely to be visited by a random walker than the blue node, which has the largest in-degree.

To handle the problem of disconnected components, sink nodes, and source nodes, in addition to following random links on the graph structure, occasionally, the random walker jumps to some arbitrary node in the system. The latter needs not necessarily to be directly connected to the walker's current position. In fact, at every time step, the agent continues the walk along the graph with probability $d$ and jumps to some random node in the network with probability $(1 - d)$. In the steady state, the probability of finding the random walker at node $i$ is given by

$$p_i^{\text{PR}} = d \sum_{j=1}^{N} T_{ji}^{\text{PR}} p_j^{\text{PR}} + \frac{1-d}{N}. \qquad (2.54)$$

For $d = 1$ we have the ordinary random walk as discussed above. $d = 0$ corresponds to the case where all sites are visited uniformly at random, independent of the graph structure. Empirical evidence suggests values around $d \approx 0.85$. The solution of the stationary distribution in Eq. (2.54) is referred to (aside from some normalization) as the node's *PageRank* or *PageRank centrality*.

The concept of the PageRank is illustrated in the example in Fig. 2.9. An alternative interpretation of the idea is that links of important vertices have higher impact than those of less important ones. The impact of a vertex, however, is itself determined by the sum of impacts from its incoming links.

**Markov Chain Monte Carlo Sampling**

Random walks also appear in the context of all kinds of sampling processes, e.g. when sampling from the probability distribution of an ERGM (see Section 2.4).

In the space of all $N$-dimensional adjacency matrices $\boldsymbol{D} \in \{0,1\}^N \times \{0,1\}^N$, $\Omega(\boldsymbol{D})$, two instances, $\boldsymbol{D}_i$ and $\boldsymbol{D}_j$, can be considered separated by one step if they can be transformed to each other by the flip of a single entry from 0 to 1 or vice versa. For example, a random walk through the space of all $2 \times 2$ adjacency matrices could be given by:

$$\begin{pmatrix} 0 & 0 \\ 0 & 0 \end{pmatrix} \to \begin{pmatrix} 0 & 1 \\ 0 & 0 \end{pmatrix} \to \begin{pmatrix} 0 & 1 \\ 1 & 0 \end{pmatrix} \to \begin{pmatrix} 0 & 0 \\ 1 & 0 \end{pmatrix} \to \begin{pmatrix} 0 & 1 \\ 1 & 0 \end{pmatrix} \to \ldots \quad (2.55)$$

If we now want to sample matrices from a distribution

$$\mathcal{P}(\boldsymbol{D}) = \frac{1}{Z} \exp\left(-\mathcal{H}(\boldsymbol{D})\right) \quad (2.56)$$

the evaluation of the partition function (see Eq. (2.32)) can be computationally expensive or even infeasible, especially for Hamiltonians which do not show dyadic independence (see page 31). However, if the desired distribution matches the steady-state distribution of the Markov chain defined by a random walk, we can simply draw from the visited matrices with uniform probability. Methods based on this principle are referred to as *Markov chain Monte Carlo (MCMC)* methods.

*2. Complex-Networks Theory*

In relation to the famous casinos of Monacco, the term *Monte Carlo* reflects the randomness involved in the sampling process.

A sufficient, but not necessary, condition for a distribution to be stationary is known as *detailed balance*. Detailed balance requires the net probability flow between any pair of nodes to be zero,

$$p_i T_{ij} = p_j T_{ji} \quad , \quad \forall\, i,j \in \{1, 2, ..., N-1, N\}, \qquad (2.57)$$

and therefore guarantees the whole probability distribution of the system to be stationary. When detailed balance is satisfied, the probability of the random walker to enter node $i$ from node $j$ is equal to the probability of it entering node $j$ from node $i$. Hence, detailed balance implies the Markov chains to be reversible in time. An example of a steady state that does not obey detailed balance would be a random walk on a unidirectional ring, $A \to B \to C \to A$, that is clearly not reversible in time.

In order to find the transition rates between states $i$ and $j$ such that detailed balance is valid for the probability distribution of the ERGM in Eq. (2.56) we need

$$\frac{\mathcal{P}(\boldsymbol{D}_1)}{\mathcal{P}(\boldsymbol{D}_2)} = e^{-[\mathcal{H}(\boldsymbol{D}_1) - \mathcal{H}(\boldsymbol{D}_2)]} \stackrel{!}{=} \frac{T_{21}}{T_{12}} \qquad (2.58)$$

to hold, where the partition function cancelled out. Since Eq. (2.58) specifies only the ratio of the rates, there are multiple ways of guaranteeing detailed balance. Among the most common realizations are the *Metropolis-* [96] and the *Gibbs-sampling* [112, 132] algorithm.

In Section 3.3.2 we will use MCMC methods to obtain ensembles of randomized networks which preserve both the in and out degree of every single node in the networks.

## 2.5.2. General First-Order Dynamical Processes

Besides random walks jumping between nodes, there are much more intricate dynamical processes associated with the vertices of graph structures. Non-equilibrium chemical reactions can be modeled with the vertices representing substances and the edges indicating conversions between them. In the brain, the interaction of neurons via synapses generates complex spiking patterns in the time evolution of their electrical potentials. Many of these dynamical processes acting on the vertices of graph structures are highly non-linear – sometimes even chaotic – and are thus difficult to analyze.

Let us consider a dynamical process acting on the $N$ nodes of a graph, $\mathcal{G}$, with adjacency matrix $\boldsymbol{A}$. The state of each vertex $i$ be described by a vector of independent dynamical variables, $\vec{x}_i(t)$. The equations of motion shall be coupled only along the edges of $\mathcal{G}$. A general first order set of equations can then be written as

$$\begin{aligned}\frac{d\vec{x}_i(t)}{dt} &= \vec{f}[\vec{x}_i(t)] + \sum_j A_{ji}\, w_{ji}\, \vec{h}\,[\vec{x}_i(t), \vec{x}_j(t)] \\ &= \vec{f}[\vec{x}_i(t)] + \sum_j G_{ji}\, \vec{h}\,[\vec{x}_i(t), \vec{x}_j(t)]\end{aligned} \quad (2.59)$$

where the function $\vec{f}$ describes the intrinsic dynamics acting on the vertices, even if no links are present. The coupling between two vertices $i$ and $j$ is modeled by the function $\vec{h}$ and the strength of the influence on node $i$ is contained in the coupling matrix $G_{ij} = A_{ij}\, w_{ij} \geq 0$. It is assumed that all vertices follow the same intrinsic dynamics and the same coupling function.

In the case of a random walk $\vec{x}(t)$ has only one component indicating the probability of the random walk to be at node $i$ at time $t$. With $\vec{f} = 0$ and $\boldsymbol{G}$ being the transition matrix, $\boldsymbol{T}$, we obtain the continous-time version of Eq. (2.48). However, Eq. (2.59) is much more general and can describe a wide range of dynamics, e.g. interacting oscillators, neurons, or the evolution of infection dynamics.

### 2.5.3. Synchronization

An important aspect of the temporal behavior of dynamical systems is the phenomenon of *synchronization* [37, 74, 78, 119]. The system

*2. Complex-Networks Theory*

described in Eq. (2.59) is said to be *(completely) synchronized* if the time evolution is the same for all vertices, i.e.

$$\vec{x}_i(t) = \vec{s}(t) \quad , \quad \forall\, i \in \{1, 2, ..., N\}. \tag{2.60}$$

It shall be mentioned that – depending on the context – there are various other kinds of synchronization, e.g. *phase synchronization* or *generalized synchronization* (see e.g. [119]), but for now we will concentrate on *complete synchronization*.

For complete synchronization to be a solution of Eq. (2.59) the coupling matrix must have the same row sum for all vertices. For convenience we choose $\boldsymbol{G}$ *column stochastic*, i.e.

$$\sum_i G_{ij} = 1, \tag{2.61}$$

since any multiplicative constant can be absorbed in the function $\vec{h}$. In the synchronized state, the time evolution of every node is then obtained from

$$\frac{d\vec{x}_i(t)}{dt} = \frac{d\vec{s}(t)}{dt} = \vec{f}[\vec{s}(t)] + \vec{h}[\vec{s}(t), \vec{s}(t)]. \tag{2.62}$$

We will now show that the stability of complete synchronization depends essentially on the topology of the graph structure. To keep things clear, we will consider dynamics with only one dynamical variable, $x_i$, per vertex and the coupling function $h$ only depending on the state of the connected vertex,

$$\frac{dx_i(t)}{dt} = f[x_i(t)] + \sum_j G_{ji}\, h[x_j(t)]. \tag{2.63}$$

The state of the whole system is then described by the $N$-dimensional vector $\vec{X}(t) = (x_1(t), x_2(t), ..., x_{N-1}(t), x_N(t))^T$. In case of complete synchronization, the trajectory is restricted to a single direction, $\vec{X}_s(t) = s(t)(1,1,...,1)^T$. However, if there are only slight perturbations transversal to the synchronized state, the synchronization potentially can be destroyed if it is unstable. Considering infinitesimally small perturbations, $\epsilon_i(t)$, to the synchronized trajectories of the $x_i(t)$, we can linearize Eq. (2.60) and obtain

$$\frac{d\epsilon_i(t)}{dt} = f'[s(t)]\epsilon_i(t) + \sum_j G_{ji}\, h'[s(t)]\epsilon_j(t) + \mathcal{O}(\epsilon^2) \qquad (2.64)$$

and in matrix form, neglecting higher order terms,

$$\frac{d\vec{\epsilon}(t)}{dt} = \left(f'[s(t)]\,\mathbb{1} + h'[s(t)]\,\boldsymbol{G}^T\right)\vec{\epsilon}(t). \qquad (2.65)$$

Since any eigenvector of $\boldsymbol{G}^T$, $\vec{\mu}_r = \mu_r\,\hat{e}_{\vec{\mu}_r}$, is also an eigenvector of the identity matrix, $\mathbb{1}$, it is convenient to analyze perturbations in their directions and thus to rewrite Eq. (2.65) as

$$\hat{e}_{\vec{\mu}_r}\frac{d\mu_r(t)}{dt} = \hat{e}_{\vec{\mu}_r}\left(f'[s(t)] + h'[s(t)]\gamma_r\right)\mu_r, \qquad (2.66)$$

where $\gamma_r$ is the eigenvalue corresponding to $\vec{\mu}_r$. Since $\boldsymbol{G}$ is a stochastic matrix, all eigenvalues are real and $|\gamma_r| \leq 1$. Hence, the synchronization is stable towards a perturbation in the $\vec{\mu}_r$ direction if

$$\left\langle f'[s(t)]\right\rangle_t + \left\langle h'[s(t)]\right\rangle_t \gamma_r < 0. \qquad (2.67)$$

$\langle\cdots\rangle_t$ indicates the time average over a time span which is large compared to the typical time scale of $s(t)$. For the system to stay synchronized, Eq. (2.67) has to hold for all $\gamma_r$ except for the one corresponding to the direction of the synchronized trajectory, $\vec{X}_s(t)$. We will assume the eigenvalues, $\gamma_i$, to be ordered by decreasing value, i.e. $\gamma_1$ be the largest positive, $\gamma_N$ the most negative eigenvalue. By construction of the stochastic coupling matrix, $\boldsymbol{G}$, it is $\gamma_1 = 1$ which is the eigenvalue corresponding to the synchronization manifold, $\vec{X}_s(t)$.

Inequalities, relating information about the network topology – in terms of the associated eigenvalues – to the nodes' function, are known as *master stability functions (MSF)* [118] and are a powerful tool for linear stability analysis. Using MSFs it can e.g. be shown that systems where the input of adjacent units is transmitted with a delay time which is large compared to the time scale of the intrinsic dynamics, synchronization is only possible, if no other eigenvalue than $\gamma_1$ has magnitude one [39]. This means that the *spectral gap*,

$$\Delta = \gamma_1 - \max\big(|\gamma_2|, |\gamma_N|\big), \tag{2.68}$$

is larger than zero.

For a more detailed treatment of synchronization on networks see e.g. the textbooks by Newman [108] or Pikovsky et al. [119].

### 2.5.4. Infection Dynamics

Many diseases spread over the connections of different kinds of network structures. For influenza, Ebola, and SARS it can be sufficient to physically contact an infected person. HIV and syphilis spread foremost via sexual interaction between individuals.

Although the biological nature and medical implications of diseases may be highly complex and entirely different, their spreading can be modeled in similar ways by means of infection dynamics. In a rather basic abstraction one can assume that the nodes of a network can be in either of three states: healthy but susceptible (S) to a disease, infected (I), or recovered and thus immune against the infection (R). This simplified approach may most likely not help to find cures for individuals infected with a disease. However, it may be supportive to evaluate the empidemic risk of an outbreak and help to identify the initially infected individual. Furthermore, it can be used to rate the prospects of different vaccination strategies in case an immunization of the whole population is infeasible.

In the so called *SIR model*, the state of an individual $i$, $\vec{x}_i$, is described by the probabilities of being infected, susceptible, or recovered, i.e. $\vec{x}_i = (S_i, I_i, R_i)^T$. For a better readability, the time dependence of the variables $S_i$, $I_i$, and $R_i$ will not be stated explicitly below. Suppose that infected individuals spread the disease to susceptible ones they interact with with probability $\beta$ per time step. Assuming further that an infected agent recovers at a rate $\gamma$, we can formulate the dynamic equations of an SIR model as

$$\begin{aligned}
\frac{dS_i}{dt} &= -\beta S_i \sum_j A_{ji} I_j \\
\frac{dI_i}{dt} &= \beta S_i \sum_j A_{ji} I_j - \gamma I_i \\
\frac{dR_i}{dt} &= \gamma I_i.
\end{aligned} \qquad (2.69)$$

This set of equations is also a realization of the general dynamics we defined in Eq. (2.59) with $\vec{f}(\vec{x}) = (0, -\gamma x_2, \gamma x_3)^T$ and $\vec{h}(\vec{x}, \vec{y}) = (-\beta x_1 y_2, \beta x_1 y_2, 0)^T$.

Analyzing the SIR model, e.g. in terms of numerical simulations, allows to test whether a disease with fixed infection and recovery rates may eventually become pandemic, i.e. affecting major parts of the system. It moreover allows to test at which nodes an outbreak of an infection will be most severe.

There are multiple variations of models involving $S$, $I$, and $R$ nodes, e.g. the *SIS model* where nodes become susceptible again after an infection, or the *SIRS model* where nodes become susceptible after a period of being immune. For a more comprehensive introduction to the different kinds of epidemic models, again, see the textbook by Newman [108] and the references therein.

# 3. Triadic Substructures in Complex Networks

The main focus of this dissertation is on the role triadic substructures play in complex networks. In Section 2.2.4 we have already learned about clustering coefficients – measures to quantify the abundance of closed triangles in networks. We will now present how clustering coefficients in different kinds of real-world networks typically look like. Subsequently, Section 3.2 gives a brief introduction to the sociological theory of social balance which is also based on triadic relationships. Moreover, in Section 3.3 we will review the commonly used methodology to identify overabundant subgraph patterns, also known as *motifs*. In the case of triadic subgraphs this can be considered a generalization of the concept of clustering coefficients. We will present results found in real datasets and we will discuss how they suggest that links may form conditionally dependent on each other. Subsequently, in Section 3.4, we will give an overview of existing models to generate synthetic networks with specified triadic structure. Section 3.5 discusses existing work on the implications of those substructures on dynamical processes.

## 3.1. Clustering Coefficients in Real Networks

In Section 2.2.4 we have introduced clustering coefficients – measures for the transitivity in networks. Table 3.1 summarizes the observed clustering coefficients for datasets of various areas. We see that the two common definitions of the clustering coefficient, $C_{\text{glob}}$ and $\langle C \rangle$, yield quite different results. The more they differ from each other,

*3. Triadic Substructures in Complex Networks*

| Dataset | $N$ | $C^{\mathrm{ER}} = p$ | $C_{\mathrm{glob}}$ | $\langle C \rangle$ | Refs. |
|---:|---:|---:|---:|---:|---:|
| Film actors | 449 913 | $2.5 \times 10^{-4}$ | 0.20 | 0.78 | [9, 149] |
| Math coauthorship | 253 339 | $1.6 \times 10^{-5}$ | 0.15 | 0.34 | [35, 55] |
| Physics coauthorship | 52 909 | $1.8 \times 10^{-4}$ | 0.45 | 0.56 | [105, 106] |
| Email address books | 16 881 | $4.0 \times 10^{-4}$ | 0.17 | 0.13 | [110] |
| Student dating | 573 | 0.003 | 0.005 | 0.001 | [20] |
| Websites `nd.edu` | 269 504 | $2.1 \times 10^{-5}$ | 0.11 | 0.29 | [3, 17] |
| Internet | 10 697 | $5.6 \times 10^{-4}$ | 0.035 | 0.39 | [28, 44] |
| Power grid | 4 941 | $5.4 \times 10^{-4}$ | 0.10 | 0.08 | [149] |
| Metabolic network | 765 | 0.01 | 0.09 | 0.67 | [69] |
| Protein interactions | 2 115 | 0.001 | 0.072 | 0.071 | [68] |
| Marine food web | 134 | 0.03 | 0.16 | 0.23 | [66] |
| Neural network | 307 | 0.03 | 0.18 | 0.28 | [149] |

**Table 3.1.:** Clustering coefficients observed in real datasets as collected in Table 8.1 in Ref. [108]. The networks consist of $N$ nodes and have a link density $p$. Their global and average local clustering coefficients are $C_{\mathrm{glob}}$ and $\langle C \rangle$, respectively. For details on the datasets see the cited references.

the more heterogeneous the local clustering coefficients are distributed. However, for most systems, both $C_{\mathrm{glob}}$ and $\langle C \rangle$ are several orders of magnitude larger than expected for an Erdös Rényi graph with the same density.

A counterexample in Table 3.1 is the student-dating network. This can be explained by the fact that it has primarily bipartite structure, i.e. for the major part students would date people of the opposite sex. Therefore, triangles appear extremely rarely.

The high clustering coefficient of most real networks supports the hypothesis that they have not formed purely at random. For some systems, the rather high clustering coefficients can be explained by considering null models with arbitrary degree distribution (e.g. for food webs [108]). Yet, for finite $\langle k \rangle$ and $\langle k^2 \rangle$, their clustering coefficient is still expected to vanish as $N \to \infty$ (compare Eq. (2.25)) which is not the case for many of the systems displayed in Table 3.1. The strong deviation from the random expectation of the null model indicates that

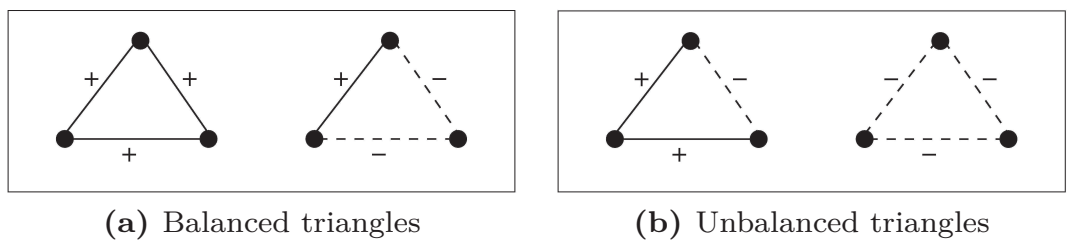

**(a)** Balanced triangles  **(b)** Unbalanced triangles

**Figure 3.1.:** Balanced and unbalanced triangles [136].

further processes, not captured by the density or the degree distribution alone, are at work. In a social context, this is plausible on a local level: suppose Bob has two friends Alice and Charlie who do not know each other yet. It is more likely for them to also get in touch with each other, compared to the situation in which they do not have a common acquaintance. This kind of process is known as *triadic closure*. The awareness of such mechanisms is useful for making forecasts of the future evolution of networks and to predict links that are unknown so far.

## 3.2. Social Balance

Moreover, triadic relationships are considered important in *signed* social networks (see page 9), in which positive edges represent friendship and negative ones animosity. In the 1940s, the social psychologist Fritz Heider introduced the theory of *social balance*, sometimes also referred to as *structural balance* [60, 61].

Triangular relationships in signed social networks are called *balanced* if either all relationships have a positive sentiment, or one relationship is positive and the other two are negative as shown in Fig. 3.1(a). Conversely, triangles are said to be *unbalanced* if two connections are friendly while one is antagonistic, or if all relationships are negative (see Fig. 3.1(b)). Balanced triangles fulfill the adages "the friend of my friend is my friend", "the friend of my enemy is my enemy", "the enemy of my friend is my enemy", and "the enemy of my enemy is my friend". Unbalanced triangles are assumed to induce social tension and are therefore hypothesized to be less stable than balanced ones: Just suppose Alice and Bob are in love with each other, but Alice cannot get along with Bob's friend Charlie.

The role of the configuration with all three connections being antagonistic is sometimes questioned [33, 86]. An alternative formulation by Davis agrees with Heider's balance theory for the three other configurations, but makes no prediction for the one with completely negative links [33]. This definition is often referred to as *weak balance.*

However, the symmetry induced by Heider's original formulation makes it highly attractive for quantitative analysis. In this formulation, a triangle is balanced if the product of its signs is positive and unbalanced if it is negative. This fact can be used to map the triangles of a graph to the spins of a spin-glass and to simulate the 'social dynamics' of such systems by defining transition rates between different configurations through a Hamiltonian $\mathcal{H} = -\sum_{i,j,k} G_{ij} G_{jk} G_{ki}$ [11, 12, 95]. Furthermore, the product-sign property can be used to estimate the ratio of balanced triangles in large networks without the need to iterate over all triads, but just by computing the largest eigenvalues of the matrix $\boldsymbol{G}$ [136].

A signed undirected complete graph $\mathcal{G}(V, E)$ with vertices $V$ and edges $E$ is said to be balanced if all of its triangles are balanced. Cartwright and Harary could proof that this is equivalent to the existence of two disjoint groups [27, 59]:

**Theorem 3.2.1** *A signed undirected complete graph $\mathcal{G}(V, E)$ is balanced if and only if its vertex set $V$ is partitioned into two disjoint subsets $\mathcal{A}$ and $\mathcal{B}$, one of which may be empty, such that all lines between vertices of the same subset are positive and all lines between points of the two different subsets are negative.*

## 3.3. Network Motifs

For directed networks, we have seen on page 13 that there are 16 distinct triadic subgraph patterns. The concept of clustering coefficients can thus be generalized from measuring the relative appearance of triangles to evaluating the abundance of all triad patterns shown in Fig. 2.2. It is furthermore possible to extend the analysis to fourth and even higher order subgraphs. However, with increasing order also the number of distinct subgraph patterns increases quickly. Patterns that appear significantly more often than expected for the null model are referred to as *motifs*. Accordingly, those that are significantly underrepresented are called *anti-motifs*.

Over the last decade the systematic study of third order subnetwork structure attracted much attention [6, 100, 102] including, but not limited to, applications in the fields of neuroscience [126, 133, 142], biology [2, 5, 129], economy [113, 130, 134], and human mobility [72, 128]. We will now introduce the common methodology used for motif detection and present results obtained from real datasets which suggest potential interdependencies in the link-formation process.

### 3.3.1. Motif-Detection Procedure

At first, in order to estimate the frequency of random appearances of each pattern, we need to define a proper null model to which we compare the network under investigation. In Chapter 2 it was shown that the expected clustering coefficient, i.e. the abundance of triangles, depends e.g. on the degree distribution. This holds for the other triad patterns as well. Also the number of uni- and bidirectional links affects the occurrences of different triadic substructures.

Since we are interested in effects beyond those artifacts, the random-null model shall have the same degree distribution and the same number of both uni- and bidirectional links as the original network under investigation. Instances of this ensemble will be generated by means of a Markov chain Monte Carlo (MCMC, see page 39) switching al-

*3. Triadic Substructures in Complex Networks*

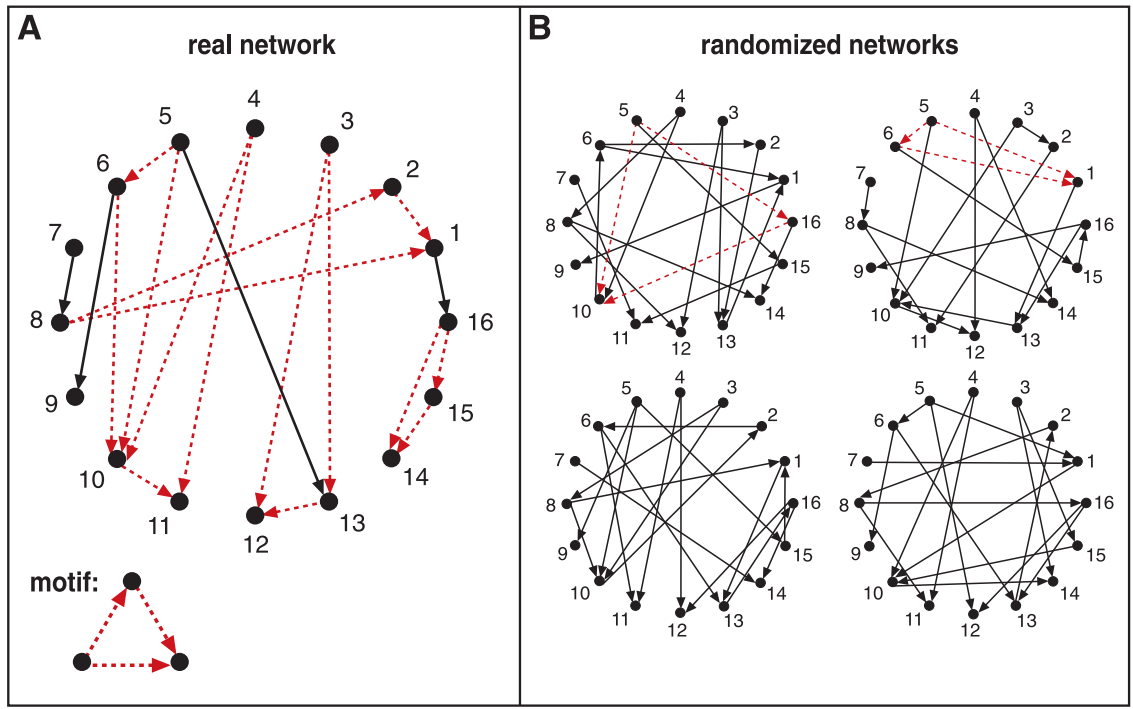

**Figure 3.2.:** Schematic view of network motif detection. Network motifs are patterns that recur much more frequently (A) in the real network than (B) in an ensemble of randomized networks. Each node in the randomized networks has the same number of incoming and outgoing edges as does the corresponding node in the real network. Dashed lines indicate edges that participate in the feedforward loop motif, which occurs five times in the real network. Reprinted with permission from AAAS[1].

gorithm [121, 124]. This is a stepwise, locally degree-preserving randomization of the original system. In particular, both the in and out degree of every single node, as well as the number of uni- and bidirectional links adjacent to each node will be conserved. An example for such a randomization of a real network is presented in Fig. 3.2. Details of the randomization procedure will be discussed in Section 3.3.2.

---

[1] From R. Milo, S. Shen-Orr, S. Itzkovitz, N. Kashtan, D. Chklovskii, and U. Alon. Network motifs: simple building blocks of complex networks. *Science*, 298(5594):824-7, 2002.

Having the ensemble of randomized networks serving as the null model at our disposal, for every pattern $i$, we can now compare the appearances $N_{\text{original},i}$ in the real system to the average number of appearances $\langle N_{\text{rand},i} \rangle$ in samples from the null model. Over- and underrepresentation of pattern $i$ is then quantified through a $Z$ score

$$Z_i = \frac{N_{\text{original},i} - \langle N_{\text{rand},i} \rangle}{\sigma_{\text{rand},i}} \qquad (3.1)$$

in which $\sigma_{\text{rand},i}$ denotes the standard deviation of $N_{\text{rand},i}$ estimated from the sample. Notice that $Z$ scores are evaluated by counting the subgraph patterns over all $\binom{N}{3}$ possible triads. Every network can be assigned a vector $\vec{Z}$ whose components comprise the $Z$ scores of all possible triad patterns. Significant patterns are referred to as *motifs* [102]. For computational reasons, it is common to consider only the $Z$ scores of connected triad patterns, i.e. those in which all three nodes are attached to an edge. Those are patterns 4 through 16 in Fig. 2.2.

Fig. 3.2 illustrates the motif detection using the example of pattern , also referred to as the *feed-forward loop (FFL)*. Visualized by dashed lines, it appears five times in the network shown in (A). Part (B) shows four samples of the ensemble of degree-preserving randomizations of the real network with only two appearances of the FFL in total, i.e. $\langle N_{\text{rand},i} \rangle = 0.5$. The estimated standard deviation is $\sigma_{\text{rand},i} = \sqrt{\frac{1}{4-1} \cdot 4 \cdot 0.5^2} = \sqrt{\frac{1}{3}}$. The $Z$ score is thus $Z = \sqrt{3}\,(5 - 0.5) \approx 7.8$ indicating the high significance of the motif. The motif-detection algorithm is implemented in the *mfinder* software which can be downloaded from Uri Alon's website [7].

Further, it shall be mentioned that one commonly refers to the normalized $Z$-score vector as the 'significance profile'

$$\vec{\text{SP}} = \vec{Z} \Big/ \sqrt{\sum_{i=4}^{16} Z_i^2}. \qquad (3.2)$$

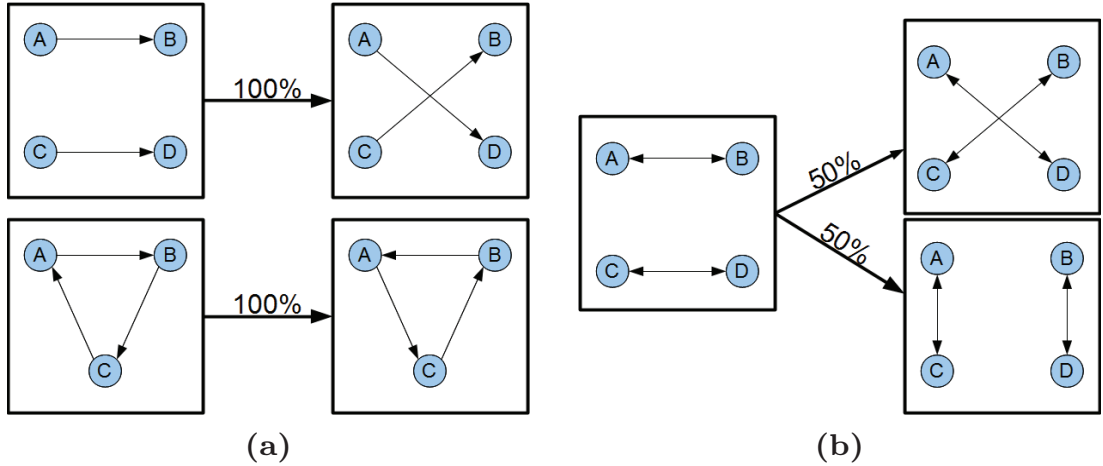

**Figure 3.3.:** Microscopic link-switchings performed to generate the randomized ensembles. **(a)** *Pair switch* and *loop switch* for unidirectional links. **(b)** *Pair switch* for bidirectional links.

This normalization makes systems of different sizes comparable, since larger systems tend to express $Z$ scores of larger magnitude [102].

### 3.3.2. MCMC Switching Algorithm for Network Randomization

Let $\mathcal{E}_{\boldsymbol{A}}$ be the ensemble of adjacency matrices in which every vertex has the same in and out degree as in $\boldsymbol{A}$ and also the numbers of both uni- and bidirectional links adjacent to every vertex are fixed. We will generate instantiations of this ensemble – the null model for motif detection – by means of a stepwise degree-preserving randomization of the original network. It can be thought of as a random walk in the state space of adjacency matrices. Two matrices are considered neighbors if they can be transformed into each other by appying a single one of the switching steps shown in Fig. 3.3 to a subset of their edges.

Fig. 3.3(a) shows the microscopic rewiring rules for unidirectional edges. For two node-disjoint links, there is exactly one way of rewiring such that the in and out degrees of all affected nodes are preserved. For

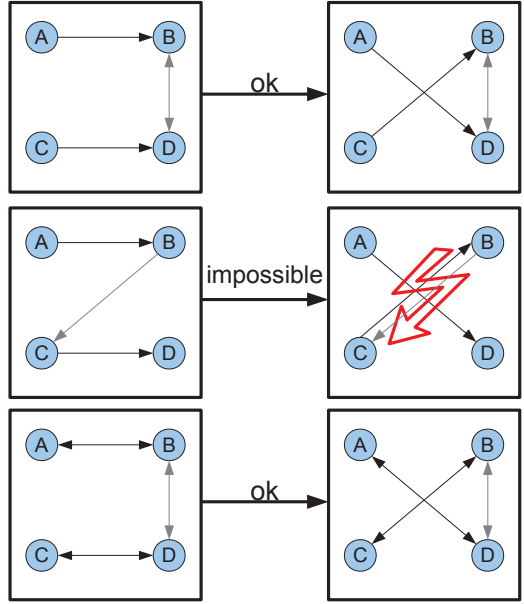

**Figure 3.4.:** Cases in which additional links affect the randomization process: only if these links interfere with the rewiring of a particular edge (middle) the step cannot be performed.

a unidirectional triadic loop all node degrees are preserved, when the directionalities of all edges are reversed. We call the former operation a *pair switch*, the latter a *loop switch*. The rewiring rules for two node-disjoint bidirectional links are illustrated in Fig. 3.3(b). Because of the symmetry of the links, there are two options for the rewiring. It shall be emphasized that for the unidirectional and bidirectional pair switches as well as for the loop switch the inverse transformation is of the same kind.

For the pair switches, besides the two links selected for the swapping, there can be additional edges between the four involved nodes. This poses a problem if there are already links present between the nodes of newly established connections. Swaps for which this is the case are forbidden, as illustrated in the middle case of Fig. 3.4.

It can be proven that any two matrices $\boldsymbol{A}_i, \boldsymbol{A}_j \in \mathcal{E}_{\boldsymbol{A}}$ can be transformed into each other by adequate applications of the rewiring steps

*3. Triadic Substructures in Complex Networks*

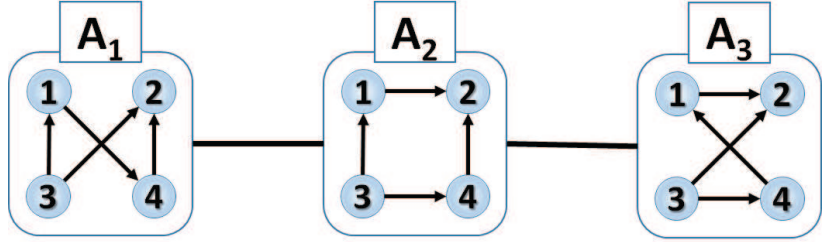

**Figure 3.5.:** State space of randomized 2×2-adjacency matrices. Those connected by edges can be transformed into each other by a single swapping step as defined in Fig. 3.3.

in Fig. 3.3 [121, 123]. Hence, a random walk in the space of matrices $A_i \in \mathcal{E}_A$ can reach any of such matrices, i.e. the system is *ergodic*. However, to serve as a reasonable null model the probability distribution over the matrices must be uniform. This is guaranteed, when the transition rates $T(A_i \to A_j)$ are specified to obey detailed balance,

$$p(A_i)\, T(A_i \to A_j) = p(A_j)\, T(A_j \to A_i). \qquad (3.3)$$

For a uniform distribution, $p(A_k) = const.$, this is true if and only if the transition rates are the same,

$$T(A_i \to A_j) = T(A_i \to A_j). \qquad (3.4)$$

Suppose, in every time step, the random walker jumps to a random neighbor in the space of adjacency matrices. It is thus $T(A_i \to A_j) = \frac{1}{g(A_i)}$ with $g(A_i)$ indicating the number of matrices in $\mathcal{E}_A$ that can be reached from $A_i$ with a single of the operations shown in Fig. 3.3. In general, the $g(A_i)$ are clearly not the same for all graphs of the ensemble. Therefore, there would be a bias in the probability distribution towards matrices with many allowed elementary switches. For the state space sketched in Fig. 3.5, for instance, the probability to sample matrix $A_2$ would be twice as high as for $A_1$ or $A_3$.

This problem can be solved by allowing the random walker to rest at different $A_i$ for different periods. Increasing the residence time at states with few possible transitions, and thus modifying their rates $T(\boldsymbol{A}_i \to \boldsymbol{A}_j)$, will then compensate for their lower probability to be reached by the Markov chain. The residence time can be increased by allowing for time steps in which no links whatsoever are getting switched.

In fact, the randomization algorithm proceeds as follows. In every time step, a link is selected randomly. Depending on its type (uni- or bidirectional) we draw a second one of the same type.

- If both edges are unidirectional and node disjoint – supposed it is allowed – we perform a pair switch according to the rules shown in Fig. 3.3(a) and thus obtain a new adjacency matrix.

- If they share a node and are part of a triadic loop, a loop switch is performed.

- For bidirectional edges one of the two possible pair switches shown in Fig. 3.3(b) is executed with a probability of 50% (supposed it is permitted).

- If no switch was successfully performed the random walk will remain at its current state for another time step.

If $\boldsymbol{A}_i$ and $\boldsymbol{A}_j$ are transformed into each other by a unidirectional pair switch this happens with rate

$$T\left(\boldsymbol{A}_i \stackrel{\text{uni}}{\to} \boldsymbol{A}_j\right) = T\left(\boldsymbol{A}_j \stackrel{\text{uni}}{\to} \boldsymbol{A}_i\right) = \frac{2}{M} \frac{1}{M_{\text{uni}} - 1} \qquad (3.5)$$

in which $M_{\text{uni}}$ indicates the number of unidirectional edges and $M = M_{\text{uni}} + M_{\text{bi}}$ is the sum of unidirectional and bidirectional links. There is exactly one elementary link swap, involving two specific unidirectional edges, that transforms between $\boldsymbol{A}_i$ and $\boldsymbol{A}_j$. The probability to select one of the relevant edges by the first draw is $\frac{2}{M}$, the likelihood (see appendix A.1) of also selecting the second one is then $\frac{1}{M_{\text{uni}}-1}$. Since the inverse process from $\boldsymbol{A}_j$ to $\boldsymbol{A}_i$ is also a unidirectional pair switch

*3. Triadic Substructures in Complex Networks*

and the right-hand side of Eq. (3.5) is independent of $i$ and $j$, the detailed-balance condition, Eq. (3.4), is met.

Supposed $\boldsymbol{A}_i$ and $\boldsymbol{A}_j$ are connected by a loop switch, the probability of selecting one of the relevant edges initially is $\frac{3}{M}$. The likelihood of subsequently selecting another of the two remaining links participating in the triadic loop is $\frac{2}{M_{\text{uni}}-1}$, i.e.

$$T\left(\boldsymbol{A}_i \stackrel{\text{loop}}{\to} \boldsymbol{A}_j\right) = T\left(\boldsymbol{A}_j \stackrel{\text{loop}}{\to} \boldsymbol{A}_i\right) = \frac{3}{M} \frac{2}{M_{\text{uni}} - 1}. \qquad (3.6)$$

Finally, if $\boldsymbol{A}_i$ and $\boldsymbol{A}_j$ are mapped to each other by a bidirectional pair switch, we have a joint probability of $\frac{2}{M} \frac{1}{M_{\text{bidir}}-1}$ to select the two specific edges participating in the swapping. Since in the bidirectional case either of the two rewiring options is attempted with probability $1/2$, it is

$$T\left(\boldsymbol{A}_i \stackrel{\text{bidir}}{\to} \boldsymbol{A}_j\right) = T\left(\boldsymbol{A}_j \stackrel{\text{bidir}}{\to} \boldsymbol{A}_i\right) = \frac{2}{M} \frac{1}{M_{\text{bidir}} - 1} \frac{1}{2}. \qquad (3.7)$$

Although the rates for the three distinct randomization steps – as displayed in Eqs. (3.5), (3.6), and (3.7) – differ, all of them satisfy the constraint $T(\boldsymbol{A}_i \to \boldsymbol{A}_j) = T(\boldsymbol{A}_j \to \boldsymbol{A}_i)$ which implies detailed balance for the uniform distribution. It is crucial to allow the random walk to rest at states for more than one time step, since this enables the definition of the transition rates, Eqs. (3.5), (3.6), and (3.7), independently from the number of adjacent states of the $\boldsymbol{A}_i$. The randomization process is summarized in Algorithm 1.

Besides the switching algorithm, one can generate the ensemble for the null model via a stubs method as in the configuration model, or a strategy called 'go with the winners' [4, 54]. However, the stubs method may suffer from nonuniform sampling and the 'go with the winners' method is rather slow [101]. Furthermore, exponential random graph models (ERGMs) allow for a faster generation of randomized networks [34], however they come with the limitation to fix only the *expectation* for individual node degrees, not necessarily the actual

---
**Algorithm 1** Degree-preserving randomization of a graph
---
  **function** Randomize(Graph $\mathcal{G}(V, E)$, no. of required steps)
      s = 0
      **while** s < number of required rewiring steps **do**
          pick a random link $e_1 \in E$
          **if** $e_1$ is unidirectional **then**
              pick a 2nd unidirectional link $e_2 \in E$ at random
          **else**
              pick a 2nd bidirectional link $e_2 \in E$ at random
          **end if**

          **if** $e_1$ and $e_2$ do not share a node **then**
              rewire according to the pair-switch rules in Fig. 3.3
              **if** one of the new links already exists **then**
                  undo the rewiring
              **end if**
          **else if** $e_1$ and $e_2$ participate in a loop **then**
              rewire according to the loop-switch rule in Fig.3.3(a)
          **end if**

          s++
      **end while**
      **return** randomized instance of $\mathcal{G}$
  **end function**
---

*3. Triadic Substructures in Complex Networks*

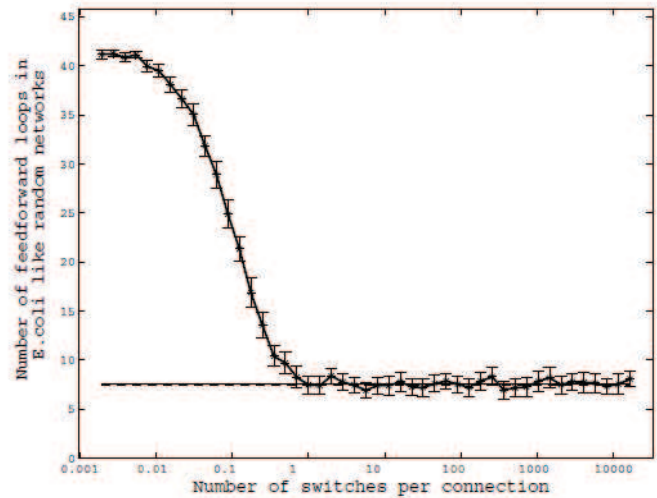

**Figure 3.6.:** Reprinted from [101]. Average number of feed-forward loops vs. average link switches per connection. Each point is an average over 100 repetitions. Error bars indicate three standard deviations. The equilibrium value is reached around one switch per edge. Similar results are reported for other networks and other patterns. Notice the linear-logarithmic scale.

values. For applications of the methodology to big data, ERGMs may serve as an alternative to generate the random null models, yet going along with a loss of accuracy.

An issue of the switching method is to estimate the appropriate number of microscopic iteration steps to reach the equililibrium. Starting from the original transcription network of E. coli, Fig. 3.6 shows appearances of pattern △ – also referred to as the *feed-forward loop* – in the randomized networks vs. the number of link-switches per edge in the system. There is a clear drop to the equilibrium value, when every link is switched once on average. According results were found for other systems and other patterns [101]. Therefore, the number of rewiring steps should be chosen proportionally to the number of links in the graph. Since all randomization steps of the switching algorithm preserve individual node degrees as well as the number of both unidi-rectional and bidirectional links, these quantities are also conserved on the macroscopic level.

### 3.3.3. Evidence for Interdependencies between Links

In 2002, Shen-Orr et al. investigated the abundance of multiple subgraph patterns in the gene regulatory network of E. coli compared to the null model introduced in the previous section. They found the triadic feed-forward loop pattern to be a motif of the network [129]. In the same year, Milo et al. extended the analysis to other systems such as neural networks, food webs and websites linking to each other, where they found a multitude of triadic motifs [102].

As a possible conclusion from their observations they suggested that – in contrast of dyadically independent link formation – motifs may serve as the actual building blocks of complex-network structure [6, 102, 129]. The two papers sparked a vast interest in subgraph analysis – together they were cited more than 6000 times by the end of 2014. However, already before motif analysis became popular in complex-networks research, triadic relationship patterns have been studied in the social sciences under the key phrase *triad census* supporting the hypothesis of *interdependent* rather than *independent* link formation [64, 146, 147].

Two years after their seminal publications, Milo, Shen-Orr et al. showed that networks of various disciplines exhibit characteristic triad-significance profiles. They suggested that systems can be grouped into just a few 'super families', as shown in Fig. 3.7 [100]. It was conjectured that there might be potential analogies between systems of very diverse origin which may help understand their evolutionary development and their functional design. Since networks with similar tasks expressed similar motifs, those were hypothesized to play a key role for the proper functioning of the systems. Evolutionary optimizations of synthetic networks with respect to a variety of goals have supported this hypothesis [73, 77]. Particularly the role of the 'feed-forward loop', 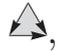, has been discussed intensively in the field [6, 93, 94, 129]. The pattern has been presumed to play a key role for systems to reliably perform

*3. Triadic Substructures in Complex Networks*

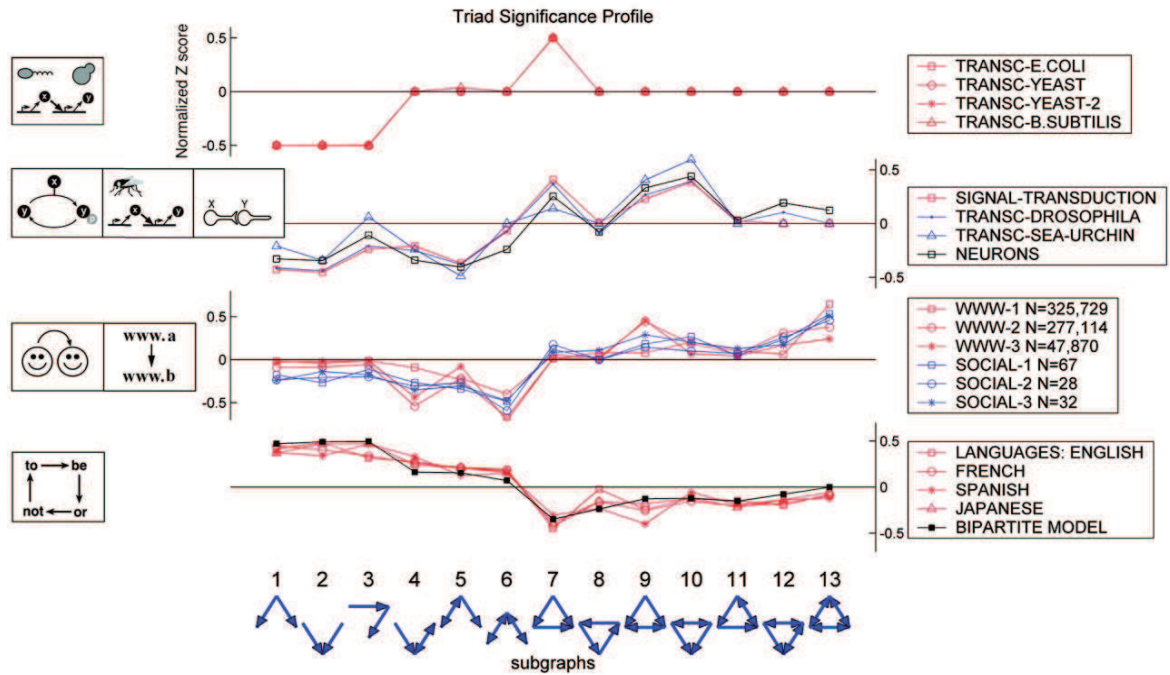

**Figure 3.7.:** TSPs of networks from various disciplines grouped into superfamilies. (i) Transcription interactions of E. coli (TRANSC-E.COLI), B. subtilis (TRANSC-B.SUBTILIS), and S. Cerevisiae (TRANsC-YEAST and TRANSC-YEAST-2). (ii) Signal-transduction interactions in mammalian cells (SIGNAL-TRANSDUCTION), transcription networks of fruit fly (TRANSC-DROSOPHILA) and sea urchin (TRANSC-SEAURCHIN), and network of neurons in C. elegans (NEURONS). (iii) hyperlinks between websites (WWW-1, WWW-2, WWW-3) and social networks, including inmates in prison (SOCIAL-1), sociology freshmen (SOCIAL-2), and students in a course about leadership (SOCIAL-3). (iv) Word-adjacency networks of a text in different languages and a bipartite model . For details on the datasets see Ref. [100]. Reprinted with permission from AAAS. From: R. Milo, S. Itzkovitz, N. Kashtan, R. Levitt, S. Shen-Orr, I. Ayzenshtat, M. Sheffer, and U. Alon. Superfamilies of evolved and designed networks. *Science*, 303(5663):1538-42, 2004.

information-processing tasks.

Over the last twelve years, motif analysis, in particular the analysis of triadic subgraph patterns, has become a standard tool in complex-network analysis and many real-world systems of diverse origin have been examined and were found to show characteristic TSPs [40, 72, 126, 134]. Many of them even fall into the four super families of Fig. 3.7.

However, already from the very beginning, the expressive power of the common motif-detection procedure has been questioned. It was suggested that spacial constraints could impose the characteristic local structure [14]. Furthermore, it was shown that global and local structure are mutually dependent on each other [143], similar to the social balance case, in which local balance implies the emergence of two homogeneous groups on the global level and vice versa (see Theorem 3.2.1). Likewise, it was found that disregarding potentially present hierarchical structure or block structure in the null model may lead to undesirable artifacts in the detection of motifs and anti motifs [21, 47, 122]. For example, many aspects of the TSP of the neural network of C. elegans could be explained by ERGMs which model both the degree distribution and mesoscopic block structure of the analyzed system, although they assume dyadic independence of link formation [122]. Nevertheless, the majority of triad significance profiles cannot be fully explained by spacial constraints or group structure alone [99], serving as evidence that – not only in signed-social networks – links do not form entirely independent from each other.

## 3.4. Modeling Triadic Structure

In order to test the functional relevance of triangles and other triadic subgraph patterns, it is necessary to generate synthetic networks exhibiting such structure. It will then be possible to simulate dynamical processes on such networks to investigate how they are affected by the abundance of three-node patterns.

A number of growth models exist which are capable of determining

*3. Triadic Substructures in Complex Networks*

the clustering coefficents by explicitly formulating 'triadic closure' processes. Starting from an initially unclustered network, one searches for edges with a common neighbor and then connects them successively to form triangles [10, 15, 65, 70, 82, 107]. Yet, the calculation of their properties is limited to numerical approaches [107].

An elegant way of parameterizing ensembles of networks with certain triadic structure is using the framework of ERGMs as defined in Section 2.4. The *Strauss model* [135], for instance, is defined in terms of the number of (undirected) edges, $M$, and the number of triangles, $T$. Its Hamiltonian is given by

$$\begin{aligned}\mathcal{H}^{\text{Str.}} &= \theta_1 \, M - \theta_2 \, T \\ &= \theta_1 \sum_{i<j} A_{ij} - \theta_2 \sum_{i<j<k} A_{ij} A_{jk} A_{ki}\end{aligned} \quad (3.8)$$

i.e. every additional edge in the system has a cost $\theta_1$ while, on the other hand, every triangle brings a gain of $\theta_2$. In principle, with an appropriate choice of $\theta_1$ and $\theta_2$, it is possible to adjust the interplay between these two mechansims to tune the expectation values $\langle M \rangle$ and $\langle T \rangle$. However, it was found that – for a wide range of parameters – although it is possible to specify the averages of the statistics correctly, the typical instantiations of the ensemble differ tremendously from the expectations. The sampled graphs are typically either very sparse or fully connected and accordingly the number of triangles is either $T \approx 0$ or $T \approx \binom{N}{3}$. This phenomenon is called *degeneracy* [56, 58, 115, 116]. The following considerations shall give some intuition on the emergence of this degeneracy. An analytic mean-field solution can be found in Ref. [116]. The cost associated with the creation of edges poses a barrier in the sampling process and thus many samples have a rather low density. However, once this barrier is overcome and a certain number of edges already exist, by adding only a few more at the right places, one can obtain large rewards by the second term in $\mathcal{H}^{\text{Str.}}$. In particular, it is possible to complete multiple triangles by a single additional link such that eventually large cliques emerge, potentially spanning the whole network. Fig. 3.8(a) displays the phase diagram of the Strauss model as calculated by Park and Newman. The area shaded in gray corresponds to the coexistence region in which both high-density and low-density graphs are sampled. Fig. 3.8(b) shows a scan of $\theta_2$ through the coexistence region.

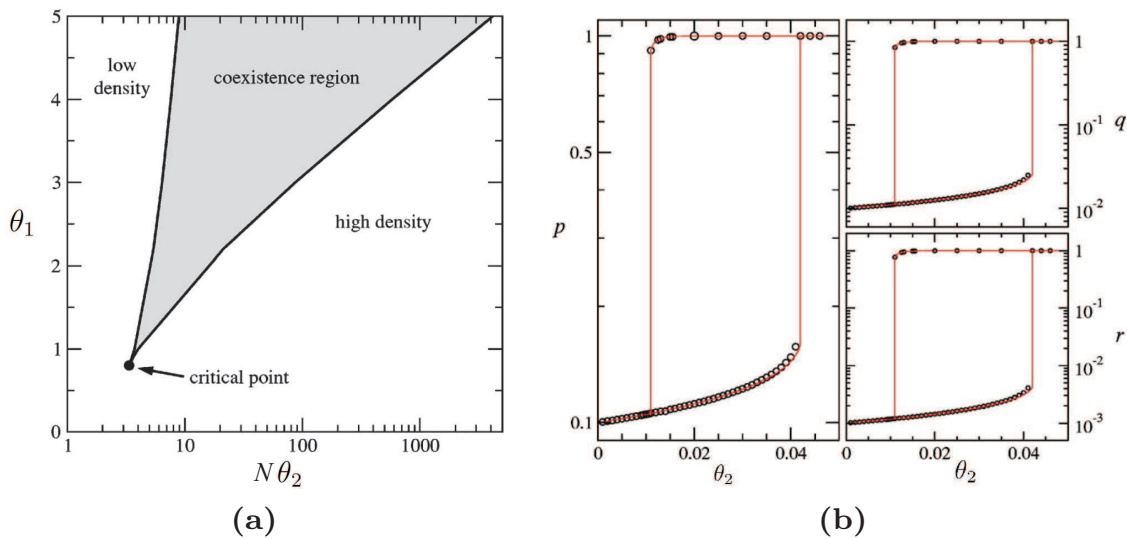

**Figure 3.8.:** **(a)** Phase diagram of the Strauss model. The shaded area corresponds to the coexistence region in which the system can be in either of two stable states, one of high density and one of low density. **(b)** Comparison of analytic mean-field solution (solid lines) and Monte Carlo simulation results (circles) for $p = \langle A_{ij} \rangle$, $q = \langle A_{ij} A_{jk} \rangle$, and $r = \langle A_{ij} A_{jk} A_{ki} \rangle$, for a system of $N = 500$ vertices and $\theta_1 = 2.2$. Adapted from Ref. [116], © 2005 by the APS.

Since $\mathcal{H}^{\text{Str.}}$, as defined in Eq. (3.8), involves *products* of entries in the adjacency matrix it does not fall into the class of dyadic-independence models. Therefore, in addition to the problem of degeneracy, it is difficult to sample from ERGMs of this kind.

Other alternatives suggested by Newman and Karrer generate networks in which both the number of links attached to every node and the number of certain subgraphs the nodes participate in are spec-ified initially [75, 107]. In analogy to the configuration model (see Section 2.3.2), the models yield networks, drawn uniformly at random from the set of all possible matchings of 'subgraph stubs'. With this generalization of random-graph models, it is possible to compute analytically component sizes, the existence and size of a giant component, and percolation properties. The model yields an unbiased ensemble of networks with clustering. However, attempting to specify the probabilities for all possible three-node subgraphs simultaneously poses a problem.

In Chapter 4 we will suggest a novel approach to model triadic structure, considering three-node subgraphs as the basic units of modeling.

## 3.5. Triad Motifs and Dynamical Processes

One of the main hypotheses in terms of the functional role of network motifs, is their relevance for controlling and stabilizing dynamical processs [5, 6, 129, 131]. It was suggested that, similar to the logic gates in a computer, "*Network motifs can be thought of as recurring circuits of interactions from which the networks are built.*" [6, page 1] and "*that each network motif can carry out specific information-processing functions.*" [6, page 1] The ultimate goal of motif analysis would therefore be to "*understand the dynamics of the entire network based on the dynamics of the individual building blocks.*" [5, page 27]

Klemm and Bornholdt studied the reliability of information processing on isolated triadic subgraphs in the presence of noise [81]. They found that certain patterns, e.g. the FFL  enhance reliability, whereas others, e.g. the loop  are detrimental to it. They further found that reliable patterns are overrepresented in the TSPs of many biological networks – particularly those of super families 1 and 2 in Fig. 3.7 – whereas unreliable three-node subgraphs are often supressed.

Similarly, Prill, Iglesias, and Levchenko studied the stability of a steady state on isolated patterns [120]. They observed correlations of the extent of stability for the isolated patterns and the abundance of these subgraphs in real-world networks. In analogy, Lodato, Boccaletti, and Latora studied the stability of synchronization processes [91], finding no obvious relation between the abundance of triadic subgraphs in real data and the synchronizability of these subgraphs considered in isolation.

It shall be emphasized that Refs. [81, 91, 120] studied the characteristics of *isolated* triad patterns and they related their observations to the frequency of occurrence of these patterns in real-world networks, which naturally consist of many vertices. Yet, it is not guaranteed that the dynamics on a pattern will be maintained when being embedded in a complex network in which many other nodes interact with the nodes of the patterns. Hence, in Chapter 7 we will investigate the influence of *non-isolated* motifs on dynamical processes.

Another approach is to optimize synthetic networks in terms of some dynamical process – e.g. with respect to the robustness against the failure of nodes or links – and eventually to study the motifs of the resulting networks [25, 71, 73]. Optimizing model gene-regulatory networks, Burda et al. discovered motifs that were also detected in real gene networks [25]. As a model for biological signal transduction, Kaluza et al. optimized flow networks with respect to their stability under link removal and found triadic $Z$-score profiles that almost perfectly resemble the second super familiy in Fig. 3.7 [71, 73].

Although these discoveries support the conjecture that motifs are important for systems to successfully perform their task, all of them suggest that *form follows function.* It remains open though, whether *form implies function.* In opposition to this hypothesis, it was argued that – even on a single, isolated motif pattern – the dynamics may show qualitatively very different behavior, depending on the choice of parameters [67]. In order to shed light on this open question, it is necessary to generate networks with appropriately adjusted abundances of motifs. In Chapter 4 we will suggest a novel class of network models that may help to accomplish this goal.

# 4. Generative Network Models Based on Steiner Triple Systems

*Parts of the content presented in this chapter have been prepublished in* M. Winkler and J. Reichardt. Motifs in Triadic Random Graphs Based on Steiner Triple Systems. *Phys. Rev. E, vol. 88, no. 2*, p. 022805, 2013. © *2013 by the American Physical Society (APS).*

Generally, dyadic relations between nodes are considered the fundamental building blocks of complex networks and hence also the fundamental unit when modeling a network. Erdös-Rényi graphs [41, 42], the configuration model [103, 104], stochastic block models [63, 112, 132, 145] and degree corrected block models [76] all fall into the class of dyadic models. The basic assumption underlying dyadic models is that dyads are *conditionally independent* given the model's parameters.

However, in Chapter 3 we have learned about evidence suggesting that the dyadic independence assumption may not be valid for many real systems. These show high clustering coefficients and non-vanishing triad significance profiles, i.e. some triad patterns appear significantly more frequently than in an ensemble of random graphs with the same degree distribution as the real networks (see Fig. 3.2). To our knowledge, to date, no general model exists that can appropriately model those triad significance profiles observed in many real-world networks, since it is necessary to test their functional relevance. In this chapter we will suggest a class of generative probabilistic models whose building blocks are not dyadically independently established edges, but rather triad patterns.

*4. Generative Network Models Based on Steiner Triple Systems*

Specifying generative models has proven difficult. Using the Strauss model [135], it is possible to generate systems with – on average – predefined link and triad appearance (see page 64). However, Park and Newman could show that the average does not describe the properties of a typical system generated by the model [116]. In fact, there is a large degenerate phase in which most instances of networks tend to be either fully connected or empty (see Fig. 3.8 on page 65).

In general, when trying to reproduce triad structures, models formulated in terms of dyads face the difficulty that each dyad influences an extensive number of triads. On the other hand, directly modeling all triad structures is impossible, as not all local triad configurations may be specified independently from each other. In Section 4.2 we will suggest a model which is based on triads which actually *can* be specified independently from each other, so-called Steiner triple systems [79].

Starting from the framework of Steiner triple systems, it will be possible to define a whole class of triadic exponential random graph models. In Chapter 5 we will discuss the most basic of such models: it assumes *the same* probability distribution of triadic subgraph configurations on all Steiner triples. This can be considered the triadic analogon to Erdös-Rényi graphs on dyadic models, in which the probability of an edge to be present is likewise *the same* for all dyads. In Section 4.1 we will introduce the concept of Steiner triple systems, discuss the prerequisites for their existence, and show how to construct them for various system sizes. Subsequently, in Section 4.2 we will define a class of generative network models based on Steiner triples.

## 4.1. Steiner Triple Systems

In a network of $N$ nodes there are $T = \binom{N}{3}$ distinct triads. Yet, it is not possible to specify all their triadic-subgraph configurations independently of each other; e.g. consider the network in Fig. 4.1. Suppose we set the relations in the three-node subgraph of nodes 1, 2, and 3, denoted as $(1, 2, 3)$, such that they adopt pattern . Further, we spec-ify the triads $(1,4,5)$ and $(4,6,2)$ such that they assume patterns △ and △, respectively. Consequently, with the choices for the discussed three triads in Fig. 4.1, the subgraph of $(4,1,2)$ is already determined to take the pattern △ implicitly. This derives from the fact that $(4,1,2)$ contains dyadic relations which have already been assigned in the other three triads.

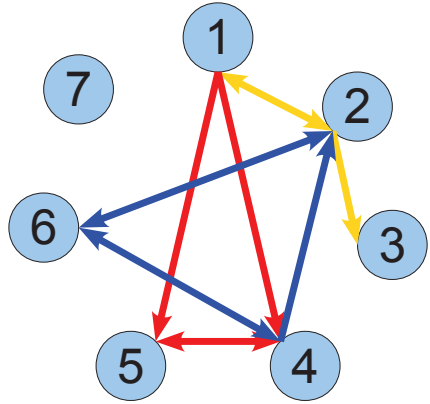

**Figure 4.1.:** Only a few triad configurations can be specified independently of each other: e.g. a specification of the triads $(1,2,3)$, $(1,4,5)$, and $(2,4,6)$ fully determines the configuration of $(1,2,4)$ [152].

Since there are only $D = \binom{N}{2}$ dyads in a network and every triad comprises three dyadic relations, there is an upper bound to the number of triads which are dyad-disjoint and thus can be set without over-determining the system:

$$\text{No. of dyad-disjoint triads} \leq \frac{D}{3} = \frac{N(N-1)}{6} \ll T \qquad (4.1)$$

Networks for which the upper bound is exactly met can be partitioned into triples such that every pair of nodes in the system is part of *exactly*

*4. Generative Network Models Based on Steiner Triple Systems*

*one* of them. Such systems are called Steiner triple systems (STSs)[1]. Accordingly, the triples to which the dyads are assigned are referred to as Steiner triples (STs). STSs consisting of $N$ vertices are called *Steiner triple systems of order $N$*, or STS($N$).

There are two necessary and sufficient requirements for the existence of an STS($N$),

$$\begin{aligned} N \bmod 2 &= 1 \\ N(N-1) \bmod 3 &= 0. \end{aligned} \quad (4.2)$$

To motivate the first constraint, regard that every node $i$ is part of $N-1$ dyads. Since there are exactly two of these dyads in every Steiner triple $i$ participates in, $N-1$ must be even and thus $N$ be odd. Moreover, for the upper bound in Eq. (4.2) to be exactly met, one requires $D \bmod 3 \stackrel{!}{=} 0$. However, this is equivalent to $N(N-1) \bmod 3 \stackrel{!}{=} 0$, which is the second constraint in Eq. (4.2). That these conditions are even sufficient for the existence of STSs will be proved in Section 4.1.1 (see also [57, page 277ff] or [141, page 205ff]). The problem was originally solved by Kirkman in 1847 [79].

In fact the proof is constructive, i.e. it even shows how to design a system of size $N$ obeying Eqs. (4.2). After the rather abstract proof, in Section 4.1.2 we will illustrate the construction by explicitly generating some Steiner triple systems. The reader who is not interested in technical details, but rather willing to get an idea of the concept of Steiner triple systems may therefore skip Section 4.1.1. However, to be able to implement an algorithm to construct STSs, Section 4.1.1 is essential. An implementation of an STS-constructor is made publicly available at [151].

---

[1]Steiner triple systems are a special case of the more general $t(v, k, \lambda)$ or $S_\lambda(t, k, v)$ designs, in which $v$ denotes the number of points and $k$ denotes the cardinality of the blocks (three for triangles). For any set $T$ of $t$ points, there are exactly $\lambda$ blocks incident with all points in $T$. Thus, Steiner triple systems are $2(v, 3, 1)$ or $S_1(2, 3, v)$. For more details see e.g. [141].

### 4.1.1. Existence of Steiner Triple Systems

We will now present a proof that Eqs. (4.2) are indeed sufficient. The proof is in major parts adapted from [57, page 279ff].

**Theorem 4.1.1** *If there are Steiner triple systems of orders $N_1$ and $N_2$, there is a Steiner triple system of order $N = N_1 N_2$ containing subsystems isomorphic to those of orders $N_1$ and $N_2$.*

**Proof:** Let A be an STS of order $N_1$ and B be an STS of order $N_2$, respectively. Further, let $(a_i, a_j, a_k)$ be any triple of A, and $(b_r, b_s, b_u)$ be any triple of B with $i, j, k \in \{1, 2, \ldots, N_1\}$ and $r, s, u \in \{1, 2, \ldots, N_2\}$.

Form a new system C with elements $c_{ij}$, $i \in \{1, 2, \ldots N_1\}$, $j \in \{1, 2, \ldots N_2\}$. Then $(c_{ir}, c_{js}, c_{ku})$ with $i \leq j \leq k$ is taken as a triple of system C if either

1. $(a_i, a_j, a_k)$ is a triple of A and $r = s = u$; or
2. $i = j = k$ and $(b_r, b_s, b_u)$ is a triple of B; or
3. $(a_i, a_j, a_k)$ is a triple of A and $(b_r, b_s, b_u)$ is a triple of B

One can think of the system as having the coarse structure of A with every vertex possessing a substructure according to B or vice versa. Those triples with $r = s = u$ form $N_2$ subsystems of C isomorphic to A and those with $i = j = k$ form $N_1$ subsystems isomorphic to B. ∎

An application of Theorem 4.1.1 for the construction of an STS(63) from an STS(7) and an STS(9) will be shown in section 4.1.2 and figure 4.3.

**Theorem 4.1.2** *If there is a Steiner triple system of order $N_2$ containing a subsystem of order $N_3$ (or $N_3 = 1$), and if there is a system of order $N_1$, we can construct a system of order $N = N_3 + N_1 (N_2 - N_3)$ containing $N_1$ subsystems of order $N_2$ and one of order $N_1$ and order $N_3$.*

*4. Generative Network Models Based on Steiner Triple Systems*

**Proof:** The $N = N_3 + N_1 (N_2 - N_3)$ elements of the Steiner triple system that we wish to construct can be arranged in $(N_1 + 1)$ rows:

$$
\begin{array}{cccccc}
S_0: & a_1 & a_2 & \ldots & a_{N_3} & \\
S_1: & b_{11} & b_{12} & \ldots & \ldots & b_{1s} \\
S_2: & b_{21} & b_{22} & \ldots & \ldots & b_{2s} \\
\ldots & \ldots & \ldots & \ldots & \ldots & \ldots \\
S_{N_1}: & b_{N_1 1} & b_{N_1 2} & \ldots & \ldots & b_{N_1 s}
\end{array}
\tag{4.3}
$$

The first row contains $N_3$ elements, each of the $N_1$ following rows contains $s \equiv N_2 - N_3$ elements.

An STS($N$) is defined from these elements by the following three rules:

1. By definition there is a Steiner triple system of order $N_3$. We accept triples $(a_i, a_j, a_k)$ if $(i, j, k)$ is a triple of the latter. Thus, all dyads including two elements $a_i$ are covered exactly once.

2. Combine the elements of $S_0$ and any $S_i$. Together they have $N_3 + s = N_3 + N_2 - N_3 = N_2$ elements: $a_1, a_2, ..., a_{N_3}, b_{i1}, b_{i2}, ..., b_{iN_s}$. By definition there is an STS($N_2$) with a subsystem of order $N_3$. Be $m \in \{1, 2, ..., N_3\}$, indicating the node labels of the subsystem of order $N_3$. Further, be $j, k$, and $r \in \{N_3 + 1, N_3 + 2, ..., N_2\}$, indicating the labels of the remaining nodes. We accept triples $(a_m, b_{ij}, b_{ik})$ if $(m, j, k)$ is a triple of the STS($N_2$) and we accept triples $(b_{ij}, b_{ik}, b_{ir})$ if $(j, k, r)$ is a triple of the latter. Triples including more than one $a_i$ cannot be accepted because they were already captured by 1.

3. Accept all triples $(b_{jx}, b_{ky}, b_{rz})$ if $(j, k, r)$ is a triple of the STS($N_1$) **and** if $x + y + z \equiv 0 \mod s$.
One can convince oneself that (for fixed $(i, k, r)$) every dyad between groups $i, k$, and $r$ is considered exactly once. Pick, e.g.,

arbitrary $x$ and $y$; the constraint $x + y + z \equiv 0 \mod s$ uniquely defines the corresponding $z = (2s - x - y) \mod s$.

In summary, 1 defines a subsystem of order $N_3$ consisting of the nodes in the first row of (4.3). 2 defines $N_1$ subsystems of order $N_2$, while the triples defined in 1 are part of each of these subsystems but, of course, for the STS($N$) are considered only once. With the rules 1 and 2, all triples including any $a_i$ are defined. Furthermore, all triples including more than one distinct element from any row in (4.3) are specified. 3 accounts for Steiner triples consisting of nodes from three different rows. In all cases, any dyad of nodes is uniquely mapped to a single Steiner triple. Thus, 1, 2, and 3 define an STS($N$). ∎

**Theorem 4.1.3** *If $N = 6\,t + 1$ or $N = 6\,t + 3$, there is a Steiner triple system of order $N$.*

**Proof:** In Theorem 4.1.2 we saw how to construct a STS($N = N_3 + N_1\,(N_2 - N_3)$) supposed we know the structure of an STS($N_1$), an STS($N_2$), and an STS($N_3$).

Let us now consider the following choices of $N_1$, $N_2$, and $N_3$ to construct an STS($N$) given an STS($N'$):

$$
\begin{array}{llllll}
(A) & N_1 = N' & N_2 = 3 & N_3 = 1 & N = 2N' + 1 & N' \geq 3 \\
(B) & N_1 = 3 & N_2 = N' & N_3 = 1 & N = 3N' - 2 & N' \geq 3 \\
(C) & N_1 = 3 & N_2 = N' & N_3 = 3 & N = 2N' - 6 & N' \geq 7 \\
(D) & N_1 = N' & N_2 = 9 & N_3 = 3 & N = 6N' + 3 & N' \geq 3 \\
(E) & N_1 = 3 & N_2 = N' & N_3 = 7 & N = 3N' - 14 & N' \geq 15 \\
(F) & N_1 = N' & N_2 = 7 & N_3 = 1 & N = 6N' + 1 & N' \geq 3 \\
\end{array} \quad (4.4)
$$

These rules allow us to construct systems recursively depending on their residue of $N \mod 36$ (see Table 4.1):

*4. Generative Network Models Based on Steiner Triple Systems*

| $N$ | Rule | $N'$ |
|---|---|---|
| $36\,\tau + 1$ | (B) | $12\,\tau + 1$ |
| $36\,\tau + 3$ | (A) | $18\,\tau + 1$ |
| $36\,\tau + 7$ | (F) | $6\,\tau + 1$ |
| $36\,\tau + 9$ | (D) | $6\,\tau + 1$ |
| $36\,\tau + 13$ | (E) | $12\,\tau + 9$ |
| $36\,\tau + 15$ | (A) | $18\,\tau + 7$ |
| $36\,\tau + 19$ | (F) | $6\,\tau + 3$ |
| $36\,\tau + 21$ | (D) | $6\,\tau + 3$ |
| $36\,\tau + 25$ | (B) | $12\,\tau + 9$ |
| $36\,\tau + 27$ | (A) | $18\,\tau + 13$ |
| $36\,\tau + 31$ | (A) | $18\,\tau + 15$ |
| $36\,\tau + 33$ | (C) | $12\,\tau + 13$ |

**Table 4.1.:** Construction of STS($N$) from a STS($N'$) grouped by the residue of $N$ mod $36$.

E.g. for $N = 36\,\tau + 1$ and application of rule (B) in Eq. 4.4 we get $36\,\tau + 1 \stackrel{!}{=} 3N' - 2$ and thus $N' = 12\,\tau + 1$. For an according choice of $\tau$, the left column in Table 4.1 can contain any value $N = 6\,t + 1$ or $N = 6\,t + 3$. Hence, if we can construct a system of any size in the left column we have proven Theorem 4.1.3. It shall be further mentioned that all entries in the $N'$ column are also congruent to either 1, 3, 7, 9, 13, 15, 19, 21, 25, 27, 31, or 33 mod 36 and therefore comply with Eqs. 4.2.

The Steiner triple system of order three is trivial:

$$(1, 2, 3) \,. \tag{4.5}$$

The STS of the next possible order, seven, can be constructed deductively. This is done in Section 4.1.2 in which the result is shown in Eq. (4.6). Theorem 4.1.1 allows to construct an STS of order $9 = 3 \times 3$. Furthermore, Theorem 4.1.2 enables the generation of systems with $N = 13 = 7 + 3 \times (9 - 7)$ and $N = 15 = 3 + 3 \times (7 - 3)$.

Having STSs of orders 3, 7, 9, 3, and 15, we are able to construct Steiner triple systems of any order smaller than 36 by setting $\tau = 0$ in Table 4.1. However, for any choice of $\tau$, Table 4.1 tells us how to construct all possible Steiner triple systems of order $N \in [36\,\tau; 36\,\tau + 35)$ requiring solely the knowledge of an STS($N'$) with $N' < 36\,\tau$. Hence, we have proven Theorem 4.1.3 inductively. ∎

From Eq. (4.2) we can conclude that, by approximation, systems of arbitrary size can be decomposed into Steiner triples. All there is to do is either to add up to three 'dummy' nodes to the system, or to ignore up to three nodes including their relations.

### 4.1.2. Construction of Steiner Triple Systems

In order to clarify the idea behind a decomposition into Steiner triples, Fig. 4.2 shows the partition of a Steiner triple system of order seven into its Steiner triples. Due to the rather small amount of vertices, it is possible to derive the STS deductively: Without loss of generality, we start with node 1. Since 1 is part of six dyads – one with every remaining node – it has to be part of three Steiner triples. The first one shall be (1,2,3) (color coded in yellow in Fig. 4.2), the second one (1,4,5) (red), and the third one (1,6,7) (cyan). Now each dyadic relation 1 participates in is covered by exactly one Steiner triple. We continue with the dyads of node 2: those with nodes 1 and 3 are already contained in $(1, 2, 3)$. 4 and 5 are already part of ST $(1, 4, 5)$ and therefore need to be assigned to different Steiner triples. We choose 6 to be in the Steiner triple with 2 and 4 (blue), and thus we have implicitly specified Steiner triple $(2, 5, 7)$ (green). Continuing with node 3, the dyads with nodes 4, 5, 6, and 7 need to be assigned to Steiner triples. 4 is already assigned to Steiner triple systems with 5 (red) and 6 (blue). Thus, the two remaining Steiner triples are $(3, 4, 7)$ (magenta) and $(3, 5, 6)$ (orange).

Hence, from the $\binom{7}{3} = 35$ possible triads of a network of order seven, $D/3 = 7 \cdot 6/6 = 7$ can be specified independently from each other.

*4. Generative Network Models Based on Steiner Triple Systems*

**Figure 4.2.:** Schematic presentation of a Steiner triple system of order seven. The Steiner triples are set to be (1,2,3), (1,4,5), (1,6,7), (2,4,6), (2,5,7), (3,4,7), and (3,5,6), as indicated by the colors of the matrix elements. Every matrix element is assigned to exactly one Steiner triple [152].

Eq. (4.6) shows all triads for a network of seven nodes and highlights the choice of an STS as constructed above, with colors corresponding to Fig. 4.2.

$$
\begin{array}{lllll}
(\mathbf{1,2,3}) & (1,3,6) & (\mathbf{1,6,7}) & (2,4,7) & (\mathbf{3,5,6}) \\
(1,2,4) & (1,3,7) & (2,3,4) & (2,5,6) & (3,5,7) \\
(1,2,5) & (\mathbf{1,4,5}) & (2,3,5) & (\mathbf{2,5,7}) & (3,6,7) \\
(1,2,6) & (1,4,6) & (2,3,6) & (2,6,7) & (4,5,6) \\
(1,2,7) & (1,4,7) & (2,3,7) & (3,4,5) & (4,5,7) \\
(1,3,4) & (1,5,6) & (2,4,5) & (3,4,6) & (4,6,7) \\
(1,3,5) & (1,5,7) & (\mathbf{2,4,6}) & (\mathbf{3,4,7}) & (5,6,7)
\end{array}
\qquad (4.6)
$$

Of course, for larger system sizes it is not practical to construct Steiner triple systems the way described above. Though, larger Steiner triple systems can be constructed by merging smaller ones. For example, knowing the structure of an $\mathrm{STS}(N_1)$ and an $\mathrm{STS}(N_2)$, one can construct an $\mathrm{STS}(N_1 \cdot N_2)$, as implied by Theorem 4.1.1. Since we know the trivial Steiner triple system of order three, STS(3) = $\{(1,2,3)\}$, we can construct an STS(9) as follows. Consider three subsystems: 1, 2, and 3. Each subsystem $\mathcal{I}$ consists of three nodes: $\mathcal{I}_1$, $\mathcal{I}_2$, and $\mathcal{I}_3$. For the first three Steiner triples we choose $(1_1, 1_2, 1_3)$, $(2_1, 2_2, 2_3)$, and $(3_1, 3_2, 3_3)$ as indicated by the red, dashed lines in Fig. 4.3(a). In the next step, we choose $(1_i, 2_i, 3_i)$ for $i = 1, 2, 3$ (Fig. 4.3 (b)-(d)). Subsequently, we select the even permutations in the subsystems: $(1_1, 2_2, 3_3)$, $(1_2, 2_3, 3_1)$, and $(1_3, 2_1, 3_2)$ (Fig. 4.4 (a)-(c)). And eventually the odd permutations: $(1_1, 2_3, 3_2)$, $(1_2, 2_1, 3_3)$, and $(1_3, 2_2, 3_1)$ (Fig. 4.4 (d)-(f)). Since an STS(9) has $\frac{9 \cdot 8}{6} = 12$ Steiner triples, we have defined all of them. With $1_1 \equiv 1$, $1_2 \equiv 2$, $1_3 \equiv 3$, $2_1 \equiv 4$, ..., and $3_3 \equiv 9$, the STS(9) reads

$$\begin{array}{lll} (1,2,3) & (4,5,6) & (7,8,9) \\ (1,4,7) & (2,5,8) & (3,6,9) \\ (1,5,9) & (2,6,7) & (3,4,8) \\ (1,6,8) & (2,4,9) & (3,5,7)\,. \end{array} \qquad (4.7)$$

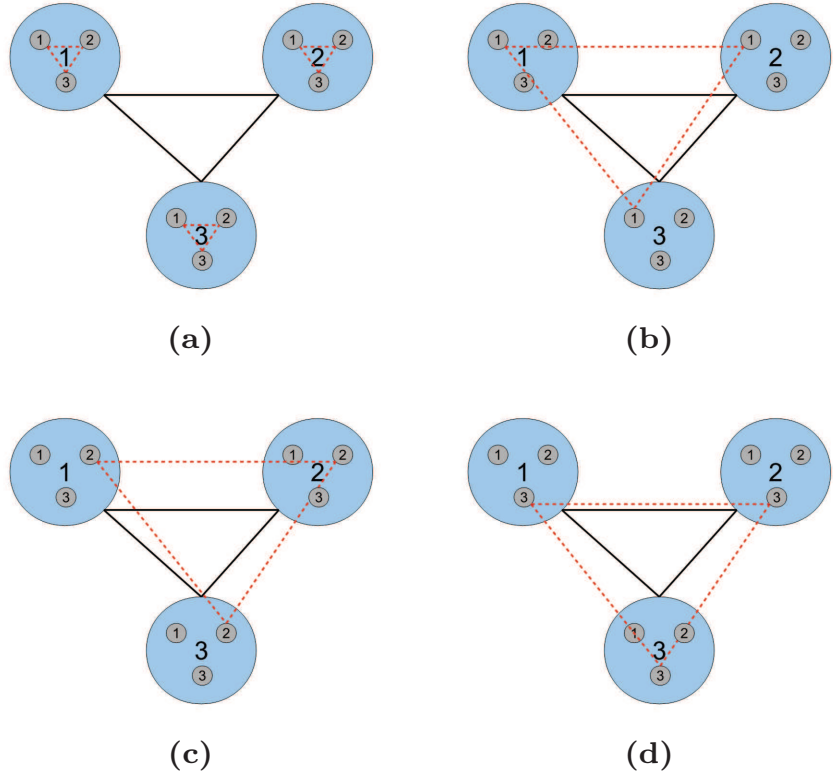

**Figure 4.3.:** Construction of a Steiner triple system of order nine by merging three systems of order three. The red, dashed lines indicate the choices of STs in the respective steps. Note that the lines do not mean that there is a link between the corresponding nodes. They rather indicate that they are part of the same ST!

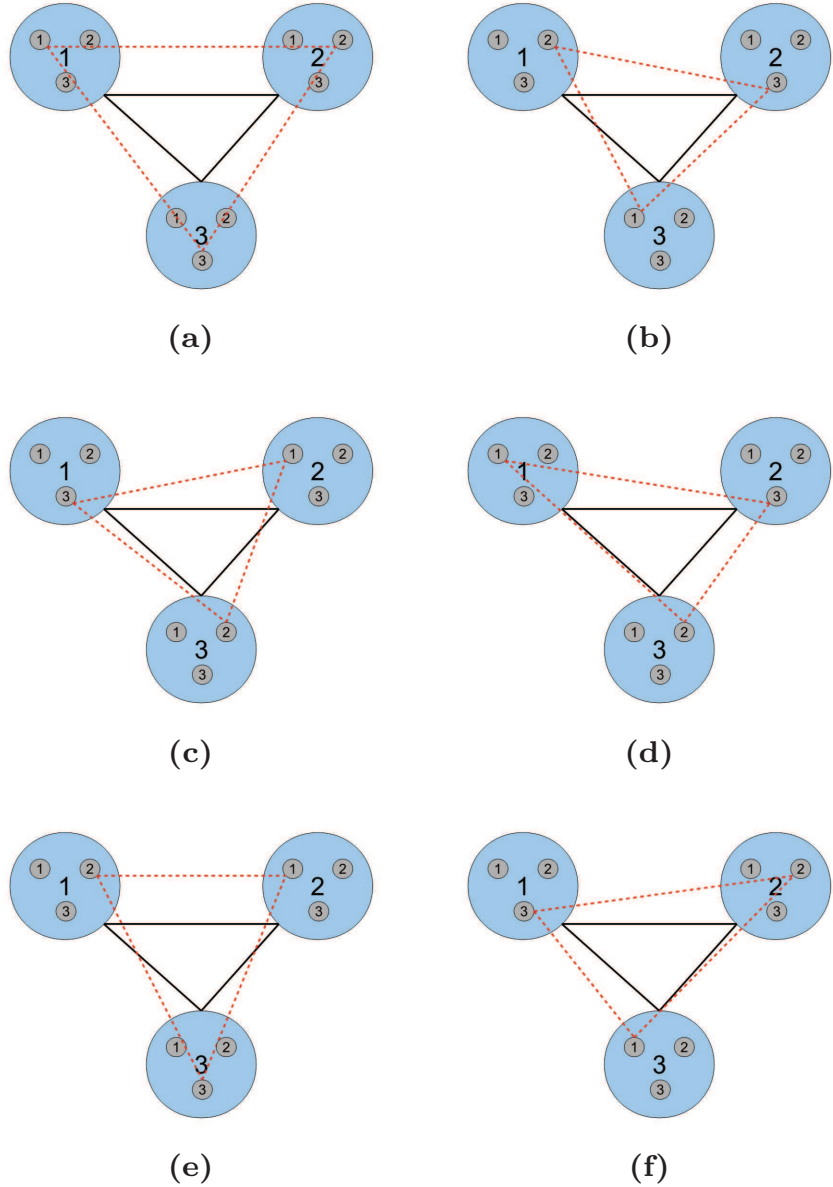

**Figure 4.4.:** Construction of a Steiner triple system of order nine by merging three systems of order three. The red, dashed lines indicate the choices of STs in the respective steps. Note that the lines do not mean that there is a link between the corresponding nodes. They rather indicate that they are part of the same ST!

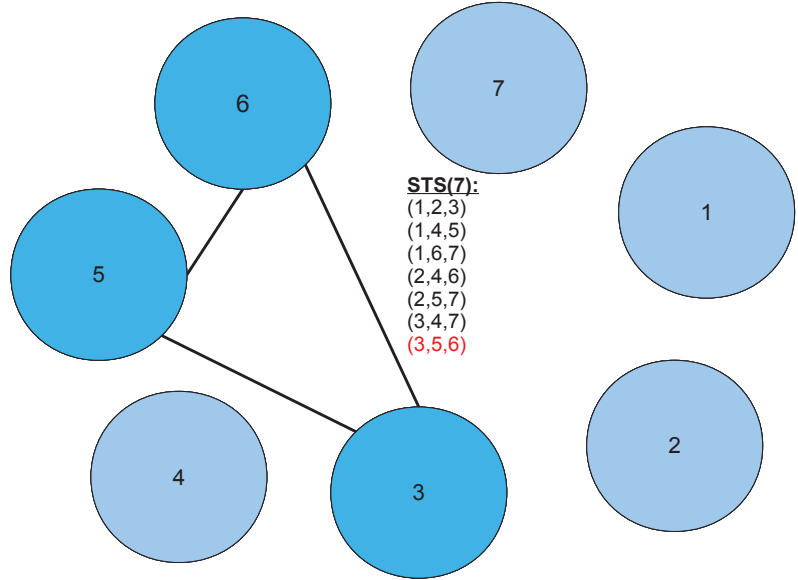

**Figure 4.5.:** Construction of an STS(63) from seven STS(9). The construction algorithm iterates over all Steiner triples of the STS(7) on the coarse level.

Finally, the combination of more complex systems shall be illustrated with an STS($63 = 7 \cdot 9$) obtained from seven STS(9) (see Fig. 4.5). In total, the STS(63) must have $\frac{63 \cdot 62}{6} = 651$ STs. Within each STS(9) there are 12 STs, i.e., there are $651 - 7 \cdot 12 = 567$ missing. Moreover, we can define an STS(7) on the coarser level. Be $(\mathcal{I}, \mathcal{J}, \mathcal{K})$ an ST of the STS(7) on the coarser level, then we choose $(\mathcal{I}_i, \mathcal{J}_i, \mathcal{K}_i)$ as a Steiner

*4. Generative Network Models Based on Steiner Triple Systems*

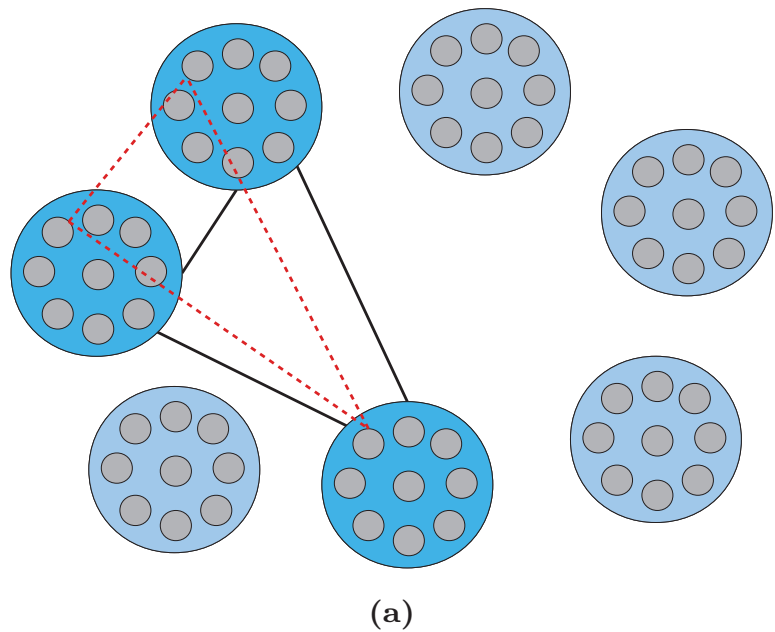

**(a)**

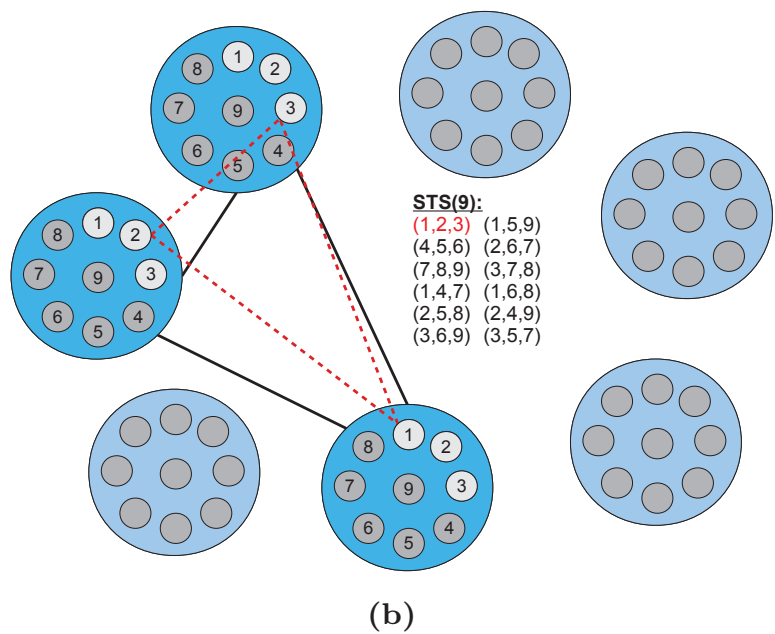

**(b)**

**Figure 4.6.:** Construction of an STS(63) from seven STS(9). **(a)** STs are selected such that the nodes within the STS(9) are at the same position. **(b)** Example of a permutation of the triple (1,2,3) at the lower STS(9) level.

triple of the STS(63) (see Fig. 4.6(a)). Doing this for the whole STS(7) and for $i = 1, 2, 3, ..., 9$ we have selected addditional $7 \cdot 9 = 63$ ST, i.e., there are $567 - 63 = 504$ left. Finally, for every $(\mathcal{I}, \mathcal{J}, \mathcal{K}) \in \text{STS}(7)$ and every $(i, j, k) \in \text{STS}(9)$ we take the three even, as well as the three odd permutations (analogously to Fig. 4.4 (a)-(c) and Fig. 4.4 (d)-(f), respectively). Proceeding like this, we finally obtain our $7 \cdot 12 \cdot 6 = 504$ missing STs and have constructed an STS(63).

For the STS(7) – as well as for the STS(3) and the STS(9) – the partitions described above are unique, apart from relabeling nodes. For STSs of higher orders, there are multiple nonisomorphic ways to partition the nodes into Steiner triples. For $N = 13$, one obtains exactly two nonisomorphic solutions, which differ in terms of four triads from each other, while for $N = 15$ there are 80 distinct Steiner triples [57]. For $N = 19$ there are already $11,084,874,829$ distinct Steiner triple systems [30].

STSs provide us with sets of triads which can actually be configured without overdetermining dyadic relations. On the other hand, *all* dyadic relations are captured by a Steiner triple. STSs can thus be considered a basis to express the adjacency matrix: instead of specifiying the configurations between all pairs of nodes, we can specify the configurations of all STs to determine the graph structure. The fact that the partition into STs is not unique should not pose a problem in this context as it is sufficient to know *one* of the possible decompositions into triads whose configurations can be specified independently of each other. However, when learning parameters on models defined in terms of STs, the inferred values may vary for different choices of STSs.

## 4.2. Model

In order to account for substructures of higher than dyadic order, our goal is now to define models based on triadic rather than dyadic entities. Since Steiner triple systems assign every dyadic relation, i.e. every pair of nodes, to exactly one triad, the specification of the configurations

## 4. Generative Network Models Based on Steiner Triple Systems

of all Steiner triples is equivalent to specifying an adjacency matrix $\boldsymbol{A}$. To demonstrate that a formulation of a network in terms of Steiner triples is equivalent to a formulation in terms of dyads, consider a directed unweighted graph with $N$ vertices. There are $\binom{N}{2}$ dyads. Each dyad $(i,j)$ may adopt four distinct configurations. Thus, in total there are $4^{\binom{N}{2}} = 2^{2\binom{N}{2}}$ possible states of the system, i.e. distinct adjacency matrices. On the other hand, there are $\binom{N}{2}/3$ distinct Steiner triples. Each of those triples may assume $2^6 = 64$ distinct configurations (see Fig. 4.7). Therefore, again we obtain $64^{\binom{N}{2}/3} = 2^{6\binom{N}{2}/3} = 2^{2\binom{N}{2}}$ distinct possible states, i.e. the number of degrees of freedom is the same. The argument for undirected graphs is analogous.

Let us recall that dyadic ERGMs assume that the likelihoods for the presence of two edges are conditionally independent of each other. Further, let the matrix $\boldsymbol{D}$ with components $D_{ij} \in \{0,1\}$ denote the random variables corresponding to the entries of an adjacency matrix. Then, the assumption of independence implies for the likelihood of observing an adjacency matrix $\boldsymbol{A}$ (compare Eq. (2.37))

$$\begin{aligned} \mathcal{P}\left(\mathbf{D} = \mathbf{A}\,|\,\vec{\theta}\right) &= \prod_{i=1}^{N-1} \prod_{j=i+1}^{N} \mathcal{P}\left(D_{ij} = A_{ij}, D_{ji} = A_{ji}|\vec{\theta}\right) \\ &= \prod_{i=1}^{N-1} \prod_{j=i+1}^{N} \mathcal{P}\left(\vec{D}_{(i,j)} = \vec{A}_{(i,j)}|\vec{\theta}\right) \end{aligned} \quad (4.8)$$

where $\vec{\theta}$ includes all parameters of the model. The vector notation on the right hand side accounts for the fact that in directed unweighted networks, there are four possible dyadic relations[2]. They can be combined in a four dimensional indicator vector $\vec{A}_{(i,j)}$ with all components being zero, except for one being one.

We will now employ the concept of Steiner triple systems to define the triadic analogon to Eq. (4.8). Now, instead of assuming the likelihoods of *dyads* to be conditionally independent of each other, we suppose the likelihoods for the configurations on *Steiner triples* to be conditionally independent. With this assumption, the likelihood of observing an adjacency matrix $\boldsymbol{A}$ factorizes as follows:

$$\mathcal{P}\left(\mathbf{D} = \mathbf{A} \,|\, \vec{\theta}\right) = \prod_{\sigma=1}^{N(N-1)/6} \mathcal{P}\left(\vec{D}_\sigma = \vec{A}_\sigma | \vec{\theta}\right)$$
$$= \prod_{\sigma=1}^{N(N-1)/6} \vec{\mathcal{P}}\left(\vec{D}_\sigma | \vec{\theta}\right) \cdot \vec{A}_\sigma \quad (4.9)$$

where $\sigma$ denotes the Steiner triples of an STS($N$), $\vec{D}_\sigma$ is an indicator variable for the configuration of Steiner triple $\sigma$, and $\vec{A}_\sigma$ is a value of this variable. Analogously to Eq. (4.8), for each of the vectors exactly one component is unity, while all others are zero, which is equivalent to the fact that a triad cannot be in multiple configurations at the same time. For undirected networks it is $\vec{D}_\sigma \in \{0,1\}^8$, for directed ones it is

---

[2]Those four dyadic relations are: $A_{ij} = 0 \wedge A_{ji} = 0$, $A_{ij} = 0 \wedge A_{ji} = 1$, $A_{ij} = 1 \wedge A_{ji} = 0$, and $A_{ij} = 1 \wedge A_{ji} = 1$.

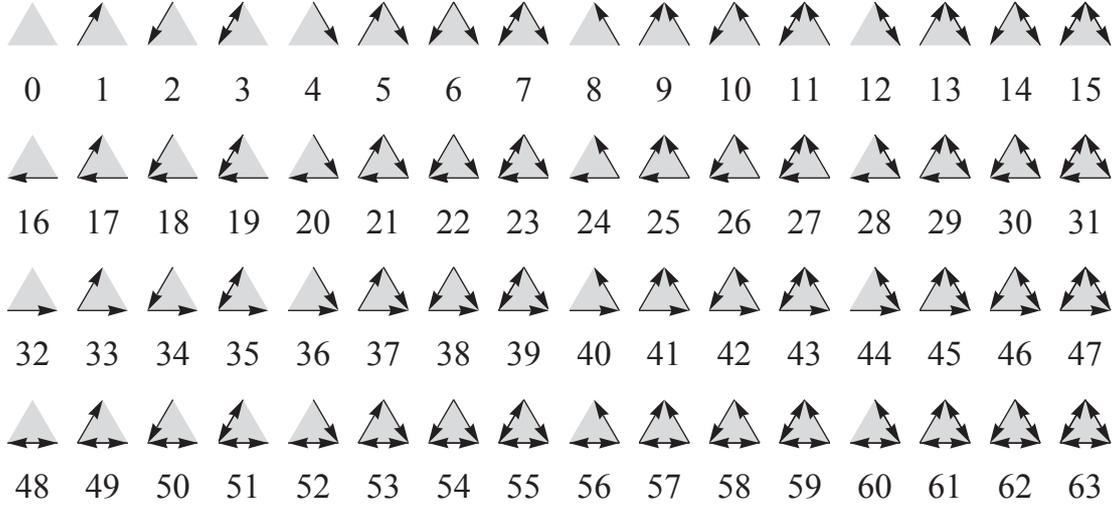

**Figure 4.7.:** All possible triad *configurations* in directed unweighted networks.

## 4. Generative Network Models Based on Steiner Triple Systems

$\vec{D}_\sigma \in \{0, 1\}^{64}$. Accordingly, it is $\vec{\mathcal{P}}\left(\vec{D}_\sigma | \vec{\theta}\right) \in [0, 1]^8$ or $[0, 1]^{64}$, respectively with the sums of the elements normalized to one. By defining Eq. (4.9) we make the assumption that the likelihoods of Steiner triple configurations factorize, i.e. they are conditionally independent of each other.

Eq. (4.9) defines the class of triadic-independence models based on Steiner triple systems. It shall be emphasized that it does not yet define the shape of the probability distributions $\vec{\mathcal{P}}\left(\vec{D}_\sigma | \vec{\theta}\right)$. The simplest case one can think of is the probability distribution to be the same for all STs, i.e. $\vec{\mathcal{P}}\left(\vec{D}_\sigma | \vec{\theta}\right) = \vec{\mathcal{P}}\left(\vec{D} | \vec{\theta}\right)$ for all $\sigma$. In order to better get to know the features of this type of generative network model, we will investigate it in detail in the following chapter.

# 5. The Triadic Random Graph Model

*The content presented in this chapter has been prepublished in* M. Winkler and J. Reichardt. Motifs in Triadic Random Graphs Based on Steiner Triple Systems. *Phys. Rev. E, vol. 88, no. 2*, p. 022805, 2013. © *2013 by the APS.*

In Chapter 2 we have learned that in directed unweighted networks, there are 16 non-isomorphic triadic subgraph patterns (see Fig. 5.1). We will now investigate how a probabilistic distribution of these triad patterns on Steiner triples affects the triad significance profiles of the corresponding networks. With this work, we provide for a new type of generative models – which we term *triadic random graphs* – capable of modeling structure of higher than dyadic order. In fact, we will show that they enable us to generate networks with non-vanishing $Z$ scores for the different triad patterns. Furthermore, we will unravel correlations in the abundance of triad patterns which we believe, occur solely for statistical reasons and discuss their implications for the functional

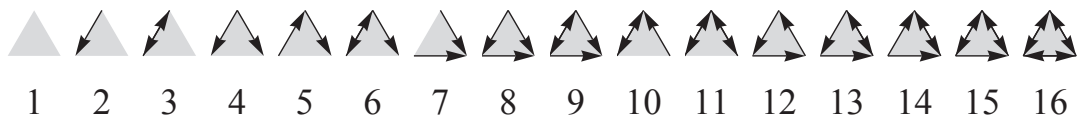

**Figure 5.1.:** All 16 possible non-isomorphic triadic subgraphs (subgraph patterns) in directed unweighted networks.

## 5. The Triadic Random Graph Model

interpretation of motifs. Moreover, we will calculate the degree distribution of triadic random graphs analytically.

Eq. (4.9) describes the most general formulation of models based on conditionally independent Steiner triples (STs). We will now study the properties of a particular realization of this class of models. The simplest such model has the same likelihood distribution for the triad configurations on all Steiner triples $\sigma$, $\vec{\mathcal{P}}\left(\vec{D}_\sigma|\vec{\theta}\right) = \vec{\mathcal{P}}\left(\vec{D}|\vec{\theta}\right)$. This can be regarded the triadic analogon to dyadic Erdös-Rényi (ER) graphs (see Section 2.3.1), in which the likelihood for the existence of an edge is the same for all dyads. We will refer to the model as the *triadic random graph model (TRGM)*. Since the ordering of the nodes in a Steiner triple is arbitrary, there is no need to distinguish between isomorphic triad configurations. For example, the likelihoods of the three configurations of subgraph 4, shown in Fig. 2.3, will be the same. Thus, the triadic random graphs (TRGs) have 16 parameters, each of them indicating the probability of a Steiner triple to assume one of the subgraphs shown in Fig. 5.1. Of course, their values need to sum up to unity.

Given the parameters $p(i)$ for each pattern $i$, the probability disitribution of each Steiner triple is given by

$$\vec{\mathcal{P}}\left(\vec{D}_\sigma|\vec{\theta}\right) = \boldsymbol{M}\vec{\mathcal{P}}$$
$$= \boldsymbol{M}\Big(p(\triangle), p(\triangle), p(\triangle), p(\triangle),$$
$$p(\triangle), p(\triangle), p(\triangle), p(\triangle), p(\triangle), p(\triangle),$$
$$p(\triangle), p(\triangle), p(\triangle), p(\triangle), p(\triangle), p(\triangle)\Big)^T. \quad (5.1)$$

The matrix $\boldsymbol{M}$ maps each of the 16 non-isomorphic triadic subgraph patterns in Fig. 5.1 to their corresponding isomorphic configurations

(see Fig. 4.7 on page 85) with equal probability,

$$\boldsymbol{M} = \begin{pmatrix} 1 & 0 & 0 & 0 & 0 & 0 & 0 & 0 & 0 & \cdots \\ 0 & \frac{1}{6} & 0 & 0 & 0 & 0 & 0 & 0 & 0 & \cdots \\ 0 & \frac{1}{6} & 0 & 0 & 0 & 0 & 0 & 0 & 0 & \cdots \\ 0 & 0 & \frac{1}{3} & 0 & 0 & 0 & 0 & 0 & 0 & \cdots \\ 0 & \frac{1}{6} & 0 & 0 & 0 & 0 & 0 & 0 & 0 & \cdots \\ 0 & 0 & 0 & 0 & \frac{1}{6} & 0 & 0 & 0 & 0 & \cdots \\ 0 & 0 & 0 & \frac{1}{3} & 0 & 0 & 0 & 0 & 0 & \cdots \\ \vdots & \vdots & \vdots & \vdots & \vdots & \vdots & \vdots & \vdots & \vdots & \ddots \end{pmatrix}_{64 \times 16}, \qquad (5.2)$$

i.e. every row of $\boldsymbol{M}$ has exactly one entry and the sums of its columns are normalized to one.

Hence, we can formulate the probability distribution of the triadic random graph model,

$$\mathcal{P}\left(\mathbf{D} = \mathbf{A} \,|\, \vec{\theta}\right) = \prod_{\sigma=1}^{N(N-1)/6} \boldsymbol{M} \vec{\mathcal{P}} \cdot \vec{A}_\sigma \qquad (5.3)$$

in which the configuration for each Steiner triple is drawn – conditionally independently of other Steiner triples – from the same probability distribution over the 16 subgraphs shown in Fig. 5.1.

If (unidirectional) links are set purely at random with probability $p$, as it is the case in ER graphs, the probabilities for the triadic subgraph

5. *The Triadic Random Graph Model*

patterns are:

$$p^{ER}_{\triangle_1} = (1-p)^6$$

$$p^{ER}_{\triangle_2} = 6\,p\,(1-p)^5$$

$$p^{ER}_{\triangle_3} = p^{ER}_{\triangle_4} = p^{ER}_{\triangle_5} = 3\,p^2\,(1-p)^4 \;,\; p^{ER}_{\triangle_6} = 6\,p^2\,(1-p)^4$$

$$p^{ER}_{\triangle_7} = p^{ER}_{\triangle_8} = p^{ER}_{\triangle_9} = 6\,p^3\,(1-p)^3 \;,\; p^{ER}_{\triangle_{10}} = 2\,p^3\,(1-p)^3 \qquad (5.4)$$

$$p^{ER}_{\triangle_{11}} = p^{ER}_{\triangle_{12}} = p^{ER}_{\triangle_{13}} = 3\,p^4\,(1-p)^2 \;,\; p^{ER}_{\triangle_{14}} = 6\,p^4\,(1-p)^2$$

$$p^{ER}_{\triangle_{15}} = 6\,p^5\,(1-p)$$

$$p^{ER}_{\triangle_{16}} = p^6.$$

The triadic random graph model allows us to deviate from this probability distribution. Therefore, we can enhance or suppress certain substructures as compared to ER graphs.

For the TRGM, the only parameters $\vec{\theta}$ are the entries of the vector $\vec{\mathcal{P}}$. We can thus denote them as $\mathcal{G}\left(N,\vec{\mathcal{P}}\right)$. Equivalently, in the limit of large system sizes, $N$, we can fix the number of appearances of each triad pattern on the STs to their expectation values, $\vec{\mathcal{T}} \approx \vec{P}N(N-1)/6$ with $\mathcal{T}_i \in \mathbb{N}$ and $\sum_i \mathcal{T}_i = N(N-1)/6$. This ensemble will be denoted as $\mathcal{G}\left(N,\vec{\mathcal{T}}\right)$.

## 5.1. *Z*-Score Profiles

In order to examine the impact of the triad distribution for the Steiner triple system (STS) on the *Z*-score profile of the total network we did extensive uniform sampling of the 16-dimensional parameter space of the distributions. Samplings were performed for both systems of size 49 and 63. For the computation of the *Z*-score profiles we used the mfinder software[1] [7] and averaged the *Z* score for each vector $\vec{\mathcal{P}}$ over multiple samples.

It shall be stressed that there are two different sampling processes involved in this investigation. Firstly, the triad-pattern distribution, $\vec{\mathcal{P}}$, is sampled uniformly from the 16-dimensional simplex of probability distributions shown in Eq. (5.1). In fact, to fix the expectation values, we sample from $\vec{\mathcal{T}}$. For instance, for a system of size 49, there are $\frac{49 \cdot 48}{6} = 392$ Steiner triples to be specified. Thus, there are $\binom{392+15}{392} \approx 8.2 \times 10^{26}$ distinct vectors $\vec{\mathcal{T}}$. For the sampling, we uniformly draw from these vectors[2].

Subsequently, for each $\vec{\mathcal{P}} = \vec{\mathcal{T}} \cdot \frac{6}{N(N-1)}$, multiple networks are sampled from Eq. (5.3) and their *Z* scores are evaluated. As a consequence, we obtain a mapping from $\vec{\mathcal{P}}$ to the averaged corresponding *Z*-score profile,

$$\vec{\mathcal{P}} \longmapsto \left\langle \vec{Z}\left(\vec{\mathcal{P}}\right) \right\rangle. \tag{5.5}$$

For computational reasons, it is common to consider only those triad configurations which have all three nodes attached to at least one edge. Thus, there are no *Z* scores for the subgraphs 1 (△), 2 (△), and 3 (△) of Fig. 5.1. Nevertheless, of course it is necessary to account for them in the input distributions for the STSs.

Fig. 5.2 displays exemplary results obtained from the sampling. The plots in the left column show the probability distributions imposed on the STSs (blue circles). This distribution already determines the link-density of the network. Suppose, e.g., 60% of the Steiner triples adopt pattern △ (which has two of the six possible links being set) and 40% adopt △ (five of the six links being set), then the density will be $p = (0.6 \cdot 2 + 0.4 \cdot 5)/6 \approx 53\%$. For comparison we also plot the distribution one would expect on the STs for a dyadic ER graph with the same link density as given by Eq. (5.4) (red squares, dashed line in Fig. 5.2).

---

[1] For the randomization we used the MCMC switching algorithm introduced in Section 3.3.2 with 100 switching steps per edge.
[2] Each of the 392 Steiner triples needs to be assigned to one of the 16 patterns. Thus, there are 15 delimiters representing the boundaries of the bins which correspond to the patterns. Hence, one can think of drawing uniformly from 407 elements, 392 Steiner Triples and 15 delimiters. We start with the first bin. Any time an ST is drawn, it is added to the current bin. Once a delimiter is drawn, we move to the next bin and so on until we have completely defined distribution $\vec{\mathcal{T}}$.

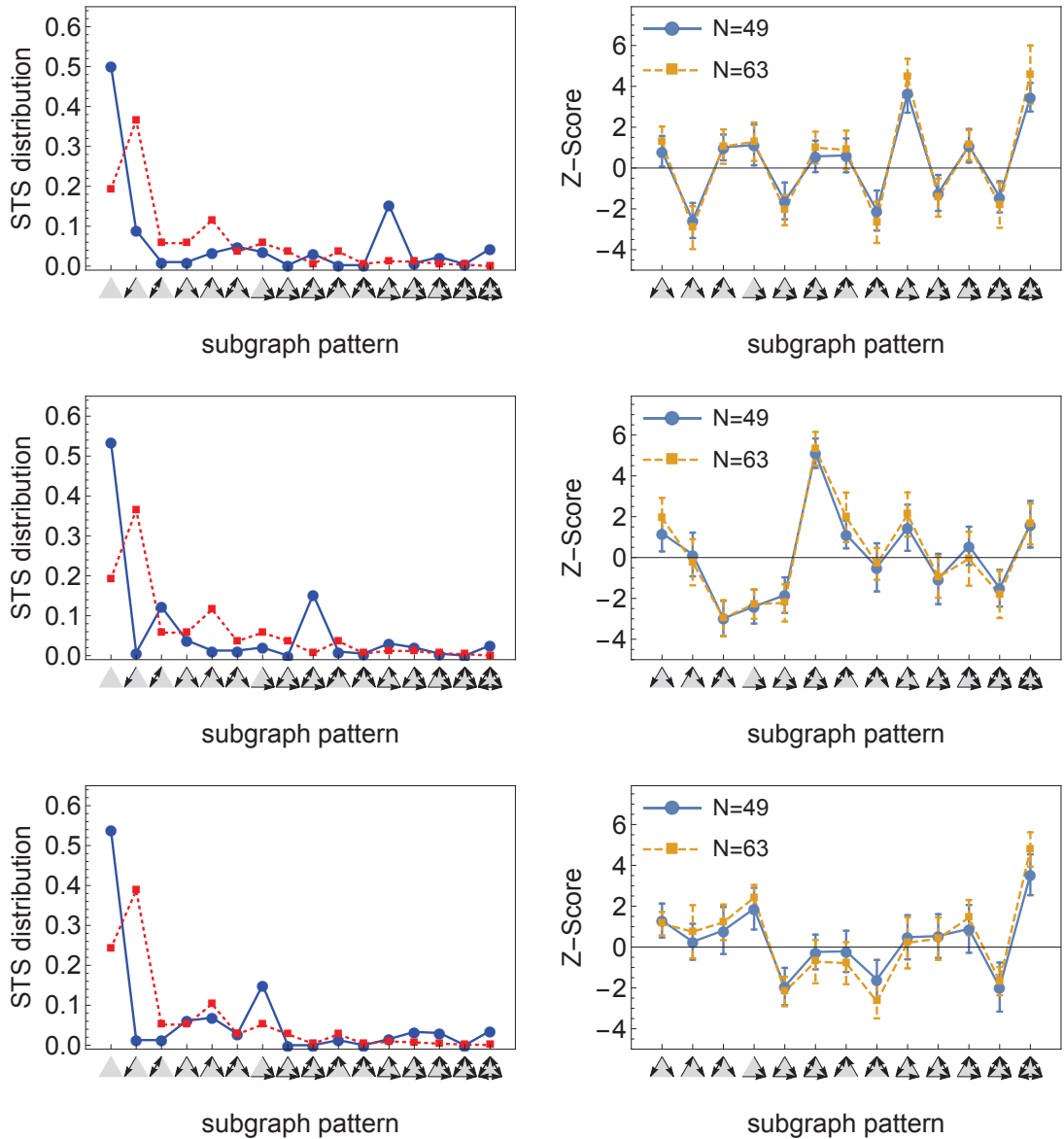

**Figure 5.2.: Left:** Distribution of triad configurations for the Steiner triples (blue circles) and expected distribution of triad configurations for ER graphs with the same link density (red squares). **Right:** $Z$ scores obtained from networks sampled from the distributions on the left for systems of size $N = 49$ (blue) and $N = 63$ (yellow), averaged over 15 sample networks. The error bars indicate one standard deviation.

The plots in the right column of Fig. 5.2 show the $Z$-score profiles obtained from the input distributions above for networks of size 49 (blue) and 63 (yellow). Displayed are the mean values averaged over 15 samples for each distribution. For systems with no higher order structure, such as ER graphs, all $Z$ scores are expected to vanish. However, for the triadic random graph model, we observe $Z$ scores with magnitudes larger than five, implying that certain motifs appear five standard deviations more frequently than expected for the randomized ensemble. Thus, triadic random graphs are capable of modelling structure of higher than dyadic order. It shall be emphasized that this higher order structure does not stem from mesoscopic group structure; all Steiner triples, and therefore all nodes, have the same parameters. In accordance with the literature [102] a larger system size results in a larger magnitude of the $Z$ scores. However, the shape of the $Z$-score profiles is size independent.

## 5.2. *Z*-Score Correlations

For the interpretation of triad significance profiles observed in real networks it is important to be aware of correlations between the $Z$ scores of pairs of triad patterns which inherently already arise solely for statistical reasons.

We did extensive uniform sampling of the 16-dimensional simplex spanned by the parameter space of the triadic random graph model (see Eq. (5.1)). In fact, we sampled more than $10^5$ distinct distribu-tions. For each of the distributions, we generated five network instances and we evaluated the average $Z$-score profiles. Using the latter, we can evaluate cross correlations between pairs of $Z$ scores over the sampled input distributions. For two patterns, $i$ and $j$, it is

$$C_{Z_i, Z_j} = \frac{\langle Z_i Z_j \rangle - \langle Z_i \rangle \langle Z_j \rangle}{\sigma_{Z_i} \sigma_{Z_j}}. \tag{5.6}$$

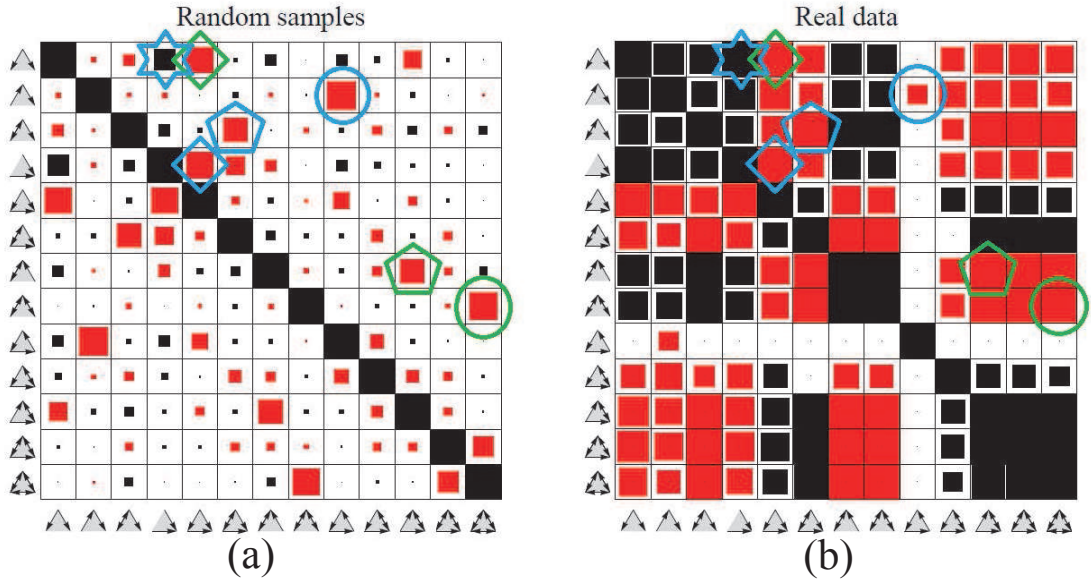

**Figure 5.3.:** **(a)** $Z$-score cross correlations in $10^5$ randomly sampled distributions on Steiner triple systems. **(b)** Correlations obtained from real datasets (Table 5.1). The length of the squares indicates the magnitude (0 to 1). Black and red shading corresponds to positive and negative values, respectively. Shown are significant entries at a level of 5%.

The averages are taken over all sampled STS distributions, $\vec{\mathcal{P}}$, considered for the evaluation of the correlation matrix. The statistical significance of the correlation is tested by means of a t-test.

Fig. 5.3 (a) shows the correlation matrix between pairs of $Z$ scores when sampling randomly. Considered are significant correlations at a level of 5%. The side lengths of the squares indicate the magnitudes of the correlation coefficients between the corresponding subgraphs. Positive values are colored in black; negative ones are colored in red. The magnitudes (zero to one) are proportional to the lengths of the squares. One can clearly see that certain $Z$ scores are strongly anti-correlated with each other, while others are positively correlated. To keep track of the impact of the link density on potential correlations, we grouped the distributions in bins of width 0.05 and evaluated seperate correlation matrices for each of the link-density ranges. It turns out that correlations and anticorrelations occur consistently between the same sets of triad patterns for all link densities sampled. Furthermore, we sampled TRGs involving exclusively unidirectional links, i.e. only the patterns ▲, ▲, ▲, ▲, ▲, ▲, and ▲ were allowed to have non-vanishing entries in the probability distribution $\vec{\mathcal{P}}$. Since the randomization algorithm of the motif-detection process preserves both the number of unidirectional and bidirectional links, there will be no bidirectional links whatsoever in the randomized ensemble and consequently no corresponding $Z$ scores. As shown in Fig. 5.4, also for exclusively unidirectional networks, correlations and anticorrelations have similar values as in the general case (Fig. 5.3 (a)).

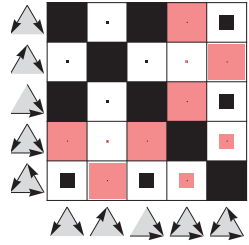

**Figure 5.4.:** *Z*-score cross correlations in $5 \times 10^3$ randomly sampled *unidirectional* distributions on STSs.

In order to distinguish between $Z$ scores which actually describe

5. *The Triadic Random Graph Model*

| Dataset |  |  |  |  |  |  |  |  |  |  |  |  |  |
|---|---|---|---|---|---|---|---|---|---|---|---|---|---|
| C. elegans | -16.5 | -6.3 | -24.2 | -12.0 | 12.4 | 24.5 | -27.0 | -16.3 | -5.0 | 2.6 | 27.3 | 13.2 | 9.6 |
| Political blogs | -76.1 | -51.3 | -49.4 | -58.2 | 55.3 | 40.3 | -54.1 | -31.2 | -2.3 | 3.0 | 47.2 | 27.1 | 24.8 |
| E. coli (v. 1.1) | -12.2 | -12.2 | 0 | -12.2 | 12.2 | 0 | 0 | 0 | 0 | 0 | 0 | 0 | 0 |
| English book | 26.1 | 13.6 | 14.4 | 22.8 | -22.5 | -10.0 | 24.7 | 13.5 | -1.4 | -6.6 | -21.8 | -13.6 | -5.5 |
| French book | 31.5 | 26.3 | 13.4 | 31.5 | -29.1 | -10.2 | 16.2 | 12.0 | -11.5 | -12.3 | -15.1 | -12.3 | -4.7 |
| Japanese book | 15.0 | 12.1 | 13.4 | 15.0 | -14.4 | -7.9 | 12.1 | 9.3 | -4.8 | -9.9 | -7.4 | -8.3 | -3.1 |
| Spanish book | 26.6 | 27.5 | 13.6 | 23.8 | -22.3 | -4.2 | 29.4 | 12.4 | -13.2 | -19.8 | -25.2 | -11.0 | -7.6 |
| leader2Inter | -2.3 | -1.2 | -2.6 | -1.2 | 0.8 | 1.3 | -3.2 | -4.5 | 0.4 | 1.2 | 2.3 | 1.8 | 3.5 |
| prisonInter | -6.1 | -3.7 | -10.1 | -9.1 | 4.3 | 7.8 | -8.3 | -13.8 | 0.4 | 2.0 | 5.4 | 7.5 | 11.9 |
| El. circ. (s208) | 1.6 | -9.6 | 0 | 1.6 | -1.6 | 0 | 0 | 0 | 11.0 | 0 | 0 | 0 | 0 |
| El. circ. (s420) | 1.6 | -17.2 | 0 | 1.6 | -1.6 | 0 | 0 | 0 | 20.7 | 0 | 0 | 0 | 0 |
| S. cerevisiae | -13.7 | -13.5 | -1.0 | -13.7 | 13.6 | -0.4 | -5.9 | 0 | -0.2 | 9.9 | 3.9 | 0 | 0 |

**Table 5.1.:** $Z$ scores observed in real-world datasets [152]. For more details on the data see page 167.

characteristics of the networks from purely statistical artifacts, we furthermore investigated $Z$-score correlations over various real-world networks. Fig. 5.3 (b) shows the correlation matrix obtained from the 16 real-world datasets shown in Table 5.1. We observe that the most pronounced correlations found in the ensemble of triadic random graphs also appear in real datasets. The attribution of functional significance to single (anti)motifs is therefore questionable and one should rather consider the $Z$-score profile as a whole. Table 5.2 displays the seven strongest cross correlations between pairs of triadic subgraph patterns which were found in our random samples of the triadic random graph ensemble together with the correlation coefficients found in real data for the respective pairs of triad patterns. Apparently, all of the top seven (anti)correlations of the statistical data are also found in the real systems. However, not all entries of the correlation matrix obtained from the triadic random graphs are reflected in Fig. 5.3 (b): e.g., patterns  and  are anticorrelated in the random ensemble, while being strongly positively correlated in the real-world data. This gives rise to the conjecture that this correlation captures valuable information about the systems' structure. On the contrary, e.g., the correlation between patterns  and  seems to stem from statistical roots.

| Rank | patterns | random samples | real data |
|---|---|---|---|
| 1 | 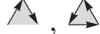, 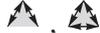 | -0.780 | -0.527 |
| 2 | 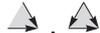, 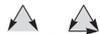 | -0.742 | -0.978 |
| 3 | 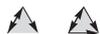, 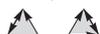 | -0.733 | -0.998 |
| 4 | 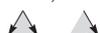, 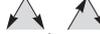 | -0.730 | -0.989 |
| 5 | 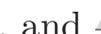, 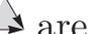 | -0.662 | -0.986 |
| 6 | 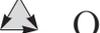, 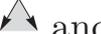 | -0.662 | -0.997 |
| 7 | 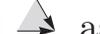, 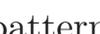 | 0.578 | 0.991 |

**Table 5.2.:** Top seven (anti)correlations between subgraph patterns found in the synthetic random samples (highlighted in Fig. 5.3(a)), as well as the corresponding ones observed in the real-world datasets of Table 5.1 (highlighted in Fig. 5.3(b)).

Investigations of correlations in the appearance of subgraph motifs have been done before by Ginoza et al. [51]. Yet, their work focuses on correlations within the randomization process of single networks. They consider motifs in two particular networks, namely in the transcriptional regulatory networks of *E. coli* and *S. cerevisiae*. One of their key results is that the abundances of patterns 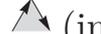, 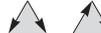, and 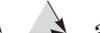 are strongly mutually correlated, while being anticorrelated with pattern 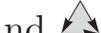. Our approach, however, considers correlations which appear over multiple network instances and is therefore complementary to the one in Ref. [51]. Again, Fig. 5.3 (a), displays our observed correlations between subgraph patterns which occur solely for statistical reasons. In accordance with Ginoza et al. we find strong correlations between patterns 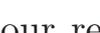 and , as well as strong anticorrelation of them with . However, the former are hardly correlated with pattern  (in fact, the correlation coefficient is even slightly negative). Although, doubtlessly, in most real networks there is a strong mutual (anti)correlation in the abundance of subgraphs , , , and , our results indicate that they do not necessariliy follow for statistical reasons. Furthermore, in addition to the findings of Ginoza et al., we also observe strong anti-correlations between △ and △, between △ and △, between △ and △, and between △ and △.

## 5.3. Degree Distributions

An important characteristic of complex networks is their degree distribution. In dyadic Erdös-Rényi graphs, node degrees are expected to be Poisson distributed (see Section 2.3.1),

$$\mathcal{P}(k = \kappa) = e^{-\langle k \rangle} \frac{\langle k \rangle^\kappa}{\kappa!}. \qquad (5.7)$$

This holds for both in and out degrees.

To derive the expected in-degree distribution for triadic random graphs, consider an arbitrary node $i$. It is part of $(N-1)/2$ Steiner triples. Now let $s_i$ be a random variable indicating the number of $i$'s Steiner triples in which a single edge is directed towards it. Further, be $d_i$ the random variable indicating the number of its Steiner triples with two links directed towards it. From the probabilities in Eq. (5.1) we can directly infer the probabilities for a single ST to contribute to $s_i$ and $d_i$, respectively:

$$\begin{aligned} p(s_i) = &\frac{1}{3}\left[p\left(\triangle\right) + p\left(\triangle\right) + p\left(\triangle\right) + p\left(\triangle\right)\right] \\ &+ \frac{2}{3}\left[p\left(\triangle\right) + p\left(\triangle\right) + p\left(\triangle\right) + p\left(\triangle\right)\right. \\ &\left. + p\left(\triangle\right) + p\left(\triangle\right)\right] + p\left(\triangle\right) + p\left(\triangle\right) \end{aligned} \qquad (5.8)$$

$$p(d_i) = \frac{1}{3}\left[p\left(\triangle\right) + p\left(\triangle\right) + p\left(\triangle\right) + p\left(\triangle\right) + p\left(\triangle\right) \right.$$
$$\left. + p\left(\triangle\right)\right] + \frac{2}{3}\left[p\left(\triangle\right) + p\left(\triangle\right)\right] + p\left(\triangle\right). \tag{5.9}$$

Since the model parameters are the same for all nodes, the expectation values for $s$ and $d$ will also be the same for all $i$:

$$\langle s \rangle = \langle s_i \rangle = \frac{N-1}{2} p(s_i)$$
$$\langle d \rangle = \langle d_i \rangle = \frac{N-1}{2} p(d_i). \tag{5.10}$$

Each of the $(N-1)/2$ Steiner triples of node $i$ has either zero, one, or two edges directed towards it. Therefore, the joint probability distribution of $s_i$ and $d_i$ is given by the multinomial:

$$p\begin{pmatrix} s_i = n_s \\ d_i = n_d \end{pmatrix} = \binom{\frac{N-1}{2}}{n_s,\, n_d,\, \frac{N-1}{2} - n_s - n_d} p(s_i)^{n_s} p(d_i)^{n_d}$$
$$\times \left(1 - p(s_i) - p(d_i)\right)^{\frac{N-1}{2} - n_s - n_d}$$
$$= \binom{\frac{N-1}{2}}{n_s,\, n_d,\, \frac{N-1}{2} - n_s - n_d} \left(\frac{2}{N-1}\right)^{n_s + n_d}$$
$$\times \langle s \rangle^{n_s} \langle d \rangle^{n_d} \left(1 - \frac{2(\langle s \rangle + \langle d \rangle)}{N-1}\right)^{\frac{N-1}{2} - n_s - n_d}. \tag{5.11}$$

For the second equality, we used Equation (5.10). For large, sparse

## 5. The Triadic Random Graph Model

systems, i.e., $\langle s \rangle, \langle d \rangle \ll N$, we find

$$
\begin{aligned}
\lim_{N \to \infty} p \begin{pmatrix} s_i = n_s \\ d_i = n_d \end{pmatrix} &= \lim_{N \to \infty} \frac{\left(\frac{N-1}{2}\right)!}{\left(\frac{N-1}{2} - n_s - n_d\right)!} \left(\frac{2}{N-1}\right)^{n_s + n_d} \\
&\quad \times \frac{\langle s \rangle^{n_s}}{n_s!} \frac{\langle d \rangle^{n_d}}{n_d!} \underbrace{\left(1 - \frac{\langle s \rangle + \langle d \rangle}{\frac{N-1}{2}}\right)^{\frac{N-1}{2}}}_{\to e^{-\langle s \rangle - \langle d \rangle}} \\
&\quad \times \underbrace{\left(1 - \frac{\langle s \rangle + \langle d \rangle}{\frac{N-1}{2}}\right)^{-(n_s + n_d)}}_{\to 1} \\
&= \lim_{N \to \infty} \underbrace{\frac{\frac{N-1}{2} \times \frac{N-3}{2} \times \ldots \times \left(\frac{N-1}{2} - n_s - n_d + 1\right)}{\left(\frac{N-1}{2}\right)^{n_s + n_d}}}_{\to 1} \\
&\quad \times \frac{\langle s \rangle^{n_s}}{n_s!} \frac{\langle d \rangle^{n_d}}{n_d!} e^{-\langle s \rangle - \langle d \rangle} \\
&= \frac{\langle s \rangle^{n_s}}{n_s!} \frac{\langle d \rangle^{n_d}}{n_d!} e^{-\langle s \rangle - \langle d \rangle}.
\end{aligned}
\tag{5.12}
$$

The in degree of node $i$ is

$$
k_i^{\text{in}} = s_i + 2d_i. \tag{5.13}
$$

The probability distribution for node $i$ to have in degree $\kappa$ is thus

$$
\begin{aligned}
p\left(k^{\text{in}} = \kappa\right) &= \sum_{n_s=0}^{\frac{N-1}{2}} \sum_{n_d=0}^{\frac{N-1}{2}} p \begin{pmatrix} s_i = n_s \\ d_i = n_d \end{pmatrix} \delta_{\kappa, n_s + 2n_d} \\
&= e^{-\langle s \rangle - \langle d \rangle} \sum_{n_d=0}^{\frac{\kappa}{2}} \frac{\langle s \rangle^{\kappa - 2n_d}}{(\kappa - 2n_d)!} \frac{\langle d \rangle^{n_d}}{n_d!},
\end{aligned}
\tag{5.14}
$$

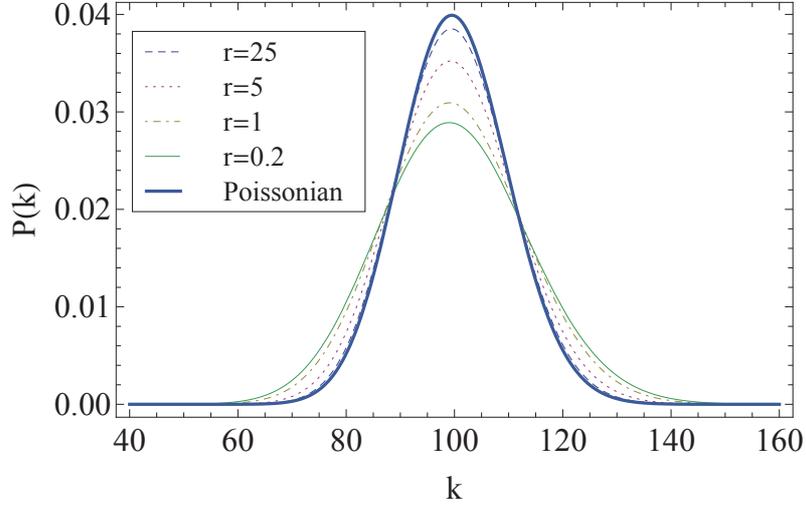

**Figure 5.5.:** Degree distributions for mean degree $\langle k \rangle = 100$ and various ratios $r = \langle s \rangle / \langle d \rangle$ [152].

where $\boldsymbol{\delta}$ is the Kronecker delta ($\boldsymbol{\delta}_{i,j} = 1$ if $i = j$, 0 otherwise). In the limit $\langle d \rangle \to 0$, the distribution is Poissonian. With $\langle d \rangle$ approaching $\frac{1}{2} \langle k^{in} \rangle$, the distribution becomes broader, implying larger deviations from $\langle k \rangle$. Fig. 5.5 shows distributions of Eq. (5.14) with fixed $\langle k \rangle = \langle s \rangle + 2 \langle d \rangle = 100$ for various ratios of $r = \langle s \rangle / \langle d \rangle$ together with the corresponding Poissonian.

The out-degree distribution can be derived analogously. In this case, only the probabilites for the triads with a single out-going edge, $p\left(s_i^{\text{out}}\right)$, and with two out-going edges, $p\left(d_i^{\text{out}}\right)$, need to be adjusted accordingly,

$$p\left(s_i^{\text{out}}\right) = \frac{1}{3}\left[p\left(\triangle\right) + p\left(\triangle\right) + p\left(\triangle\right) + p\left(\triangle\right)\right]$$
$$+ \frac{2}{3}\left[p\left(\triangle\right) + p\left(\triangle\right) + p\left(\triangle\right) + p\left(\triangle\right)\right. \quad (5.15)$$
$$\left. + p\left(\triangle\right) + p\left(\triangle\right)\right] + p\left(\triangle\right) + p\left(\triangle\right)$$

*5. The Triadic Random Graph Model*

$$p\left(d_i^{\text{out}}\right) = \frac{1}{3}\left[p\left(\triangle\right) + p\left(\triangle\right) + p\left(\triangle\right) + p\left(\triangle\right) + p\left(\triangle\right) \right. \\ \left. + p\left(\triangle\right)\right] + \frac{2}{3}\left[p\left(\triangle\right) + p\left(\triangle\right)\right] + p\left(\triangle\right). \quad (5.16)$$

## 5.4. Design of Significance Profiles

To design networks with certain triad significance profiles, it is important to understand the relationship between the distribution of triad configurations on Steiner triples and the $Z$ scores obtained from their ensembles. Therefore, we also investigated cross correlations between the Steiner-triple configurations and the obtained corresponding $Z$ scores,

$$\widetilde{C}_{\mathcal{P}_i, Z_j} = \frac{\langle \mathcal{P}_i Z_j \rangle - \langle \mathcal{P}_i \rangle \langle Z_j \rangle}{\sigma_{\mathcal{P}_i} \sigma_{Z_j}}. \quad (5.17)$$

The results are presented in Fig. 5.6(a). Of course, there is a strong correlation between the imposed triad patterns on the Steiner triples and the $Z$ scores of these patterns. Nevertheless, as for the $Z$-score-$Z$-score cross correlations, again we observe strong anticorrelations between certain patterns. As before, the observations are valid for all examined link densities. Fig. 5.6(b) shows the correlations when only permitting unidirectional links in $\vec{\mathcal{P}}$, i.e. patterns △, △, △, △, △, △, and △. Correlations between the input distributions on the STSs and the obtained overall $Z$-score profiles can be helpful in designing systems with predefined triad significance profiles (TSPs).

For a simplistic approach, we assume a linear relation between the input distribution $\vec{\mathcal{P}}$ and the corresponding significance profile, $\vec{\text{SP}}$, conveyed by the correlation matrix $\widetilde{C} \in [-1, 1]^{13} \times [-1, 1]^{16}$ (Fig. 5.6(a)),

$$\vec{\text{SP}} \propto \widetilde{C}\,\vec{\mathcal{P}}. \quad (5.18)$$

In order to design systems with predefined triad significance profiles, it is necessary to map the latter to a corresponding input distribution,

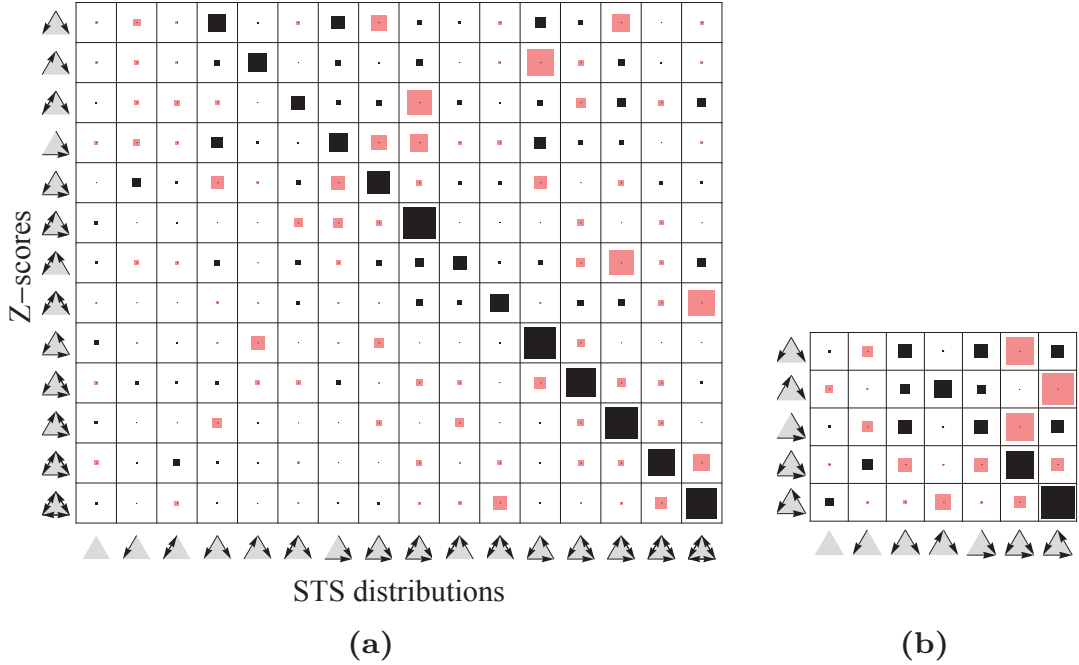

**Figure 5.6.:** Correlation matrices between triad-pattern distributions on Steiner triples and the resulting $Z$-score profiles obtained from 5000 samples. **(a)** results for general distributions [152], **(b)** results for distributions with unidirectional links only. The lengths of the squares indicate the magnitudes (0 to 1). Black and red shading corresponds to positive and negative values, respectively.

which can be realized by means of the pseudo-inverse matrix $\widetilde{\boldsymbol{C}}^{-1}$,

$$\vec{\mathcal{P}} \propto \widetilde{\boldsymbol{C}}^{-1}\, \vec{\mathrm{SP}}. \tag{5.19}$$

Fig. 5.7 shows the triad significance profiles of triadic random graphs generated from the STS-probability distributions from Fig. 5.2 (blue circles) together with the prediction obtained from Eq. (5.18) (yellow squares). The predictions agree very well with the actually observed profiles. Though, attempts to design entirely unidirectional significance-profiles often fail, for instance in the case of the first superfamily suggested by Milo et al., shown in Fig. 5.8. However, allowing only unidirectional links in the STS distribution $\vec{\mathcal{P}}$, we are able to generate triadic random graphs with TSPs matching the first superfamily.

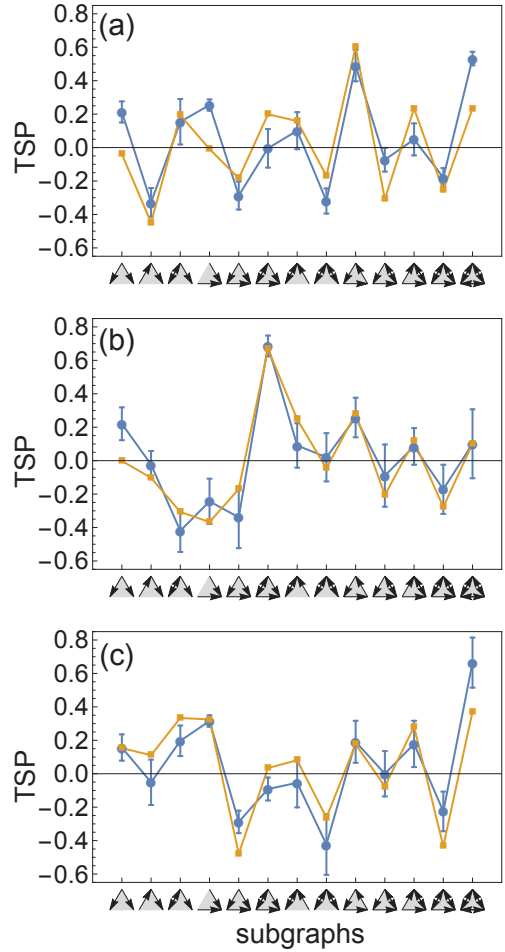

**Figure 5.7.:** Triad significance profiles corresponding to the distributions in Fig. 5.2 for systems of size 49 (blue circles). The yellow squares indicate the prediction obtained from the input distribution $\vec{\mathcal{P}}$ by assuming $\vec{\mathrm{SP}} \propto \widetilde{C}\vec{\mathcal{P}}$.

Nevertheless, it is not possible to generate graphs with arbitrary significance profiles, particularly not with the linear relation of Eq. (5.19). This may be for various reasons. On the one hand, the relationship between $\vec{\mathcal{P}}$ and the significance profile is certainly not entirely linear.

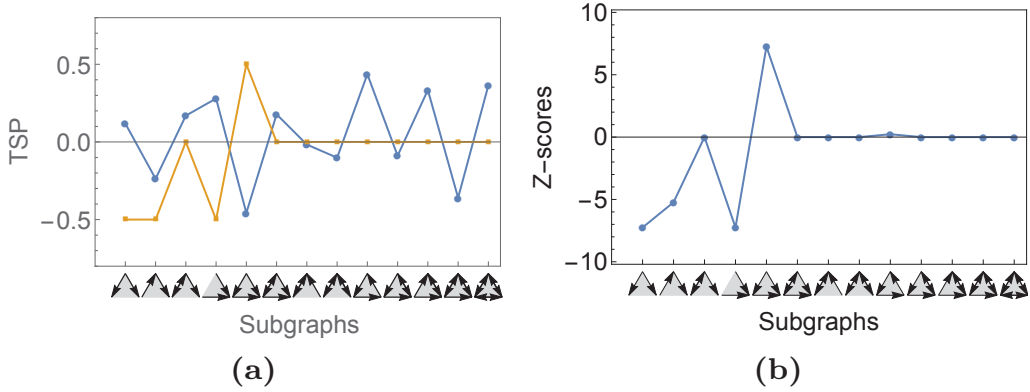

**Figure 5.8.:** **(a)** Attempt to model the significance profile of the first superfamily from Fig. 3.7, indicated by the yellow squares. However, networks with their parametrization obtained from Eq. (5.19) with the $13 \times 16$ correlation matrix, shown in Fig. 5.6(a), yield very different significance profiles (blue dots). **(b)** Considering probability distributions with exclusively unidirectional patterns, the $Z$-score profiles can be successfully modeled.

Secondly, not all significance profiles are necessarily realizable, e.g., think of a TSP with all patterns being overrepresented. Furthermore, the triadic random graph model describes the most simplistic model based on STSs, which, e.g., does not account for individual node properties. This is also reflected by the fact that the degree distributions of triadic random graphs are close to a Poissonian. A formulation of more specific models based on STSs may overcome these shortcomings. Still, these first steps open the way to efficiently generate networks in which certain motifs are over- or underrepresented and thus enable systematic investigations of the functional significance of these motifs. In Chapter 7, we will use TRGMs to test the effect of motifs on dynamical processes acting on the vertices of networks.

# 6. Node-Specific Subgraph Analysis

*Major parts of the content presented in this chapter have been prepublished in* M. Winkler and J. Reichardt. Node-Specific Triad Pattern Mining for Complex-Network Analysis. *IEEE ICDMW, Data Mining in Networks*, p. 605-612, 2014. © *2014 by the IEEE.*[1]

    Elucidating the relationship between a network's function and the underlying graph topology is one of the major branches in network science. The mining of graphs in terms of their local substructure, as introduced in Chapter 3, is a well-established methodology to analyze their topology. It was hypothesized that motifs play a key role for the ability of a system to perform its task. Yet, the framework commonly used for motif detection (Section 3.3.1) averages over the local environments of all nodes. It therefore remains unclear whether motifs are overrepresented homogeneously in the whole system or only in certain regions. If motifs were indeed critical for a network's function, but at the same time bound to specific parts of the graph, a failure of only very few important nodes could severely disable the whole system. Further, especially for larger networks composed of different functional components, there may be areas in which *one* structural pattern is of importance, whereas in different regions *other* patterns are relevant.

---

*6. Node-Specific Subgraph Analysis*

On the system level, the abundance of these patterns may average out and hence, their importance may not even be recognized.

In this chapter, we will investigate this issue in more detail by mining *node-specific* patterns. More specifically, instead of detecting frequent subgraph patterns of the whole system, we investigate the neighborhood of every single node separately, i.e., for every vertex we consider only the subgraphs it participates in. This will allow us to localize the regions of a graph in which the instances of a motif predominantly appear. Thus, it is possible to identify and remove the nodes and links which eventually make a certain pattern a motif of the network. This approach will facilitate future investigations to assess whether it is actually the presence of a motif which enables a system to perform its task or whether other structural aspects are more relevant.

After introducing node-specific $Z$ scores and the framework of node-specific triad pattern mining in Sections 6.1 and 6.2, in Section 6.3 we will investigate systems of various fields and find that, for many of them, motifs are distributed highly heterogeneously. Furthermore, our methodology provides for a new set of features for each node. We will use these features to cluster the vertices of a neural network and the international airport-connectivity network. In Section 6.4 we will further discuss an extension of the methodology from directed graphs to signed networks.

## 6.1. Node-Specific Triadic *Z* Scores

We will now introduce *node-specific*, triadic $Z$-score profiles. For every node $\alpha$ in a graph, we evaluate the abundance of all structural patterns in $\alpha$'s neighborhood. The patterns in $\alpha$'s *neighborhood* or *environment* shall be defined as those patterns in which $\alpha$ participates in. The frequency of occurrence of patterns in the system under investigation is compared to the expected frequency in a randomized ensemble of the original network. In the randomization, both individual in- and out degrees of all nodes, and the number of unidirectional and bidirectional links are the same as in the original network (see Algorithm 1 in Section 3.3.2).

In principle, the framework of node-specific subgraph mining can be realized for patterns composed of an arbitrary number of nodes, $n$. Nevertheless, with increasing $n$, the number of non-isomorphic subgraphs also increases rapidly. For the remainder of this work we will focus on triad patterns ($n = 3$).

We strive to evaluate the abundance of triad patterns from a particular node $\alpha$'s point of view. Therefore, the symmetry of most patterns shown in Fig. 5.1 is now broken and the number of connected *node-specific* triad patterns increases from 13 to 30. These are shown in Fig. 6.1. To understand the increase in the number of patterns, consider the ordinary subgraph . From the perspective of one particular node, it splits into the three node-specific triad patterns 1, 5, and 10 in Fig. 6.1.

Some patterns are included in others, e.g. pattern 1 is a subset of pattern 3. In order to avoid biased results, we do not double count, i.e. an observation of pattern 3 will only increase its corresponding count and not the one associated with pattern 1.

For every node $\alpha$ in a graph, we will now compute $Z$ scores for each

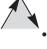

**Figure 6.1.:** All possible connected, nonisomorphic triadic subgraph patterns in terms of a distinct node (here: lower node).

*6. Node-Specific Subgraph Analysis*

of the 30 node-specific patterns $i$ shown above,

$$Z_i^\alpha = \frac{N_{\text{original},i}^\alpha - \left\langle N_{\text{rand},i}^\alpha \right\rangle}{\sigma_{\text{rand},i}^\alpha}. \tag{6.1}$$

$N_{\text{original},i}^\alpha$ is the number of appearances of pattern $i$ in the triads node $\alpha$ participates in. Accordingly, $\left\langle N_{\text{rand},i}^\alpha \right\rangle$ is the expected frequency of pattern $i$ in the triads node $\alpha$ is part of in the randomized ensemble. $\sigma_{\text{rand},i}^\alpha$ is the corresponding standard deviation.

## 6.2. Node-Specific Triad Pattern Mining (NoSPaM$_3$)

We now suggest **No**de-**S**pecific **Pa**ttern **M**ining (NoSPaM), an algorithm to compute the *node-specific Z*-score profiles suggested in Eq. (6.1). In particular, we will focus on *triad* patterns (NoSPaM$_3$).

### 6.2.1. Algorithm

The algorithm for node-specific triad pattern mining consists of three parts. The first part is the degree-preserving randomization defined in Algorithm 1 on page 59. Further, Algorithm 2 performs the counting process for the appearances of triad patterns in a graph. Because it is computationally expensive to test all triads in the system (the complexity is of order $\mathcal{O}\left(N^3\right)$), we rather iterate over pairs of adjacent edges in the graph. Since real-world networks are usually sparse, this is much more efficient.

Using the functions defined in Algorithm 1 and Algorithm 2, we can eventually formulate the routine of node-specific triad pattern mining (NoSPaM$_3$). Algorithm 3 describes its formalism. It computes the node-specific $Z$ scores as defined in Eq. 6.1. The definition of the standard deviation over all $I$ randomized instances involves the corre-

---

**Algorithm 2** Counting of node-specific triad patterns

    **function** NSPPATTERNCOUNTER(Graph $\mathcal{G}(V, E)$)
        $\mathcal{N}$: $N \times 30$-dimensional array storing the pattern counts for every node of $\mathcal{G}$
        **for** every edge $e \in E$ **do**
            $i, j \leftarrow$ IDs of $e$'s nodes with $i < j$
            $\mathcal{C} \leftarrow \{\}$ be list of candidate nodes to form triad patterns comprising $e$
            $\mathcal{C} \leftarrow$ all neighbors of $i$
            $\mathcal{C} \leftarrow$ all neighbors of $j$
            **for** all $c \in \mathcal{C}$ **do**
                **if** $i + j <$ sum of IDs of all other *connected* dyads in triad $(ijc)$ **then**
                    increase the counts in $\mathcal{N}$ for $i$, $j$, and $c$ for their respective node-specific patterns
                **end if**
            **end for**
        **end for**
        **return** $\mathcal{N}$
    **end function**

---

sponding mean value,

$$\left(\sigma_{\text{rand},i}^{\alpha}\right)^2 = \frac{1}{I} \sum_{k=1}^{I} \left(N_{\text{rand},i,k}^{\alpha} - \left\langle N_{\text{rand},i}^{\alpha}\right\rangle\right)^2. \tag{6.2}$$

Using Eq. (6.2) it would be necessary to store all $N \times 30 \times I$ values $N_{\text{rand},i,k}^{\alpha}$, since they all contribute to both $\sigma_{\text{rand}}^{\alpha}$ and $\left\langle N_{\text{rand},i}^{\alpha}\right\rangle$. However, we can easily evaluate the standard deviation in one sweep utilizing

$$\left(\sigma_{\text{rand},i}^{\alpha}\right)^2 = \left\langle \left(N_{\text{rand},i}^{\alpha}\right)^2\right\rangle - \left\langle N_{\text{rand},i}^{\alpha}\right\rangle^2 \tag{6.3}$$

and therefore saving a factor $I$ of storage consumption.

*6. Node-Specific Subgraph Analysis*

We define all operations on the arrays $\mathcal{N}_{\text{original}}$, $\mathcal{N}_{\text{rand}}$, $\mathcal{N}_{\text{sq,rand}}$, and $\mathcal{Z}$ in Algorithm 3 to be performed elementwise.

---

**Algorithm 3** Node-specific triad pattern mining (NoSPaM$_3$)

---

**function** NoSPaM(Graph $\mathcal{G}$, # required rewiring steps, # randomized instances)
 $\mathcal{N}_{\text{original}} \leftarrow$ NspPatternCounter($\mathcal{G}$)
 $\mathcal{N}_{\text{rand}} \leftarrow \{\}$
 $\mathcal{N}_{\text{sq,rand}} \leftarrow \{\}$
 **for** # randomized instances **do**
  $\mathcal{G} \leftarrow$ Randomize($\mathcal{G}$, # required rewiring steps)
  counts $\leftarrow$ NspPatternCounter($\mathcal{G}$)
  $\mathcal{N}_{\text{rand}} \leftarrow \mathcal{N}_{\text{rand}} +$ counts
  $\mathcal{N}_{\text{sq,rand}} \leftarrow \mathcal{N}_{\text{sq,rand}} +$ counts $*$ counts
 **end for**
 $\mathcal{N}_{\text{rand}} \leftarrow \mathcal{N}_{\text{rand}}/(\text{\#randomized instances})$
 $\mathcal{N}_{\text{sq,rand}} \leftarrow \mathcal{N}_{\text{sq,rand}}/(\text{\#randomized instances})$
 $\sigma_{\text{rand}} \leftarrow \sqrt{\mathcal{N}_{\text{sq,rand}} - (\mathcal{N}_{\text{rand}} * \mathcal{N}_{\text{rand}})}$
 $\mathcal{Z} \leftarrow (\mathcal{N}_{\text{original}} - \mathcal{N}_{\text{rand}})/\sigma_{\text{rand}}$
 **return** $\mathcal{Z}$
**end function**

---

### 6.2.2. Performance

The computational cost of Algorithm 1, $C_1$, scales with the number of required randomization steps per instance, which should be chosen proportionally to the number of edges $M = |E|$ in graph $\mathcal{G}$ (see Section 3.3.2), i.e. $C_1 = \mathcal{O}(M)$.

Algorithm 2 iterates over all edges of $\mathcal{G}$ and their adjacent edges. Therefore, it is $C_2 = \mathcal{O}(M \cdot k_{\max}) \leq \mathcal{O}(M^2)$ where $k_{\max}$ is the maximum node degree in $\mathcal{G}$. In real-world networks, $k_{\max}$ is usually much smaller than $M$.

Finally, the total computational cost of NOSPAM$_3$, i.e. of Algorithm 3, depends on the desired number of randomized network instances, $I$. Algorithm 2 is invoked $(1+I)$ times; Algorithm 1 is invoked $I$ times. Hence, the total computational cost is

$$C_{\text{NOSPAM}_3} = \mathcal{O}\left(I \cdot M \cdot k_{\max}\right). \tag{6.4}$$

Furthermore, NOSPAM$_3$ is parallelizable straightforwardly since the evaluations in terms of the randomized network instances can be executed independently of each other. An implementation of the pattern-mining program is made publicly available online [151].

## 6.3. Node-Specific Triad Patterns in Real-World Data

We will now present results obtained from the application of NOSPAM$_3$ to various peer-reviewed real-world datasets. All networks are directed and edges are treated as unweighted. A brief description of the datasets can be found in appendix A.2 on page 167.

### 6.3.1. Node-Specific vs. Ordinary Triadic *Z*-Score Profiles

Figs. 6.2 and 6.3 show the node-specific triadic $Z$-score profiles for various systems, averaged over 1000 instances of the randomized ensemble. Note that there is one curve for *every* node in the graph. The node-specific patterns on the horizontal axis are oriented the way that the node under consideration is the lower one.

We find that systems from similar fields have similar node-specific triadic $Z$-score profiles. Figs. 6.2(a) and 6.2(b) show biological transcriptional networks, Figs. 6.2(c) and 6.2(d) show data from a social context, specifically a social network of prisoners and the network of hyperlinks between political blogs. Figs. 6.3(a) and 6.3(b) show word-adjacency

*6. Node-Specific Subgraph Analysis*

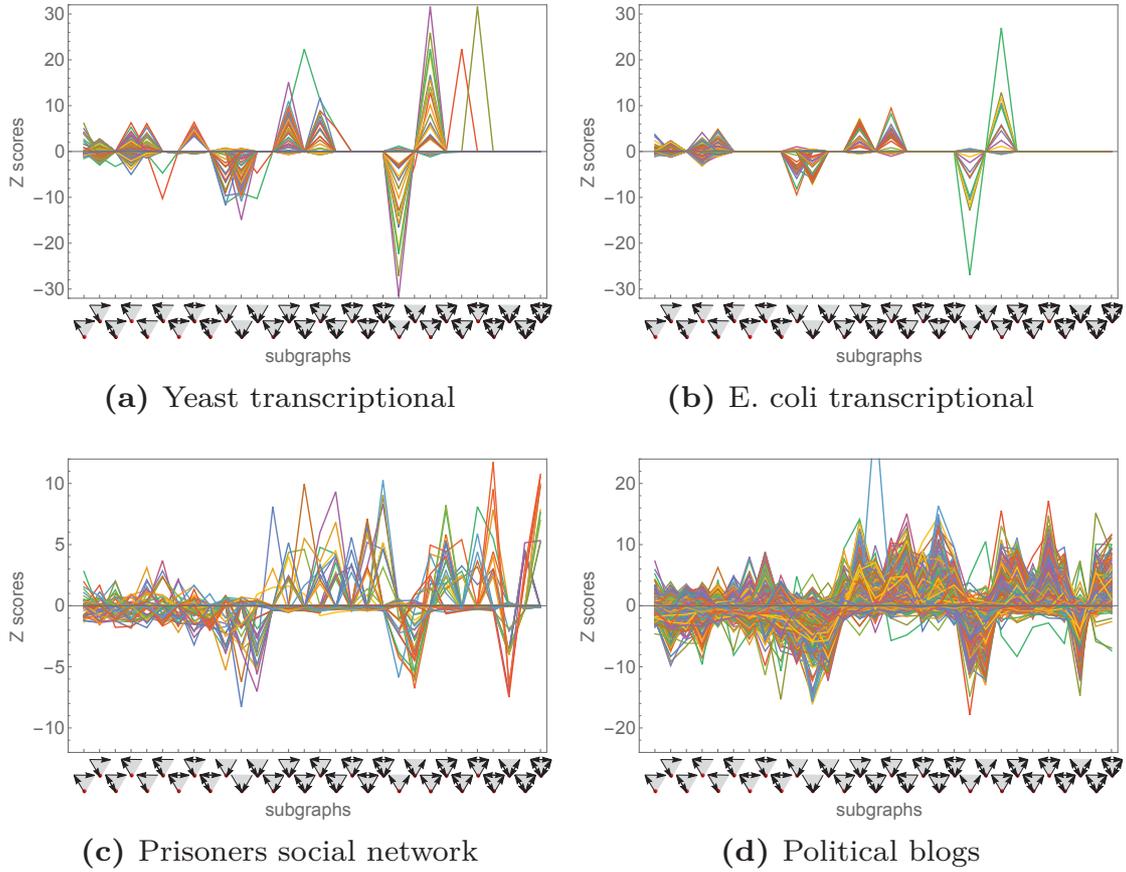

**Figure 6.2.:** Node-specific *Z*-score profiles of various real-world networks (see appendix A.2): transcriptional networks of **(a)** the yeast S. cerevisiae [7, 32] and **(b)** E. coli [7, 93] and socially related networks such as **(c)** a social network of prisoners and **(d)** hyperlinks between political blogs [1, 109]. The node-specific patterns on the horizontal axis are oriented the way that the node under consideration is the lower one. For details on the datasets see A.2.

networks in French and Spanish language, respectively. The observation that systems from a similar context exhibit similar local structural characteristics fosters the hypothesis that the latter are strongly linked to the systems' function.

The fact that NoSPaM$_3$ provides *localized* data enables us to iden-tify the areas of a graph where certain subgraph patterns primarily occur. Particularly, it allows us to test whether motifs of a system are overabundant throughout the entire network or if they are restricted to limited regions or the proximity of few nodes. In order to explore this issue, for each node, we will map its *node-specific Z* scores to a score for the *regular* triad patterns (shown in Fig. 5.1). This will be realized by taking the mean over the $Z$ scores of all node-specific triad patterns corresponding to a regular triad pattern. The mapping is shown in Table 6.1. The vector composed of the mean node-specific $Z$ scores of a node $\alpha$ is denoted as $\vec{M}^\alpha$. The measure for the contribution of a node $\alpha$'s environment to the regular pattern 14 ( ) is then for instance

$$M_{14}^\alpha \equiv \vec{M}^\alpha \left( \right) = \frac{1}{2} \left[ \vec{Z}^\alpha \left( \right) + \vec{Z}^\alpha \left( \right) \right]. \qquad (6.5)$$

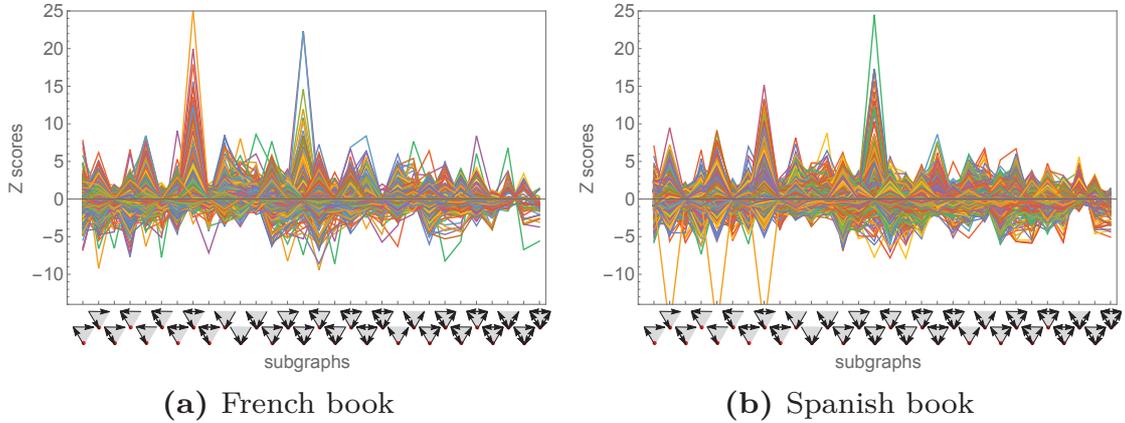

(a) French book

(b) Spanish book

**Figure 6.3.:** Node-specific $Z$-score profiles of word-adjacency networks of **(a)** French books [7, 100] and **(b)** Spanish books [7, 100]. The node-specific patterns on the horizontal axis are oriented the way that the node under consideration is the lower one. For details on the datasets see A.2.

Hence, we obtain a 13-dimensional mapped node-specific $Z$-score profile, $\vec{M}^\alpha$, for *every node* in a graph.

## 6. Node-Specific Subgraph Analysis

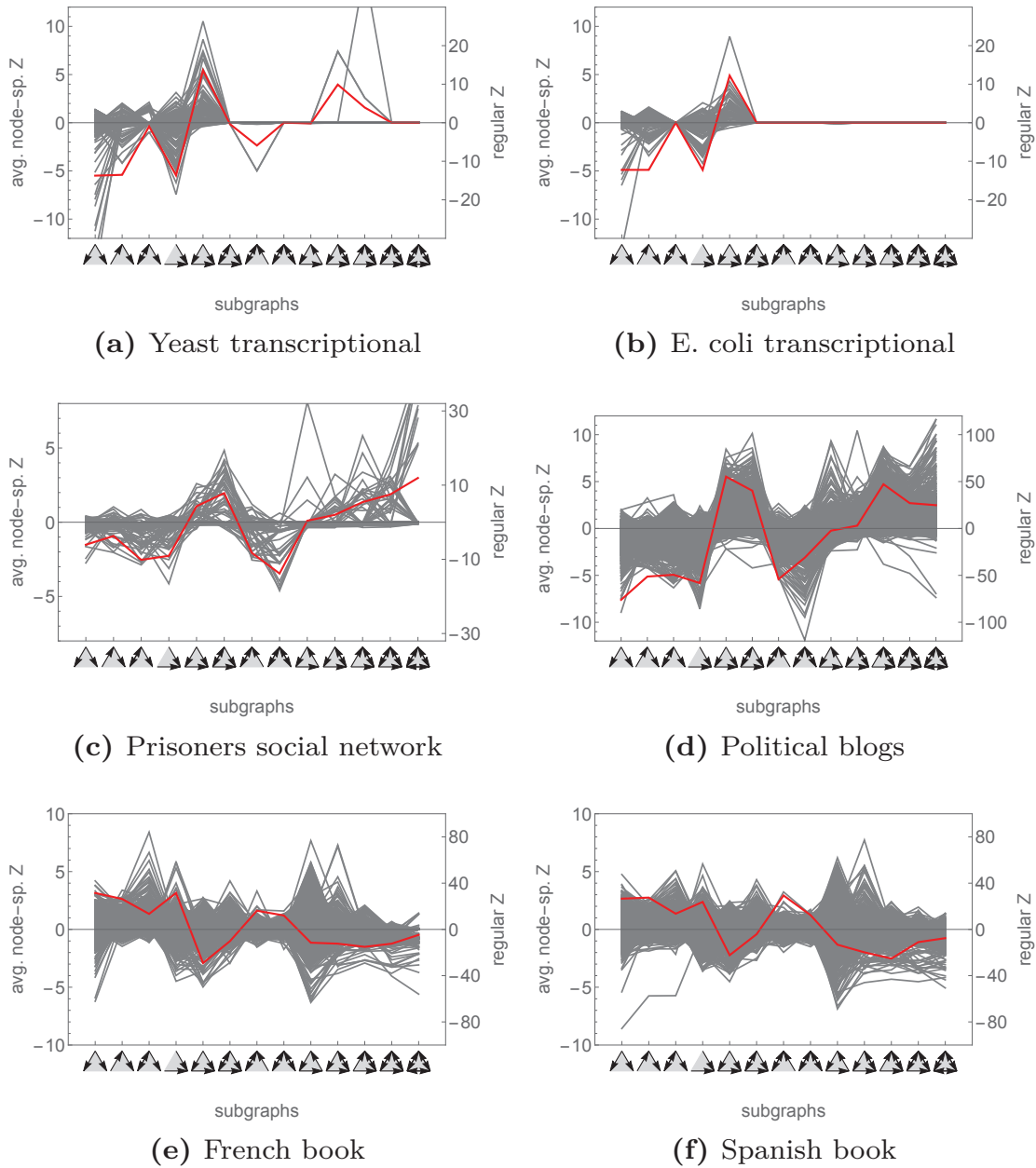

**Figure 6.4.:** Node-specific triadic $Z$ scores mapped to the patterns of Fig. 5.1. For each pattern, the average is taken over all corresponding node-specific patterns (Table 6.1). The scaling on the left corresponds to the node-specific triad patterns, the one on the right to the $Z$ scores of the ordinary triad patterns.

**Table 6.1.:** Mapping of node-specific triad patterns to their regular triad patterns.

| Node-specific triad patterns | | | | | | | | | | | | | |
|---|---|---|---|---|---|---|---|---|---|---|---|---|---|
| Regular triad patterns | | | | | | | | | | | | | |

The gray, thin curves in Fig. 6.4 show the mapped $M$ scores of each node for the networks presented in Fig. 6.2. In addition, the red, thick curve shows the regular $Z$-score profile over the whole network obtained by the commonly used motif-detection analysis (see Section 3.3). Although the gray and the red curves are not independent of each other, it shall be noticed that the regular $Z$-score profile can not be computed from the gray curves directly. In particular, it is not the mean of the latter.

It can be observed that even though a pattern may be overrepresented in terms of the system as a whole, it may still be underrepresented in the neighborhood of certain nodes. Moreover, there are patterns with a rather low regular $Z$ score, while there are both nodes with a strong positive and nodes with a strong negative contribution to the pattern. These contradictory effects seem to compensate each other on the system level. The described phenomenon can be particularly observed in the word-adjacency networks in Figs. 6.4(e) and 6.4(f), especially for the loop pattern, △.

## 6.3.2. Heterogeneous Abundance of the Feed-Forward Loop

To further investigate whether motifs appear homogenously distributed over a graph we will devote ourselves to the feed-forward loop (FFL)

## 6. Node-Specific Subgraph Analysis

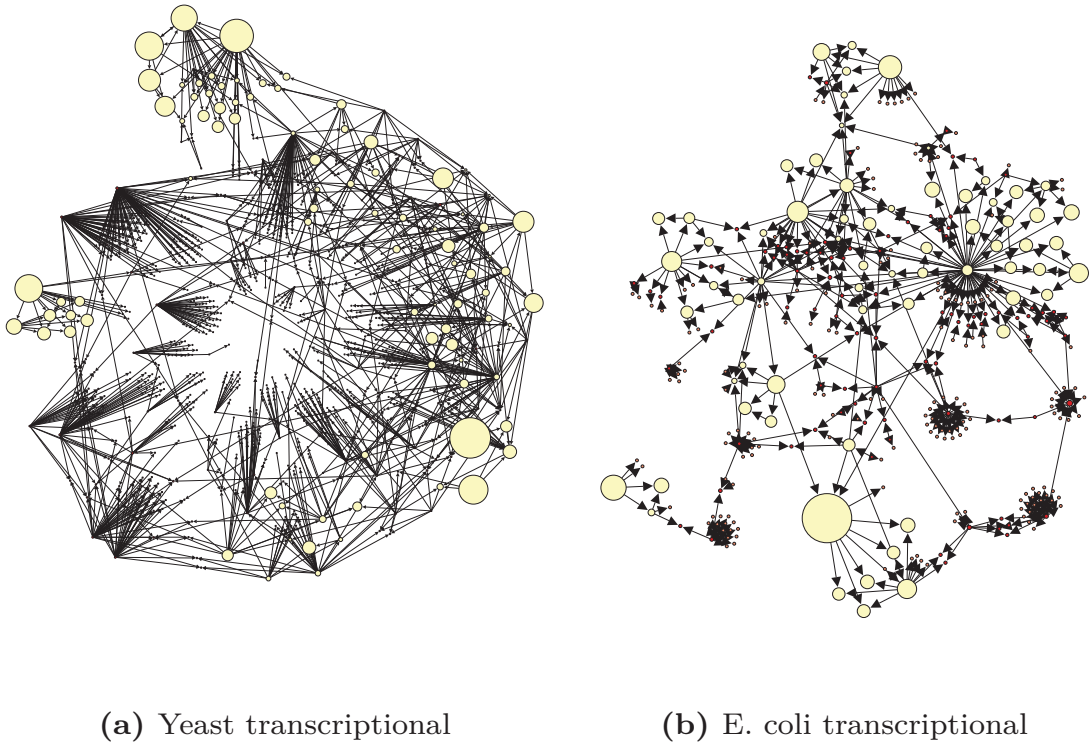

**(a)** Yeast transcriptional

**(b)** E. coli transcriptional

**Figure 6.5.:** Transcriptional regulatory networks. Vertex sizes indicate the magnitude of the mean node-specific $Z$ scores corresponding to the feed-forward loop motif (numbers 14, 16, and 23 in Fig. 6.1). Pale vertices indicate positive values, red vertices correspond to negative values (occur only with very small magnitude, hardly visible nodes). Plots were produced with Gephi [19]. Figures as appeared in [153].

pattern, ◭. The FFL is one of the patterns that has been studied most intensively in terms of its relevance for guaranteeing systems to reliably perform their functions [6, 93, 94, 129]. Specifically in transcriptional regulation networks, it was argued that the FFL pattern might play an important role for facilitating information-processing tasks [129].

Fig. 6.5 shows two of those transcriptional regulation networks. In both of them the FFL is a motif (compare Figs. 6.4(a) and 6.4(b)).

Vertex sizes are scaled by the magnitude of the averaged node-specific $Z$ scores of the three patterns corresponding to the FFL. Positive contributions are shown with bright vertices, negative ones are filled in red (do only occur with very small magnitude, i.e. very small node sizes). Apparently, there are no nodes in the networks with a significant negative contribution to the FFL. Yet, neither is the pattern homogenously overrepresented throughout the whole system, even though it is a motif. In fact, for most nodes the FFL-subgraph structure does not seem to play any role whatsoever. In contrast, there are few nodes with a rather strong contribution to the FFL eventually making it a motif of the entire system.

This effect becomes even clearer when considering histograms over the nodes' FFL contributions of the two systems. Fig. 6.6(a) shows the histogram of S. cerevisiae, Fig. 6.6(b) the one of E. coli. Both exhibit a strong peak around zero, indicating that most nodes do not participate in FFL structures more frequently or less frequently than expected at random. Only very few nodes have a large mean node-specific $Z$ score, $\vec{M}^\alpha(\triangle)$, for the patterns corresponding to the FFL.

There are two potential implications which can be derived from these observations: One conclusion could be that the FFL motif is actually not that important for the systems to work reliably. The second consequence could be that, in fact, very few nodes are critical for the systems to work the way they are supposed to. In the second case, the systems would be very prone to the failure of these crucial vertices. It may be subject of future research to further investigate these possible implications for dynamical processes on different topologies and under node failure. We will approach the functional relevance of triadic motifs in terms of dynamical processes in Chapter 7.

The feed-forward loop is also a motif in the neural network of C. elegans and the network of hyperlinks between political blogs. Histograms of their mean node-specific $Z$ scores corresponding to the FFL are shown in Figs. 6.6(c) and 6.6(d), respectively. Although their $\vec{M}^\alpha(\triangle)$ distributions also peak around zero, there are many more nodes with

## 6. Node-Specific Subgraph Analysis

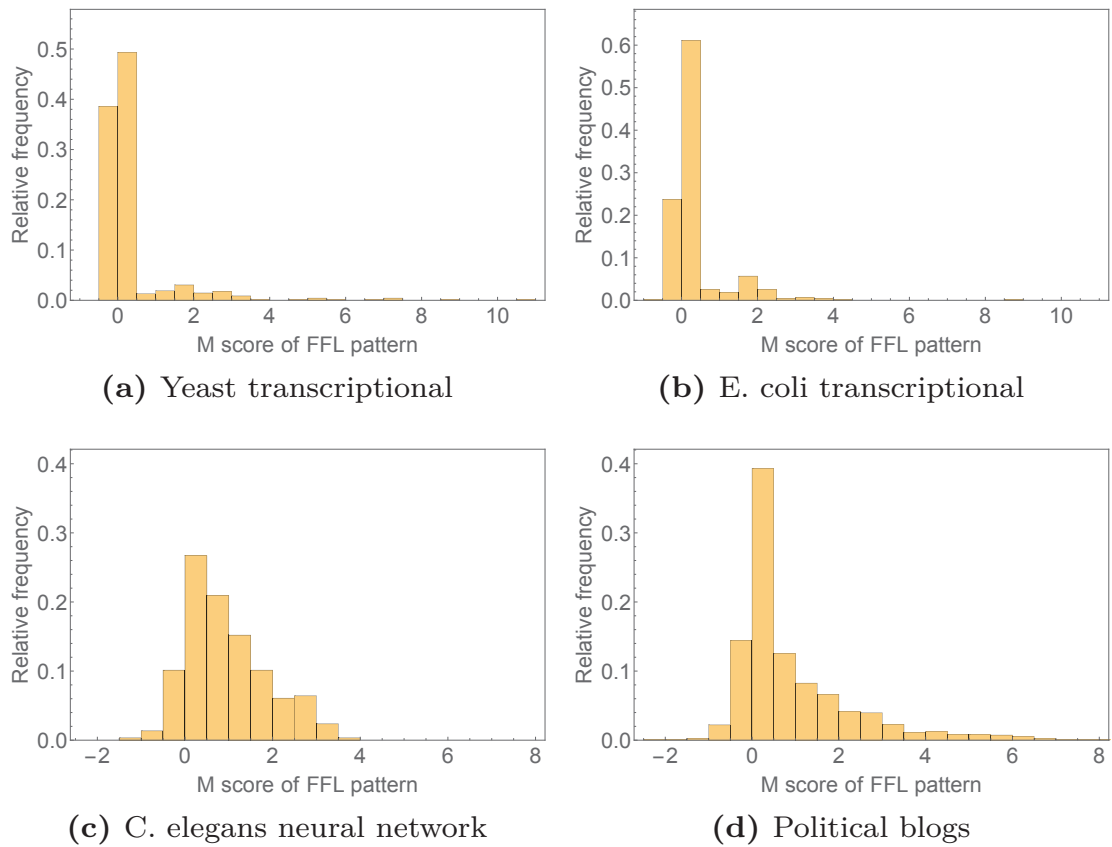

**(a)** Yeast transcriptional

**(b)** E. coli transcriptional

**(c)** C. elegans neural network

**(d)** Political blogs

**Figure 6.6.:** Histograms for the mean node-specific triadic $Z$ scores corresponding to the feed-forward loop pattern (numbers 14, 16, and 23 in Fig. 6.1). In all four networks the FFL is a *motif*.

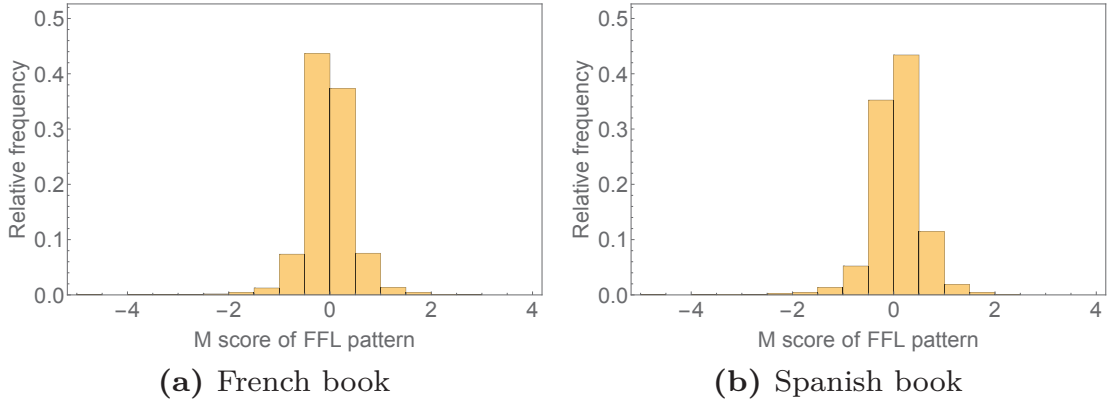

(a) French book  (b) Spanish book

**Figure 6.7.:** Histograms for the mean node-specific triadic $Z$ scores corresponding to the feed-forward loop pattern (numbers 14, 16, and 23 in Fig. 6.1). In neither of the networks the FFL is a *motif*.

positive contributions to the FFL in comparison to the transcriptional regulation networks. This suggests a more homogeneous appearance of the pattern.

Furthermore, Fig. 6.7 shows $\vec{M}^\alpha(\triangle)$ histograms of two word-adjacency networks. The feed-forward loop is a motif in neither of them. In accordance, their distributions are narrowly centered around zero.

### 6.3.3. Homogeneity and Heterogeneity of Triadic Structures Across Different Systems

As we can infer from Figs. 6.6 and 6.7, the heterogeneity of the abundance of motifs seems to vary between different systems. We will now aim at quantifying the degree of *homogeneity* in the appearance of triadic subgraphs. Moreover, it will be interesting to investigate the *homophily* of systems with respect to their triadic structure, i.e. whether vertices with similar triadic structure are more likely to be connected than others, or whether the opposite is the case. Having measures to quantify homogeneity and homophily at our disposal, we will further be able to compare different networks with each other.

## 6. Node-Specific Subgraph Analysis

**Homogeneity**

Let us define the mean correlation of the mapped node-specific $Z$ scores, $\vec{M}^\alpha$, with the regular $Z$-score profile, $\vec{Z}$, as a measure for the *homogeneity* of a graph $\mathcal{G}(V, E)$, in terms of its triadic substructure,

$$\left\langle C\left(\vec{M}^\alpha, \vec{Z}\right)\right\rangle_\alpha = \frac{1}{|V|} \sum_{\alpha=1}^{|V|} C\left(\vec{M}^\alpha, \vec{Z}\right). \tag{6.6}$$

A large homogeneity of a network indicates a similar triadic neighborhood of its vertices.

**Homophily**

Furthermore, we want to define a measure for the *homophily* in terms of a network's triadic structure, i.e. a quantity to evaluate whether connected vertices are more similar to each other than unconnected ones. The similarity between the topological triadic environment of two nodes can be quantified by the correlation of their node-specific $Z$ scores, $C\left(\vec{Z}^\alpha, \vec{Z}^\beta\right)$. Hence, we define the homophily of a graph, $\mathcal{G}(V, E)$, as the deviation of the mean pairwise correlation over the *connected pairs* of nodes, from the mean correlation over *all* pairs of nodes,

$$\left\langle C\left(\vec{Z}^\alpha, \vec{Z}^\beta\right)\right\rangle_{\alpha\mathrm{NN}\beta} - \left\langle C\left(\vec{Z}^\alpha, \vec{Z}^\beta\right)\right\rangle_{\alpha,\beta}$$
$$= \sum_{\alpha=1}^{|V|-1} \sum_{\beta=\alpha+1}^{|V|} \left(\frac{1}{\binom{|E|}{2}} A_{\alpha\beta} - \frac{1}{|V|}\right) C\left(\vec{Z}^\alpha, \vec{Z}^\beta\right). \tag{6.7}$$

A large, positive homophily indicates that *connected* nodes tend to have much more similar structural environments than unconnected ones. A strongly negative homophily means that connected nodes tend to have more dissimilar neighborhoods. A homophily of zero implies that the alikeness of the surrounding structure of two vertices does not depend on whether they are connected or not.

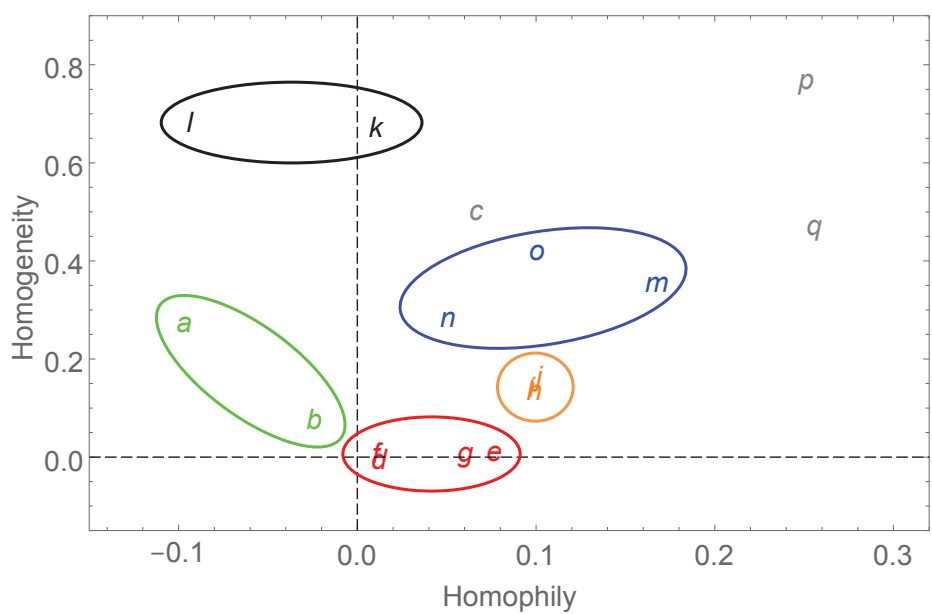

**Figure 6.8.:** Comparison of multiple systems in terms of the homogeneity and homophily of their triadic substructure. **a:** E. coli transcriptional network; **b:** yeast transcriptional network; **c:** C. elegans neural network; word-adjacency networks of **d:** English book, **e:** French book, **f:** Spanish book, and **g:** Japanese book; electronic circuits **h:** s208, **i:** s420, and **j:** s838; triadic random graphs with FFL being a motif and **k:** unidirectional links only, **l:** uni- and bidirectional links; **m:** political blogs; **n:** leadership social network; **o:** prisoners social network; **p:** scientific-citations between articles; **q:** airport-connections. For a description of the datasets see Appendix A.2.

**Results**

Analyses of a multitude of networks of diverse origin in terms of their homogeneity and homophily are presented in Fig. 6.8. The horizontal axis represents the homophily; the vertical axis indicates the homogeneity of the systems. Fig. 6.8 suggests that the position of a graph in

*6. Node-Specific Subgraph Analysis*

the homogeneity-homophily space is strongly linked to the corresponding network's function. For the transcriptional-regulatory networks, **a** and **b** (shown in green), the homogeneity is slightly positive while the homophily is slightly negative. For the word-adjacency networks, **d** - **g** (shown in red), the homogeneity is zero and their homophily is slightly positive. Both a positive homogeneity and homophily can be observed for the electronic circuits, **h** - **j** (shown in orange), and the social networks, **m** - **o** (shown in blue), with the latter being even more homogeneous. The positive homophily suggests that nodes with similar structural environment – and hence a similar structural role – are often directly linked to each other. For the triadic random graphs, **k** and **l** (shown in black) – as introduced in Chapter 5 – the homogeneity of approximately 0.7 is quite large with a slightly negative homophily. The five groups mentioned above appear clearly separated in the homogeneity-homophily space of Fig. 6.8, potentially reflecting typical structural aspects of graphs representing systems from the corresponding fields. Moreover, Fig. 6.8 indicates a rather large homogeneity for the neural network of C. elegans **c**, the citation network between scientific articles **p**, and the airport-connection network **q** (shown in gray). The last two furthermore exhibit a large homophily, i.e. connected nodes are structurally more similar than non-connected ones.

Our definition of homogeneity in Eq. (6.6) averages over the similarity of $\vec{M}^\alpha$ and the overall $Z$ score for all nodes $\alpha$ in the graph. To see whether this distribution itself is homogeneous, let us consider the standard deviation corresponding to the mean value in Eq. (6.6). Fig. 6.9 displays the two measures plotted against each other for the systems of Fig. 6.8. Again, the five groups are separated from each other in the two-dimensional plane. As in Fig. 6.8, we can draw non-intersecting borders around the instances of every single group. The standard deviation itself is a measure for the *heterogeneity* of the distribution. In accordance with this fact, with decreasing standard deviation, the systems in Fig. 6.9 show increasing homogeneity.

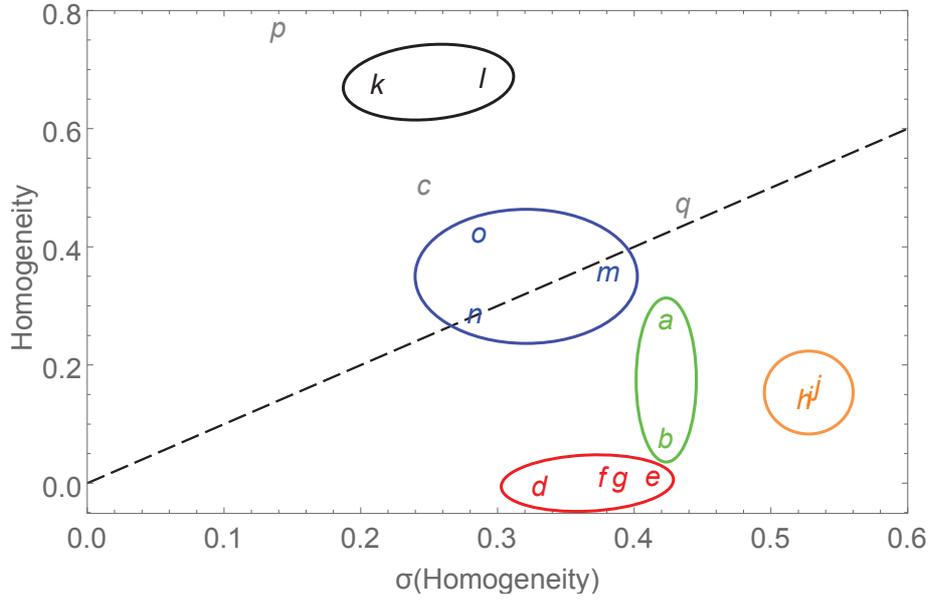

**Figure 6.9.:** Comparison of multiple systems in terms of the homogeneity of their triadic substructure. The vertical axis shows the correlation between the mapped node-specific $Z$ scores, $\vec{M}^\alpha$, with the regular $Z$-score profile, $\vec{Z}$, averaged over all nodes of the systems. The horizontal axis represents the respective standard deviation. The presented systems correspond to those of Fig. 6.8.

### 6.3.4. Clustering Based on Node-Specific *Z*-Score Profiles

Beyond allowing for a detailed investigation of the homogeneity of the appearance of motifs, with the triadic node-specific *Z*-score profiles, NoSPaM$_3$ provides for a whole new set of features for every node. This new, 30-dimensional feature vector may be used for clustering and classification purposes. We will now utilize the node-specific $Z$ scores, $\vec{Z}^\alpha$, to detect groups in the neural network of C. elegans as well as in the international airport-connection network by means of *complete-link clustering*.

**Complete-link clustering** is a hierarchical, agglomerative clustering methodology. Following a bottom-up approach, for a system consisting of $N$ units, agglomerative clustering algorithms initially assign every unit to its own cluster. Subsequently, using a *distance*, or *dissimilarity function*, $\delta(\mathcal{C}_\mu, \mathcal{C}_\nu)$, between two clusters, $\mathcal{C}_\mu$ and $\mathcal{C}_\nu$, the two most similar ones are merged. For the remaining $N - 1$ clusters, the merging process is iteratively repeated until either the desired number of clusters is reached, or the smallest dissimilarity between two clusters exceeds a certain threshold.

For clustering nodes in terms of their local triadic substructure, we use the Euclidean distance between their node-specific triadic *Z*-score profiles,

$$d(\alpha, \beta) = \sum_i \left(Z_i^\alpha - Z_i^\beta\right)^2, \quad (6.8)$$

as the dissimilarity function between two nodes $\alpha$ and $\beta$.

However, given a distance function, $d(\alpha, \beta)$, between two *nodes*, $\alpha$ and $\beta$, there are multiple ways to define a distance function, $\delta(\mathcal{C}_\mu, \mathcal{C}_\nu)$, between two *clusters*, $\mathcal{C}_\mu$ and $\mathcal{C}_\nu$, each consisting of more than one node. For complete-link clustering this distance between two clusters is defined as the maximum distance between any node in $\mathcal{C}_\mu$ and any node in $\mathcal{C}_\nu$,

$$\delta_{\text{complete-link}}(\mathcal{C}_\mu, \mathcal{C}_\nu) = \max\{d(N_\alpha, N_\beta) \mid N_\alpha \in \mathcal{C}_\mu, N_\beta \in \mathcal{C}_\nu\}. \quad (6.9)$$

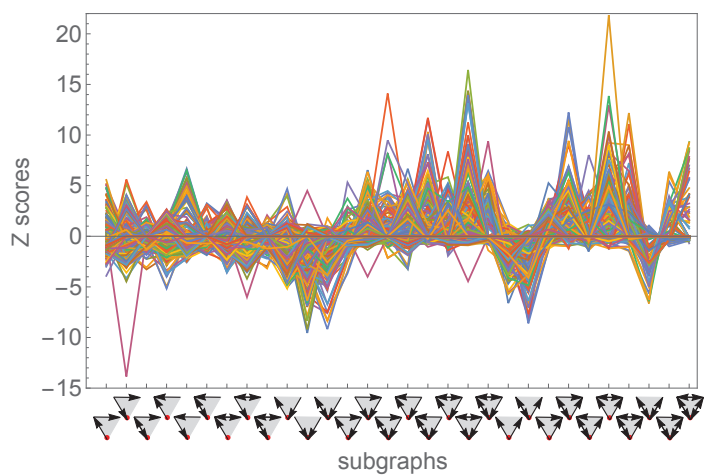

**Figure 6.10.:** Node-specific *Z*-score profile of the neural network of C. elegans. The patterns on the horizontal axis are oriented the way that the node under consideration is the lower one.

**Neural Network of C. Elegans**

The tiny roundworm *Caenorhabditis elegans (C. elegans)* has been studied intensively over the last decades and detailed data on its neural network is publicly available at www.wormatlas.org [8]. On the one hand, with approximately 300 nodes, its neural network has a manageable size and can thus be examined in great detail. On the other hand, it can serve as a model organism for more complex animals. We will now consider the largest connected component of the graph representing the somatic nervous system of C. elegans. This neural network is composed of 279 neurons and 2,194 chemical synapses between them. The neurons of C. elegans are particularly suitable for cluster analysis, since existing expert classifications are available for comparison with our findings. The neurons are, for instance, classified into motor neurons, interneurons, and sensory neurons.

All 279 node-specific triadic *Z*-score profiles of C. elegans' neural network are displayed in Fig. 6.10. There is a multitude of peaks in both positive and negative direction. Yet, not all peaks are present in the $\vec{Z}^\alpha$ profiles of *all* nodes simultaneously. Partitioning the neurons

## 6. Node-Specific Subgraph Analysis

into five groups via complete-link clustering, we obtain the mean node-specific triadic $Z$-score profiles shown in Fig. 6.11. The largest group, shown in Fig. 6.11(a), consists of 172 neurons in whose neighborhood none of the triadic patterns is significantly overrepresented or underrepresented as compared to the null model. This observation implies that the triadic environment of these nodes is primarily explained by the degree distribution of the network. The second group, displayed in Fig. 6.11(b), consists of 38 neurons with a clear overrepresentation of pattern 19, 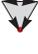. In the group shown in Fig. 6.11(c), pattern 30 (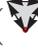) is a node-specific motif; in Fig. 6.11(d), pattern 26 (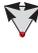) is a motif. In the last group (Fig. 6.11(e)), patterns 17 (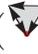), 24 (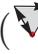), and 30 (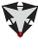) are significantly overrepresented.

Fig. 6.12 shows the adjacency matrix of the neural network, in which rows and columns are ordered according to the complete-link clustering. The groups from Fig. 6.11 are separated by lines. It shall be emphasized that the clustering algorithm does not take into account whether nodes that are grouped together are densely connected with each other. Neurons are rather assigned to the same cluster if their neighborhood possesses a similar triadic substructure. The 'o' labels on the left and on the bottom in Fig. 6.12 indicate sensory neurons, interneurons, motor neurons, and polymodal neurons. We find that the large group with no exceptional triadic structure is comprised of all types of neurons. The second group consists predominantly of motor neurons, group four is majorly composed of sensory neurons, and five is comprised mostly of interneurons. Group three is dominated by sensory neurons and interneurons.

The analysis shows that, on the one hand, our clustering is, to great extent, in agreement with expert classifications in which neurons were assigned to be motor, sensory, or interneurons based on their structural and functional embedding in the network. On the other hand, our clustering sorted out many neurons which, apparently, do not contribute strongly to the network's local structural characteristics (group 1). Moreover, we detect two distinct sets which are mainly comprised

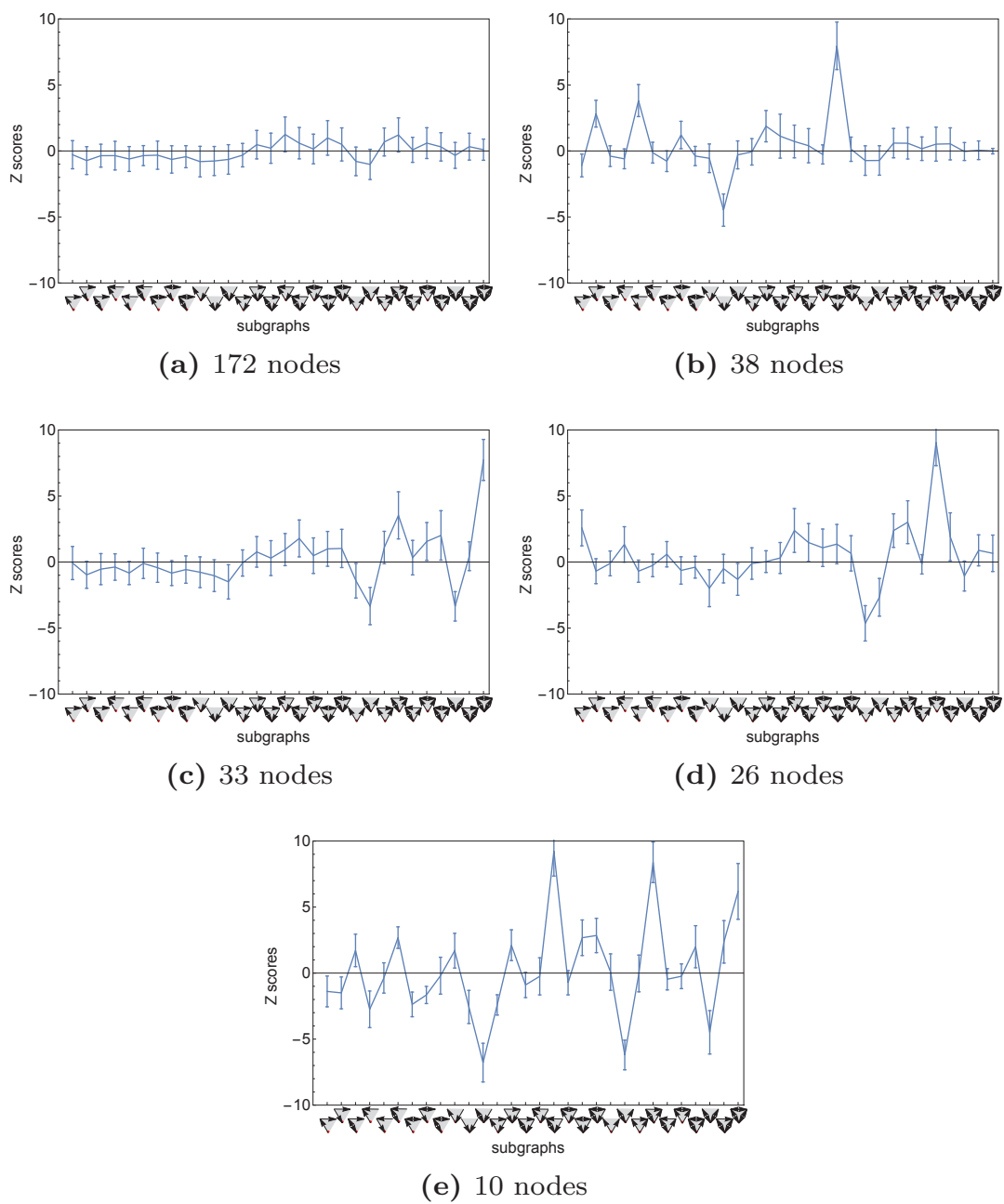

**Figure 6.11.:** Averaged *Z*-scores over the groups of neurons separated by lines in Fig. 6.12. **(a)** comprises all kinds of neuron types, the majority in **(b)** are motor neurons, **(c)** and **(e)** are dominated by interneurons, and **(d)** by sensory neurons. Error bars indicate one standard deviation.

## 6. Node-Specific Subgraph Analysis

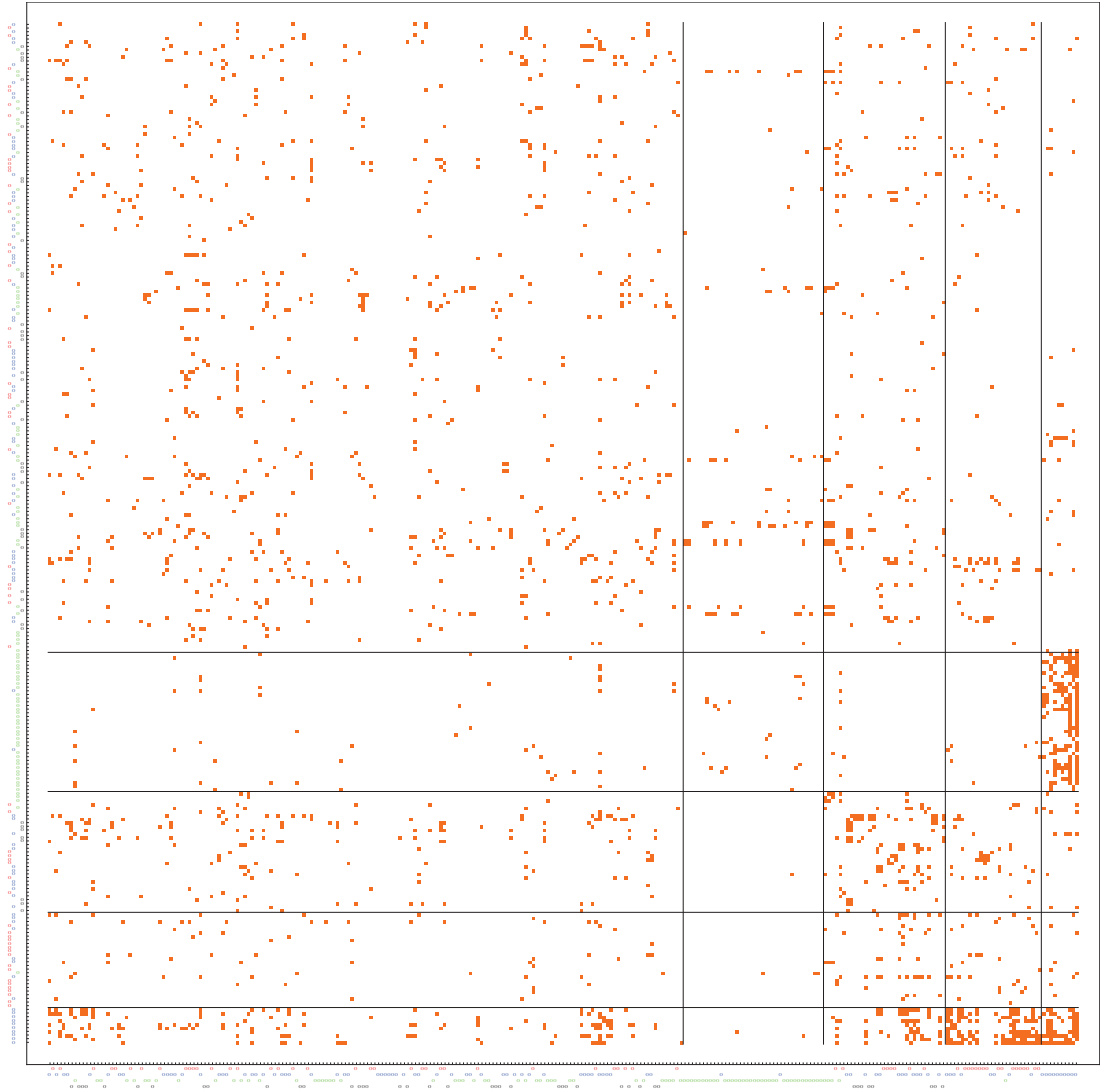

**Figure 6.12.:** Adjacency matrix of the neural network of C. elegans. Rows and columns are sorted according to the result of a complete-link hierarchical clustering of nodes with respect to their node-specific triadic Z-score profiles. The position of the 'o' labels for the different rows indicates (from left to right) sensory neurons (red), interneurons (blue), motor neurons (green), and polymodal neurons (black). Columns are marked accordingly from top to bottom.

of interneurons. Potentially, interneurons can be further divided into subgroups of different topological neighborhoods and possibly distinct functional roles. Hence, our novel network-analysis tool may provide for interesting new aspects of data which can be used by experts from the respective fields to better understand their buildup.

A branch of research closely related to this approach is the general mining of *roles* in complex networks [50, 62]. Nodes are assigned to the same role if they are similar in terms of a certain set of structural features. Both methodologies detect groups of nodes that share the same function in contrast to community-detection algorithms. The latter aim to detect groups of vertices which are closely connected within the groups, while inter-group connections are rare. In fact, our node-specific properties can serve as an input for role-extraction and mining algorithms, allowing to combine our triadic structural aspects with other features in order to improve role detection in complex networks.

**International Airport-Connection Network**

We will now consider the network of global flight connections between airports. The data is retrieved from `http://openflights.org/` and comprises 3,438 nodes and 34,775 edges.

Fig. 6.14(a) shows the entirety of the 3,438 node-specific triadic *Z*-score profiles and Fig. 6.14(b) their mappings to the regular patterns, together with the ordinary triadic *Z*-score profile. The loop and the bidirectionally closed triangle are motifs of the network. Again, we find structural groups of nodes with distinct characteristic profiles.

Fig. 6.13 shows the graph of flight connections with vertex coordinates reflecting the geographical positions of the various airports. Sizes of the vertices represent their degrees, and their colors indicate groups detected by means of a complete-link clustering. We observe that in America and Asia, most of the high-degree nodes belong to the group colored in green, while in Europe many of them are assigned to the red group, indicating structural differences in the flight connections within the continents.

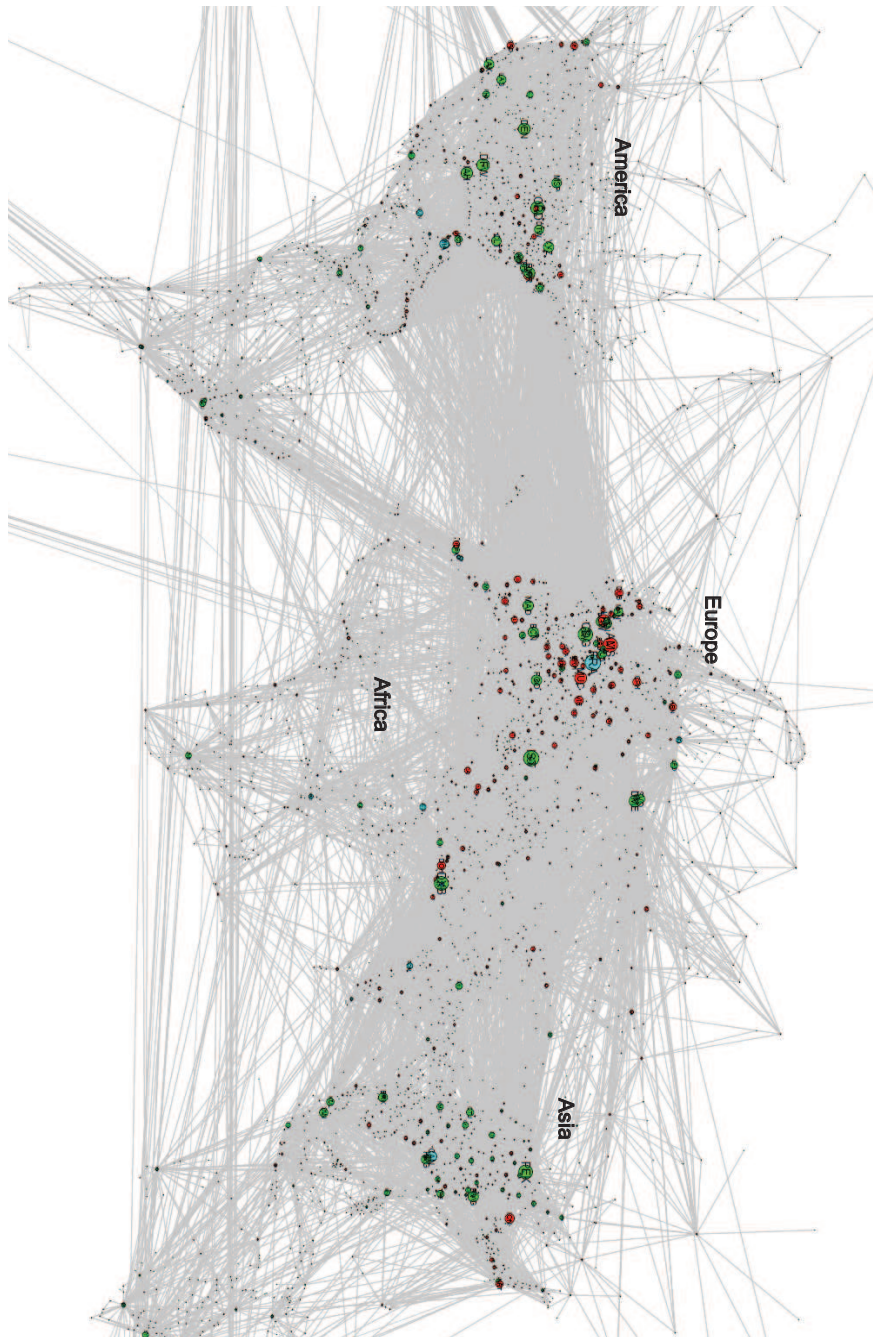

**Figure 6.13.:** Geographical map of international flight connections. Vertex sizes reflect degrees. Colors correspond to groups detected via complete-link clustering with respect to node-specific triadic $Z$-scores.

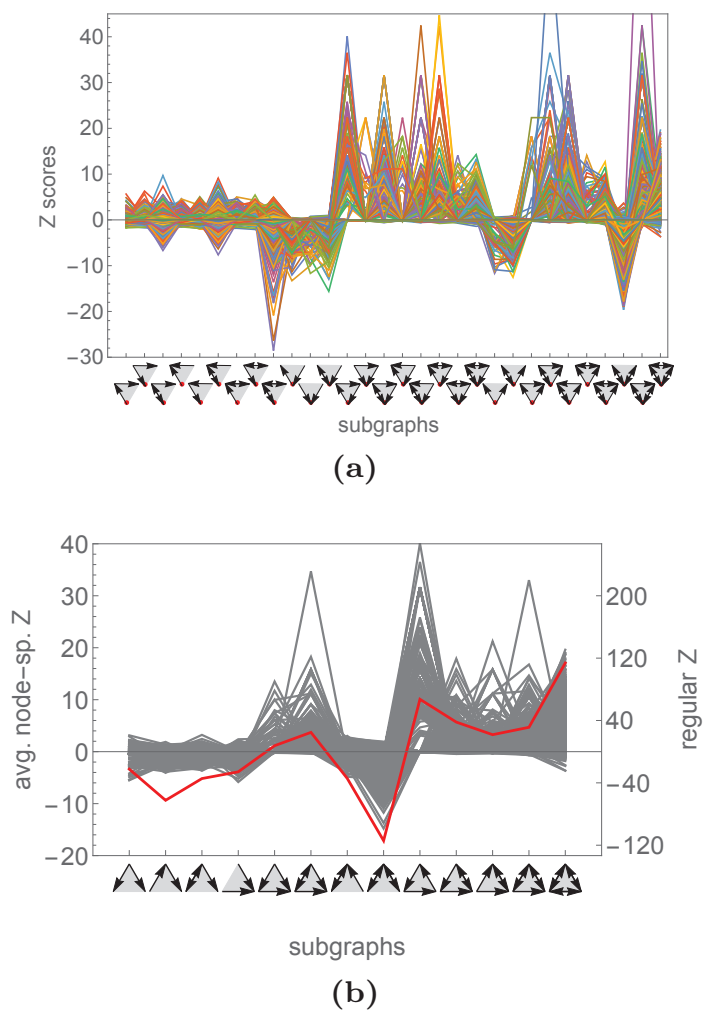

**Figure 6.14.:** **(a)** Node-specific triadic $Z$-score profiles of the airport-connectivity network. The patterns on the horizontal axis are oriented the way that the node under consideration is the lower one. **(b)** Node-specific $Z$ scores mapped to the patterns of Fig. 5.1. For each pattern, the average is taken over all corresponding node-specific patterns (Table 6.1). The scaling on the left corresponds to the node-specific contributions, the one on the right to the ordinary $Z$-score profile.

## 6. Node-Specific Subgraph Analysis

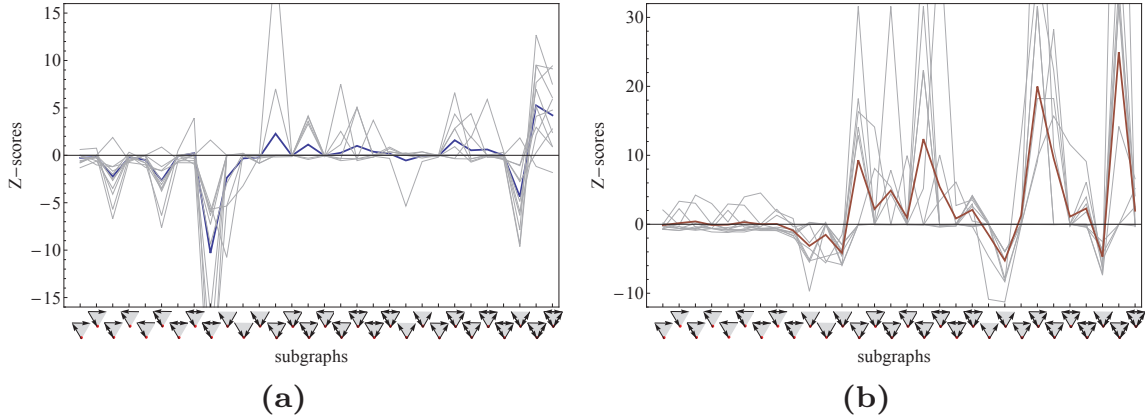

**Figure 6.15.:** Node-specific $Z$-score profiles of **(a)** the ten busiest airports in terms of passenger traffic; **(b)** the top-ten airports by magnitude of their $|Z^\alpha|$ values.

Let us further consider the structural neighborhood of the ten busiest airports in terms of passenger traffic (Fig. 6.15(a)) and the airports with the largest $\left|\vec{Z}^\alpha\right|$ values (Fig. 6.15(b)). The busiest airports have a clear antimotif, , in which they are positioned at the beginning of a bidirectional line. Pattern  and the closed triangle, , are motifs of the busiest airports, reflecting their central positions in the graph.

On the other hand, among the airports with the largest $\vec{Z}^\alpha$ vectors by magnitude, the triangle, is not a motif. In contrast, node-specific patterns involving unidirectional links – e.g. , , , or the loop,  – are overrepresented. Unidirectional connections do generally not occur between hubs, but rather involve airports with less passenger traffic, e.g. in case a single aircraft serves multiple destinations, one after the other, and eventually returns to its starting point. Considering the identities of airports with large $\left|\vec{Z}^\alpha\right|$ values[2], we find that those airports are either airports with very little passenger traffic itself, or bridging nodes which connect very small airports to the rest of the world. Fig. 6.15 shows that, by only regarding the local neighborhood of vertices it is possible to estimate their role in the whole system.

---

[2] Galena, Alaska (GAL) / Mombasa, Kenya (MBA) / Kigali, Ruanda (KGL) / Mehamn, Norwegen (MEH) / Chevak Airport, Arkansas (VAK) / Pikangikum,

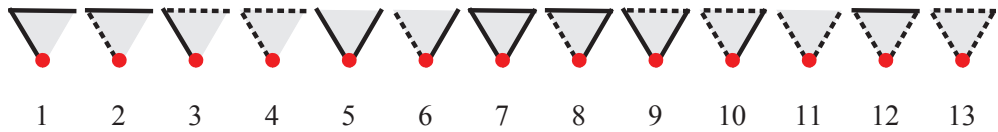

**Figure 6.16.:** All possible connected, nonisomorphic, *signed* triadic subgraph patterns with respect to a distinct node (here: lower node). Solid lines represent positive edges, dashed lines correspond to edges with a negative sign.

## 6.4. Node-Specific Triad Patterns in Signed Networks

So far we have concentrated on directed, unsigned networks for our investigation of node-specific triadic substructure. However, the analysis can be adapted to various other types of networks. We will now give an outlook on a generalization to *signed, undirected* graphs like they frequently appear in a social context (see Section 2.1.1 and Section 3.2). Positive edges describe mutual friendship, while negative ones indicate antagony.

From the perspective of a particular node, there are 13 non-isomorphic, connected triadic subgraph patterns in undirected, signed graphs. Those are displayed in Fig. 6.16.

---

Ontario (YPM) / Landsdowne House, Ontario (YLH) / Tobago, Trinidad and Tobago (TAB) / Red Lake, Ontario (YRL) / Mota Lava, Vanuatu (MTV)

## 6. Node-Specific Subgraph Analysis

### 6.4.1. Algorithm

In order to evaluate their node-specific $Z$ scores, we need to adapt NOSPAM$_3$. Therefore, the randomization in Algorithm 1 needs to be adjusted to guarantee that the number of both positive and negative edges adjacent to each node is preserved. The modified version of the link-switching is shown in Algorithm 4. The corresponding microscopic link-switching steps are illustrated in Fig. 6.17.

---

**Algorithm 4** Degree-preserving randomization of a signed, unweighted graph

---
    **function** RANDOMIZESIGNED(Graph $\mathcal{G}(V,E)$, no. of required steps)
        s = 0
        **while** s < number of required rewiring steps **do**
            pick a random link $e_1 \in E$
            **if** $e_1$ is positive **then**
                pick a 2nd *positive* link $e_2 \in E$ at random
            **else**
                pick a 2nd *negative* link $e_2 \in E$ at random
            **end if**
            **if** $e_1$ and $e_2$ do not share a node **then**
                rewire according to the pair-switch rules in Fig. 6.17
                **if** one of the new links already exists **then**
                      undo the rewiring
                **end if**
            **end if**
            s++
        **end while**
        **return** randomized instance of $\mathcal{G}$
    **end function**

---

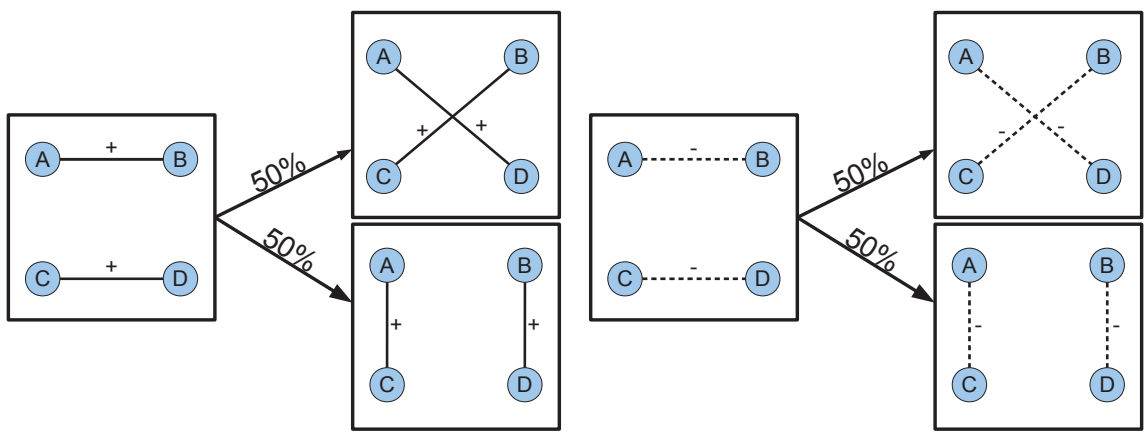

**Figure 6.17.:** Microscopic link-switchings performed to generate the randomized ensembles for signed, undirected graphs.

### 6.4.2. Political Sentiment Between Countries

As an application of NoSPaM$_3$ to a real, signed dataset we will consider the network of international relations between countries, obtained from the GDELT[3] database (see Appendix A.2.2).

Fig. 6.18(a) shows the signed, node-specific triadic $Z$-score profiles of the countries in September 2001. In Figs. 6.18(b)-(d), they are partitioned into three groups in terms of the structural similarity of their environment, i.e. the similarity of their $Z$-score profiles.

While there is no unbalanced pattern (see Chapter 3.2) overrepresented in the first two groups (Figs. 6.18(b) and 6.18(c)), the last group (Fig. 6.18(d)) has the unbalanced motif ▽. According to Heider's theory [60, 61] unbalanced motifs imply 'social tension' in the relationships of the corresponding countries. The extraordinary situation in September 2001 – shortly after the terrorist attacks on the World Trade Center – makes it particularly worth investigating the identities

---

[3] Global data on events, location, and tone [84]. `http://gdeltproject.org/`

## 6. Node-Specific Subgraph Analysis

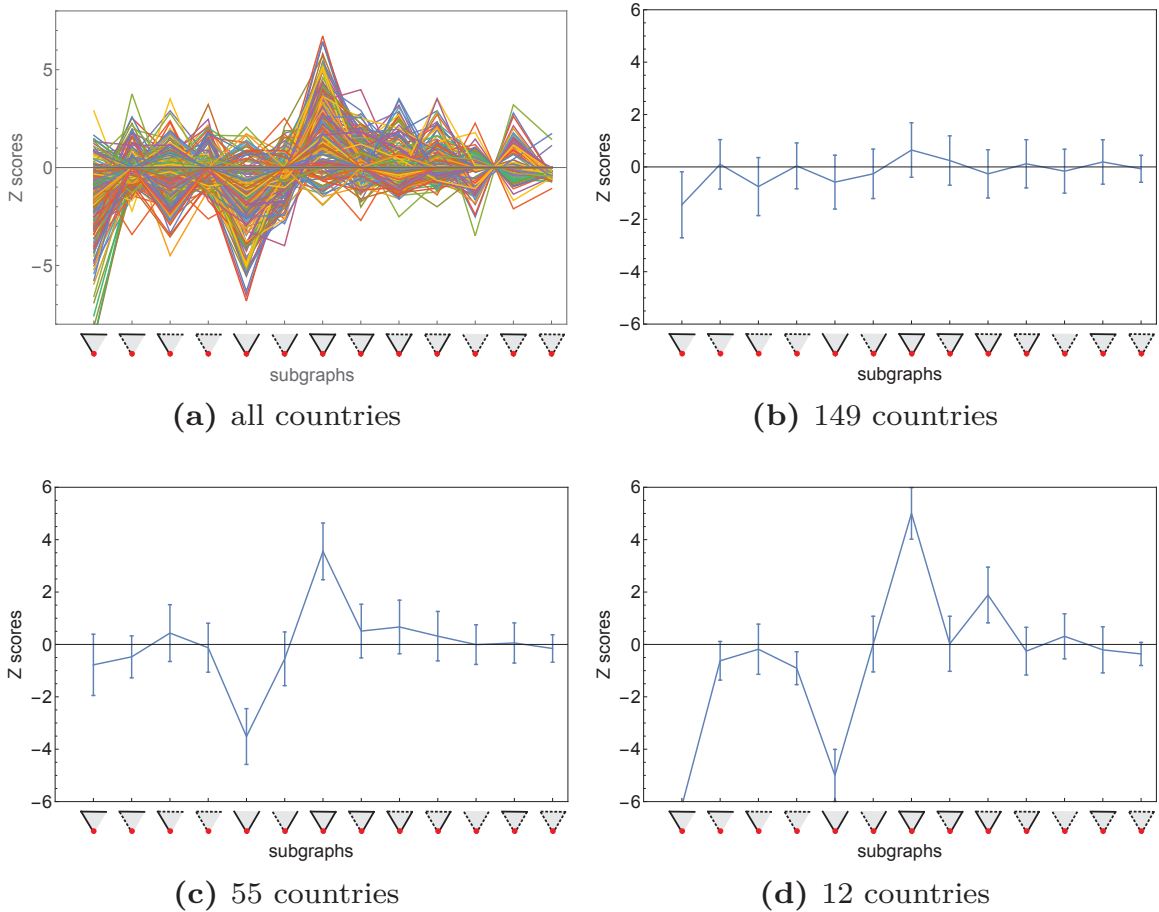

**Figure 6.18.:** **(a)** Signed node-specific $Z$ scores of the political-sentiment network of September 2001. **(b)-(d)** mean $Z$-score profiles for groups of countries clustered with respect to the similarity of their node-specific $Z$ score.

| Country | Change |
|---:|:---:|
| United Kingdom | 11 |
| France | 4 |
| Germany | 5 |
| Italy | 4 |
| Belgium | 4 |
| Europe | 6 |
| Russia | 3 |
| Egypt | 4 |
| Saudi-Arabia | 4 |
| Occupied Palestinian Territory | 1 |
| Iran | 4 |
| Afghanistan | 16 |

**Table 6.2.:** Frequency of link changes in the relationships of countries corresponding to Fig. 6.18(d) between September 2001 and October 2001. For comparison: the average number of changes over all countries in the same period was 1.4.

of the countries in the third group. Those are the United Kingdom, France, Germany, Italy, Belgium, Europe (i.e. the European Union in general), Russia, Egypt, Saudi-Arabia, the Occupied Palestinian Territory, Iran, and Afghanistan. Interestingly, nearly all of these countries are either directly or indirectly involved in the global conflicts succeeding the 9/11 terrorist attacks.

The potential instabilities in the relationships of these countries are substantiated by considering changes in the relationships of the last group compared to the other two groups. Let us quantify the change in the relationships of a country, $i$, as the number of connections which either change their sign from $+$ to $-$ or vice versa within one month. Due to the incompleteness of the data, of course there are also edges vanishing and establishing. Since we cannot make any statement on whether they changed or not, they are not taken into account. For the countries in the third group (Fig. 6.18(d)), the average change from September to October 2001 was 5.5 in contrast to an average of 0.9 and 1.8 for the members of the first and second group, respectively.

*6. Node-Specific Subgraph Analysis*

The individual changes of the twelve countries from the third group are shown in Table 6.2. While, not surprisingly, the political relationships involving Afghanistan record the largest amount of sentiment changes, nearly all of the twelve group members have their sentiments changing considerably more often than 1.4 – the average over all 216 countries. Remarkably, the sentiments affecting the Palestinian Territory change the least, possibly reflecting that the cause of their unbalanced relationships dates back further in time.

This preliminary analysis already suggests the high potential of the structural analysis of signed networks for contributing to a better understanding of global relationships and their interconnectedness. The GDELT database provides for a multitude of data to be analyzed in future research.

# 7. The Influence of Triad Motifs on Dynamical Processes

*The work presented in this chapter originates from a collaboration with Otti D'Huys. A publication in a peer-reviewed journal is currently in preparation.*

In Chapter 2.5, we have given an introduction to dynamical processes acting on the vertices of graph structures. Existing work concerning the effect of triadic subgraphs on the evolution of such processes was reviewed in Chapter 3.5. Most publications provide evidence that systems with certain dynamical behavior, typically possess a certain structure [25, 71, 73]. On the other hand, investigations on the question whether particular triadic subgraph structures imply a particular dynamical behavior have hardly been conducted. Moreover, most functional analysis of triad patterns has focused on isolated subgraphs [81, 91, 120]. It is not obvious though, that properties observed on isolated motifs are sustained in case they are embedded in a larger network and thus affected by other adjacent nodes. These shortcomings of existing research are mainly due to a lack of sound generative network models necessary to design synthetic graphs with predefined triadic structure.

However, in Chapter 5 we have suggested the triadic random graph model (TRGM) which facilitates to generate ensembles of random graphs with non-trivial triadic substrucure. We will now utilize triadic random graphs to test the abundance of triad motifs' influence on the evolution of dynamical processes. Subsequently, in Section 7.2, we will inves-

*7. The Influence of Triad Motifs on Dynamical Processes*

tigate how triadic motifs influence the spectral gap of the coupling matrix which is particularly relevant in the context of synchronization processes.

## 7.1. Triad Motifs in Networks of Coupled Oscillators

We will now use the triadic random graph model to generate random networks with predefined expected triadic substructure. TRGMs, $\mathcal{G}\left(N, \vec{\mathcal{T}}\right)$, utilize a partition of a graph's $N$ vertices into pair-disjoint triads, so called Steiner triples (STs, see Chapter 4). Since Steiner triples share no dyads, they can be specified independently of each other. A TRGM is parametrized by the distribution of triad patterns on the $N(N-1)/6$ STs. The desired frequency of occurrence of the different patterns on the STs is contained in the vector

$$\vec{\mathcal{T}} = \Big( N(\triangle), N(\triangle), N(\triangle), N(\triangle),$$
$$N(\triangle), N(\triangle), N(\triangle), N(\triangle), N(\triangle), N(\triangle), \quad (7.1)$$
$$N(\triangle), N(\triangle), N(\triangle), N(\triangle), N(\triangle), N(\triangle) \Big)^T$$

with $\sum_i \mathcal{T}_i = N(N-1)/6$. In order to sample from a TRGM, $\mathcal{G}\left(N, \vec{\mathcal{T}}\right)$, the patterns specified in $\vec{\mathcal{T}}$ are distributed – with random orientation – randomly to the Steiner triples. By adjusting the entries of $\vec{\mathcal{T}}$, it is possible to generate graphs with tuned triadic subgraph structure.

The triadic random graphs will now allow us to test whether systems with the same triadic subgraph structure show similar dynamical behavior. For this purpose, we will focus on networks of noise-driven, coupled, damped oscillators. The dynamics on each node, $j$, is governed by the following set of differential equations:

$$\frac{dx_j(t)}{dt} = (-a + i\omega)\, x_j(t) + b\, e^{i\theta} \sum_k A_{kj}\, x_k(t) + \xi_j(t). \qquad (7.2)$$

$x_j(t) \in \mathbb{C}$ is the dynamical variable corresponding to node $j$. The damping parameter, $a$, causes an exponential decay of the magnitude of $x_j(t)$, while $\omega$ is the natural frequency of an induced phase oscillation. The second term in Eq. (7.2) models the effect of the dynamical variables of all nodes adjacent to vertex $j$. $b$ controls their overall impact on the dynamics of node $j$. $\theta$ is the coupling phase, i.e. the phase shift that applies to the signals arriving at node $j$ from its neighbors. Considering only the first two terms on the right-hand side of Eq. (7.2), without any external stimuli, the network would remain in the silent state in which $x_j(t) = 0$ for all times. Therefore, we constantly impose noise, modeled by $\xi_j(t)$ to each node $j$. We consider Gaussian noise. In particular, the probability-density for $\xi_j(t)$ is given by a Gaussian distribution with zero mean, $\langle \xi_j \rangle_t = 0$, and standard deviation one, $\sqrt{\langle \xi_j^2 \rangle_t - \langle \xi_j \rangle_t^2} = 1$ with $\langle \cdots \rangle_t$ representing the time average. The dynamical equations defined by Eq. (7.2) are a realization of Eq. (2.59) on page 41, setting $\vec{x}_j = x_i$, $\vec{f}[\vec{x}_j(t)] = (-a + i\omega)\, x_j(t) + \xi_j(t)$, $\boldsymbol{G} = \boldsymbol{A}$, and $\vec{h}[\vec{x}_j(t), \vec{x}_k(t)] = b\, e^{i\theta}\, x_k(t)$.

Using the networks, generated by means of the TRGM, we will now present evidence suggesting that systems with the same triad motifs show similar dynamical properties and that, for the dynamics defined by Eq. (7.2), some behavior known for isolated motifs is maintained when embedded in a network. For varying coupling phase, $\theta$, we will investigate the *network output*,

$$\left\langle |x_j(t)|^2 \right\rangle_{j,t} = \sum_{j=1}^{N} \int_{t=t_0}^{t_0+\Delta t} |x_j(t)|^2\, dt, \qquad (7.3)$$

## 7. The Influence of Triad Motifs on Dynamical Processes

as well as the *network correlation*,

$$\left\langle \frac{\langle |x_j(t)\, x_k(t)|\rangle_t}{\sqrt{\langle |x_j(t)|^2\rangle_t \langle |x_k(t)|^2\rangle_t}} \right\rangle_{j\neq k}$$
$$= \sum_{j<k} \frac{\int_{t=t_0}^{t_0+\Delta t} |x_j(t)\, x_k(t)|\, dt}{\sqrt{\left(\int_{t=t_0}^{t_0+\Delta t} |x_j(t)|^2\, dt\right)\left(\int_{t=t_0}^{t_0+\Delta t} |x_k(t)|^2\, dt\right)}}.$$

(7.4)

The network output measures the squared magnitude of the dynamical variables, averaged over time and over all vertices in the graph. The network correlation indicates to which extent the dynamics of individual nodes affect each other. For an entirely independent time evolution, the network correlation would yield zero. If all $x_i(t)$ evolved synchronously, the correlation would be one.

In the following simulations, we will use a second-order Runge-Kutta method with a step size of 0.01 to iterate the dynamical Eqs. (7.2). Parameters are set to $a = 1$, $\omega = 2\pi$, and $b = 0.8/\gamma_1$ with $\gamma_1$ being the largest eigenvalue of the adjacency matrix $\boldsymbol{A}$. The effect of the coupling phase, $\theta$, will be examined in the following analysis, traversing its codomain of $[0, 2\pi]$. Simulations are started from random initial conditions, sampled from a Gaussian distribution with zero mean and a variance of two. After a transient of $t_0 = 2 \times 10^3$ time steps, the time averages in Eqs. (7.3) and (7.4) are taken over $5 \times 10^5$ time steps.

In Chapter 5, we found that there are some strong correlations in the $Z$ scores of different patterns and that not all $Z$-score profiles can be realized. Here, our focus will be on the overrepresentation of a single pattern at a time. Naturally, the overrepresentation of a single pattern comes along with the underrepresentation of others. Nevertheless, these underrepresentations will not play a decisive role in our analysis. We will concentrate on the two purely unidirectional triangles: the feedforward loop, 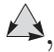, and the loop, 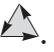.

## 7.1. Triad Motifs in Networks of Coupled Oscillators

### 7.1.1. Feed-Forward Loop Pattern

In order to generate ensembles of triadic random graphs with the feed-forward loop as a motif, we use the $\vec{\mathcal{T}}$ vectors shown in Table 7.1. For each id, we sample 50 instantiations of networks, each consisting of $N = 49$ nodes. The corresponding mean $Z$-score profiles – evaluated with the procedure introduced in Chapter 3.3 – are displayed in Fig. 7.1.

For an isolated FFL pattern, , we would expect the average pairwise correlations between the nodes' dynamical variables to have a minimum at $\theta = \pi$. For this value of $\theta$, signals directly travelling from the upper node to the lower right one are in antiphase with those traversing the lower left node and thus cancel out. On the contrary, for $\theta = 0$ we would expect the average correlation to peak, since the signals are in phase.

Fig. 7.2(a) shows results for the network correlation of the triadic random graphs composed of 49 nodes with the feed-forward loop motif. In analogy to the isolated feed-forward loop, the average correlations have their maximum at $\theta = 0$ and their minimum at $\theta = \pi$. The characteristic shape of the curves becomes clearer for sparser graphs (corresponding to a smaller id). We will now compare these observations to the random expectation for networks with the same density and the same node degrees as in the triadic random graphs. For this purpose, we perform a degree-preserving randomization (see Section 3.3.2) of the graphs and evaluate the network correlations on these randomized

| id | | | | | | | | | | | | | | | | | $M$ |
|----|-----|---|---|---|---|---|---|----|---|---|---|---|---|---|---|---|-----|
| 20 | 357 | 2 | 0 | 1 | 1 | 0 | 1 | 30 | 0 | 0 | 0 | 0 | 0 | 0 | 0 | 0 | 98  |
| 21 | 358 | 2 | 0 | 0 | 0 | 0 | 0 | 32 | 0 | 0 | 0 | 0 | 0 | 0 | 0 | 0 | 98  |
| 22 | 351 | 0 | 0 | 0 | 0 | 0 | 0 | 41 | 0 | 0 | 0 | 0 | 0 | 0 | 0 | 0 | 123 |
| 23 | 343 | 0 | 0 | 0 | 0 | 0 | 0 | 49 | 0 | 0 | 0 | 0 | 0 | 0 | 0 | 0 | 147 |
| 24 | 334 | 1 | 0 | 0 | 0 | 0 | 0 | 57 | 0 | 0 | 0 | 0 | 0 | 0 | 0 | 0 | 172 |

**Table 7.1.:** Components of the $\vec{\mathcal{T}}$ vectors used to generate triadic random graphs of size $N = 49$ with the FFL as a motif. The last column indicates the number of edges, $M$, in the networks.

*7. The Influence of Triad Motifs on Dynamical Processes*

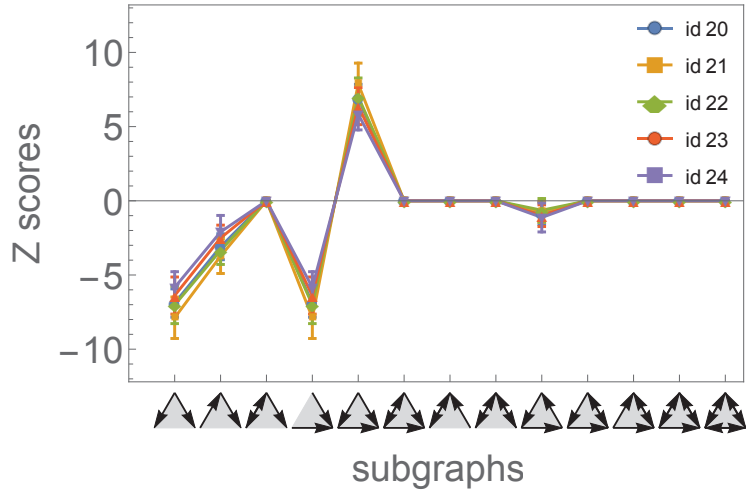

**Figure 7.1.:** *Z*-score profiles of triadic random graphs with the feed-forward loop being a motif. The $\vec{\mathcal{T}}$ vectors used for the network generation are shown in Table 7.1. For each id the average over 50 samples is taken. Error bars indicate one standard deviation.

networks. Results are presented in Fig. 7.2(b). It can be observed that the minimum of the correlation is much broader and flatter than for the triadic random graphs. Like for the graphs with the feed-forward loop motif there is a distinct maximum at $\theta = 0$. However, this peak can be observed for all networks in which signals can be transmitted from a node $i$ to a node $j$ on different paths[1], in a time span that is small compared to the typical time scale of the dynamical process (for instance the period of an oscillator). For the transmission time of the signals being similar and no phase shifts occurring along the paths, all signals will interfere constructively and therefore result in a strong correlation. The disparity in the dependence of the average pairwise correlation on the coupling phase, for systems with and those without the feed-forward loop motif, becomes particularly clear when illustrating the respective curves in the same plot (see Fig. 7.3(a)).

Similar results are obtained for the network output. For systems

---

[1] E.g. $i \to k_1 \to k_2 \to k_3 \to j$ and $i \to k_4 \to k_5 \to j$.

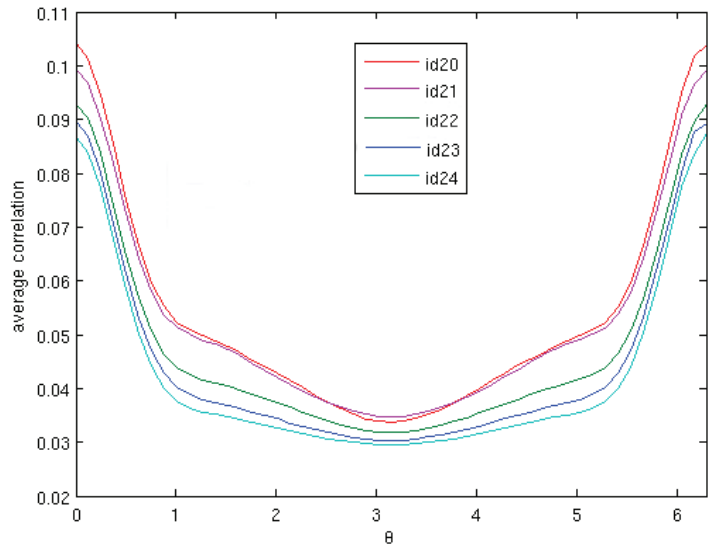

**(a)**

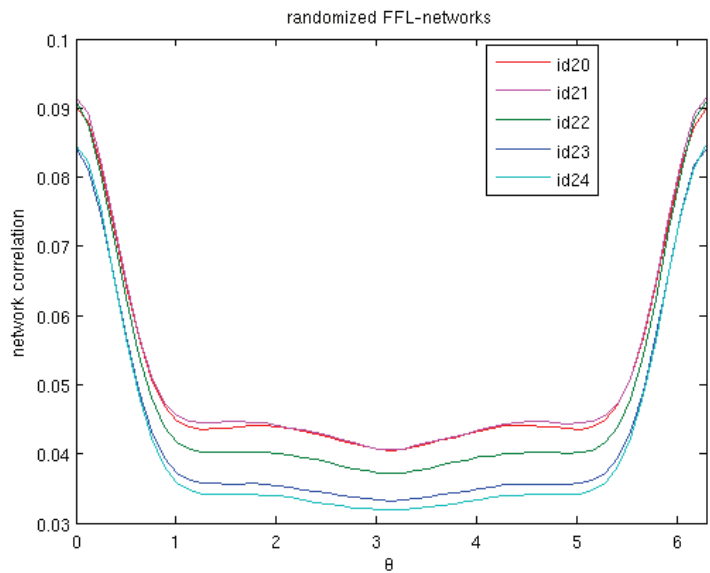

**(b)**

**Figure 7.2.:** Average pairwise correlation for varying coupling phase for **(a)** several TRGM networks of size $N = 49$ in which the FFL pattern is a motif, and **(b)** degree-preserving randomizations of the networks shown in **(a)**. Parameters are $a = 1$, $\omega = 2\pi$, and $b = 0.8/\gamma_1$.

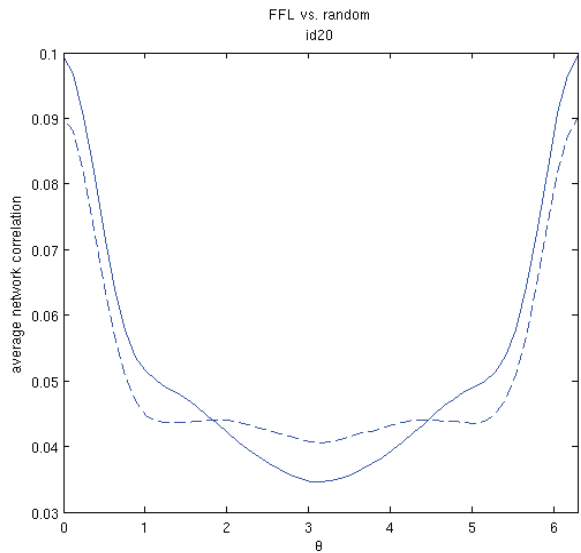

**(a)**

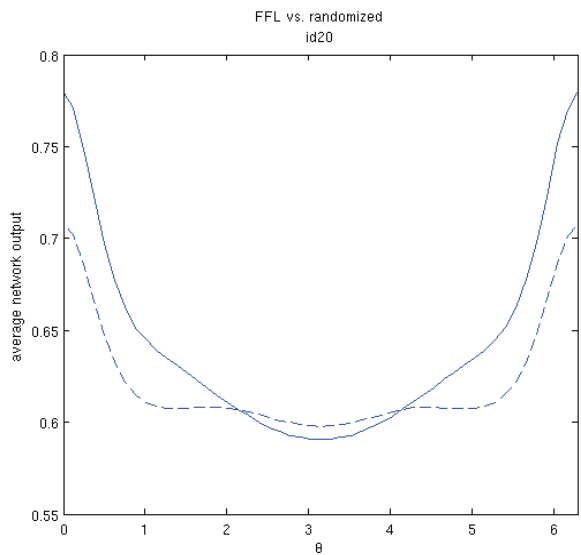

**(b)**

**Figure 7.3.:** Solid lines indicate **(a)** average pairwise correlation and **(b)** average network output for varying coupling phase, $\theta$, for triadic random graphs in which the FFL pattern is a motif (id 20 in Table 7.1). Dashed curves represent the measures for degree-preserving randomizations of the TRGs. Parameters are $a = 1$, $\omega = 2\pi$, and $b = 0.8/\gamma_1$.

*7.1. Triad Motifs in Networks of Coupled Oscillators*

with an overrepresentation of the feed-forward loop pattern, we also observe curves with their maximum at zero and their only minimum at $\pi$. Merely the shape of the curves for the output is slightly different from the one for the correlation. Fig. 7.3(b) illustrates the comparison of TRGMs with id 20, with their randomized versions in terms of their network outputs.

Figs. 7.2 and 7.3 clearly show the visibility of the FFL pattern, even though it is not investigated in isolation, but embedded in network structures composed of 49 vertices. In Fig. 7.2 we see that the FFL's signature can be recognized best for sparser graphs. This is plausible as for denser networks, the FFLs will eventually get masked, culminating in the extremal case of a complete graph in which all local structure will be lost.

### 7.1.2. Loop Pattern

The $\vec{\mathcal{T}}$ vectors used to generate triadic random graphs with the loop being a motif are displayed in Table 7.2. Again, for each id, 50 network instantiations, each consisting of $N = 49$ nodes, were sampled. The respective mean $Z$-score profiles are illustrated in Fig. 7.4.

Considering an isolated loop pattern, 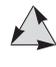, we would expect the average pairwise correlations between the nodes' dynamical variables to be minimal whenever traversing a loop of length three results in a de-

| id |  |  |  |  |  |  |  |  |  |  |  |  |  |  |  |  | $M$ |
|---|---|---|---|---|---|---|---|---|---|---|---|---|---|---|---|---|---|
| 25 | 359 | 0 | 1 | 0 | 0 | 0 | 0 | 0 | 0 | 0 | 0 | 32 | 0 | 0 | 0 | 0 | 98 |
| 26 | 355 | 0 | 1 | 0 | 0 | 0 | 0 | 0 | 0 | 0 | 0 | 36 | 0 | 0 | 0 | 0 | 112 |
| 27 | 351 | 0 | 0 | 0 | 0 | 0 | 0 | 0 | 0 | 0 | 0 | 41 | 0 | 0 | 0 | 0 | 123 |
| 28 | 343 | 0 | 0 | 0 | 0 | 0 | 0 | 0 | 0 | 0 | 0 | 49 | 0 | 0 | 0 | 0 | 147 |
| 29 | 334 | 1 | 0 | 0 | 0 | 0 | 0 | 0 | 0 | 0 | 0 | 57 | 0 | 0 | 0 | 0 | 171 |

**Table 7.2.:** Components of the $\vec{\mathcal{T}}$ vectors used to generate triadic random graphs of size $N = 49$ with the loop as a motif. The last column indicates the number of edges, $M$, in the networks.

*7. The Influence of Triad Motifs on Dynamical Processes*

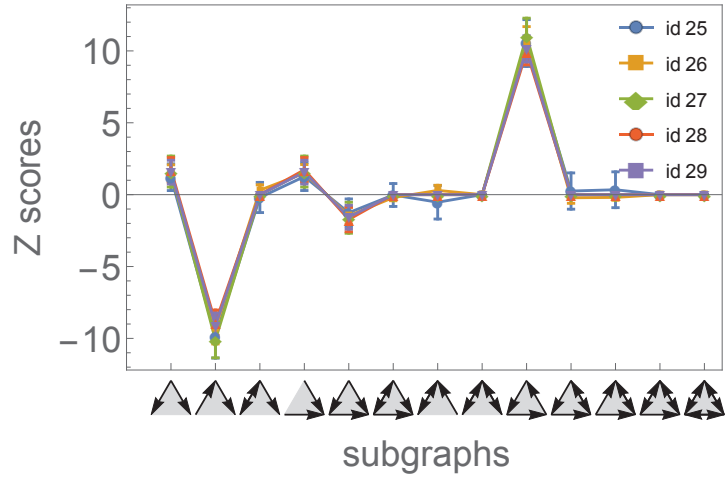

**Figure 7.4.:** Z-score profiles of triadic random graphs with the loop being a motif. The $\vec{\mathcal{T}}$ vectors used for the network generation are shown in Table 7.2. For each id the average over 50 samples is taken. Error bars indicate one standard deviation.

structive interference of the signal with itself. This is equivalent to the condition $3\theta = (2k + 1)\pi$, i.e. $\theta \in \left\{\frac{\pi}{3}, \pi, \frac{5\pi}{3}\right\}$. In analogy, we expect an amplification of the signal for $3\theta = 2k\pi$, i.e. $\theta \in \left\{0, \frac{2\pi}{3}, \frac{4\pi}{3}\right\}$.

Results of the network correlation of triadic random graphs composed of $N = 49$ nodes, with the loop pattern as a motif, are presented in Fig. 7.5(a). Both minima and maxima of the average pairwise correlations are found exactly at the values of $\theta$ for which they are expected in isolated triadic loops. Like for the FFL motif, the signature of the loop pattern is clearer for sparser graphs (smaller id). However, it is clearly pronounced for all investigated densities. Results for the randomized versions of the graphs are shown in Fig. 7.5(b). We observe a rather flat minimum between $\theta \approx \pi/3$ and $5\pi/3$ and the common peak at a coupling phase of zero. Fig. 7.6(a) shows the network correlation for systems with the triadic loop as a motif (id 25) and, for comparison, the correlation in their randomized versions.

Fig. 7.6(b) illustrates the resulting curves of the network output for id 25. Again, we observe a pronounced peak at a coupling phase of

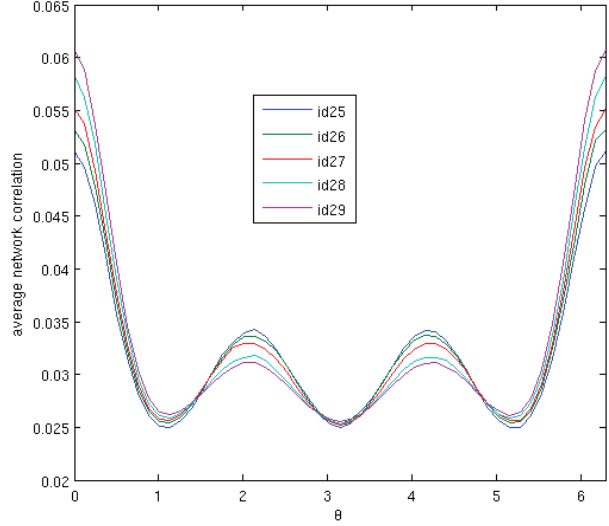

**(a)**

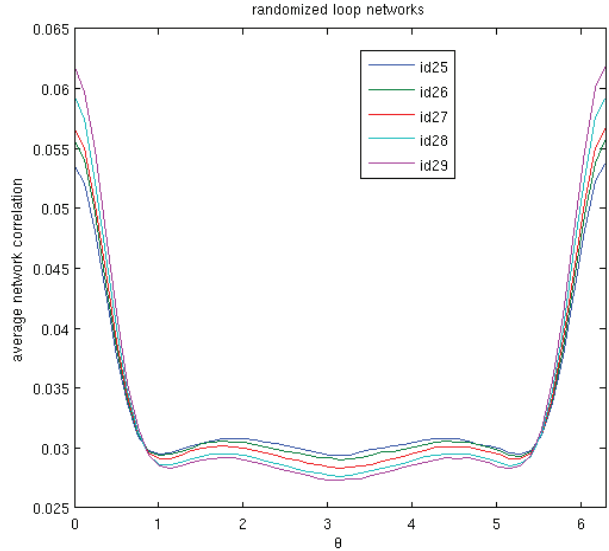

**(b)**

**Figure 7.5.:** Average pairwise correlation for varying coupling phase for **(a)** several TRGM networks of size $N = 49$ in which the loop pattern is a motif, and **(b)** degree-preserving randomizations of the networks shown in **(a)**. Parameters are $a = 1$, $\omega = 2\pi$, and $b = 0.8/\gamma_1$.

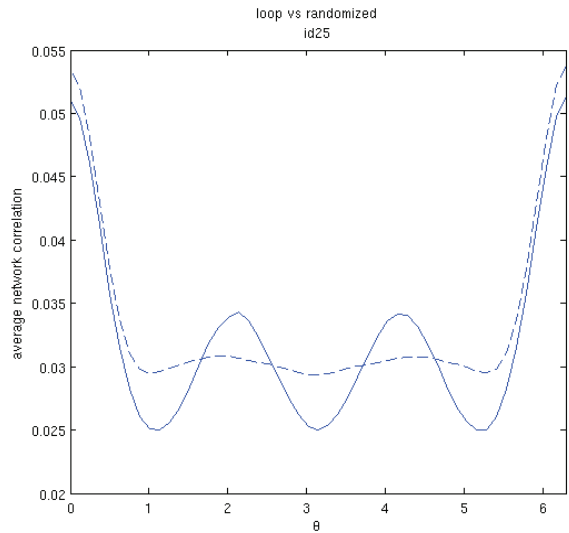

**(a)**

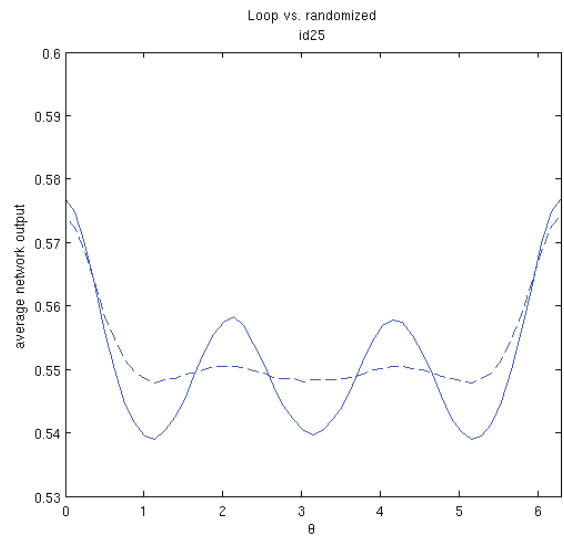

**(b)**

**Figure 7.6.:** Solid lines indicate **(a)** average pairwise correlation and **(b)** average network output for varying coupling phase, $\theta$, for triadic random graphs in which the loop pattern is a motif (id 20 in Table 7.2). The dashed curves represent the measures for degree-preserving randomizations of the triadic random graphs. Parameters are $a = 1$, $\omega = 2\pi$, and $b = 0.8/\gamma_1$.

zero, two local maxima at $\theta = \frac{2\pi}{3}$ and $\frac{4\pi}{3}$, and minima at $\theta = \frac{\pi}{3}$, $\pi$, and $\frac{5\pi}{3}$.

In summary, for systems in which the triadic loop is a motif and whose dynamics is governed by Eq. (7.2), the influence of the motif is clearly visible both for the network correlation and the network output.

## 7.2. Triad Motifs and the Spectral Gap

We will now investigate the influence of motifs in terms of synchronization processes for two biological networks, namely the neural network of C. elegans and the transcriptional network of the yeast S. cerevisiae. The concept of node-specific triad patterns, as suggested in Chapter 6, allows us to identify the vertices of a graph in whose neighborhood certain patterns predominantely occur. Hence, it is possible to consider graph topologies observed in real networks and, by means of a targeted removal of the relevant nodes, to alter the local substructure. By removing only few vertices the overall structure of the network will be maintained.

Consider, for instance, the neural network of C. elegans in which the triad patterns , , , , and  are overrepresented. Fig. 7.7(a) shows the evolution of the $Z$ score profile under consecutive removal of the top 50 vertices in terms of their mean node specific $Z$ scores corresponding to the bidirectional triangle, $\vec{M}^\alpha($$)$. The brighter the color of the curve, the more nodes – together with their adjacent edges – have been removed. While the general shape of the $Z$-score profile stays the same, the $Z$ score of pattern 16 constantly decreases from approximately 20 to zero. In analogy, Fig. 7.7(b) displays the transition of the $Z$-score profile when deleting nodes with a strong contribution to pattern 15 ().

In Section 2.5.3 we have learned about master stability functions (MSFs) relating topological aspects of a graph to the properties of dynamical processes acting on the vertices. A topological characteristic

*7. The Influence of Triad Motifs on Dynamical Processes*

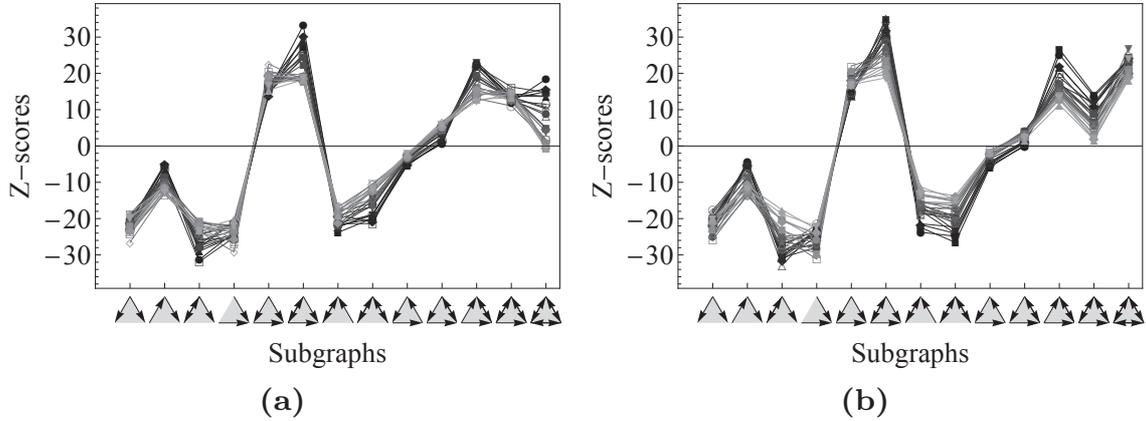

**Figure 7.7.:** *Z*-score profiles of C. elegans after consecutive removal of the 50 nodes with the strongest contribution to **(a)** pattern 16, $\vec{M}^\alpha(\triangle)$ **(b)** pattern 15, $\vec{M}^\alpha(\triangle)$. Nodes are removed in decreasing order of their contribution to the respective patterns. Curves with a brighter gray correspond to graphs in which more vertices have already been removed.

which is important for many dynamical processes is the spectral gap of the coupling matrix, $G$, with normalized row sum,

$$\Delta = \gamma_1 - \max(|\gamma_2|, |\gamma_N|), \qquad (7.5)$$

where $\gamma_1$ is the largest eigenvalue of $G$ and $\max(|\gamma_2|, |\gamma_N|)$ yields the second largest absolute value of the eigenvalues.

A non-vanishing spectral gap is, e.g., a prerequisite for many dynamical systems[2] in order to be able to synchronize. Synchronization phenomena are particularly important in neural dynamics where they are believed to be related to selective attention, learning, and information processing [48, 154], but also to medical conditions such as Parkinson's, epilepsy, schizophrenia, or Alzheimer's [52, 137, 140].

---

[2] The exact conditions may depend on details of the differential equations governing the dynamics.

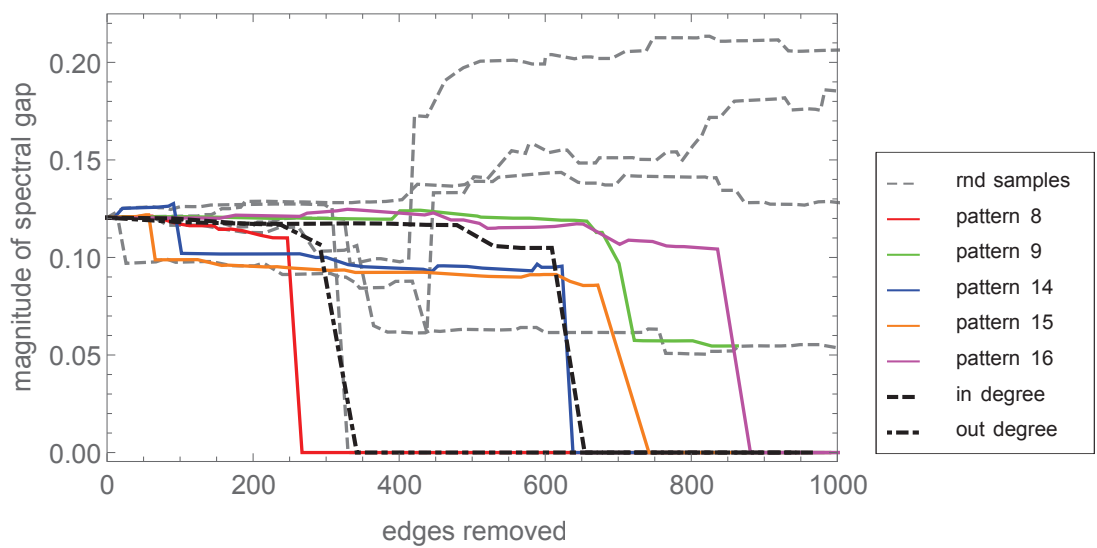

**Figure 7.8.:** Spectral gap, $\Delta$, between the largest and the second largest eigenvalue of the normalized coupling matrix representing the neural network of C. elegans. The horizontal axis indicates the number of edges removed when consecutively erasing the nodes with the strongest contribution to the network's triad-motif patterns (patterns 8, 9, 14, 15, and 16), or the largest degrees, respectively. For comparison we consider consecutive deletion of random nodes.

To investigate how the spectral gap changes under targeted node removal, we will now further consider the neural network of C. elegans. As illustrated in Fig. 7.7, there are five triad patterns which are significantly overrepresented, compared to the random expectation for a system with the same node degrees and the same number of both unidirectional and bidirectional links: pattern 8 (△), pattern 9 (△), pattern 14 (△), pattern 15 (△), and pattern 16 (△). In order to examine the influence of these patterns on the spectral gap, we consecutively remove nodes in decreasing order of their mean node-specific $Z$ scores corresponding to the respective patterns. For comparison, we also test a targeted removal of nodes with the highest degrees and a uniformly random node removal. Of course, removing nodes with high degrees

*7. The Influence of Triad Motifs on Dynamical Processes*

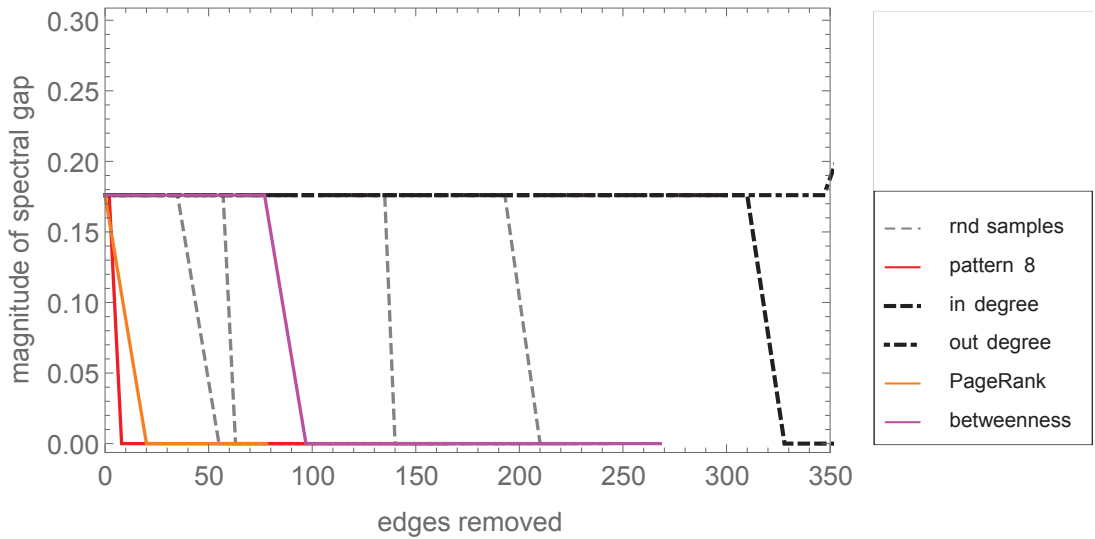

**Figure 7.9.:** Spectral gap, Δ, between the largest and the second largest eigenvalue of the normalized coupling matrix representing the transcriptional network of the yeast S. cerevisiae. The horizontal axis indicates the number of edges removed when consecutively erasing the nodes with the strongest contribution to the network's triad motif (the feed-forward loop, pattern 8), the largest degrees, the largest PageRank, or the largest betweenness, respectively. For comparison we consider consecutive deletion of random nodes.

constitutes a more drastic manipulation of the graph structure than removing those with low degrees. Therefore, we focus on the number of edges implicitly removed by the node deletions. Fig. 7.8 shows the evolution of the spectral gap plottet against the total number of edges removed. We find that a targeted removal of nodes contributing to pattern number 8, the feed-forward loop (FFL), leads to the fastest disappearance of the spectral gap. In other words, decreasing the number of FFLs in the graph strongly prevents the dynamics on the vertices from synchronizing. Note that the vanishing of the spectral gap is not due to a separation of the network structure. Fig. A.1 in the appendix shows the graph structure of the neural network before and after Δ dropping to zero, in case nodes with a strong FFL contribution are consecutively removed. A random deletion of nodes hardly leads to a drop of the spectral gap to zero, even if 1000 of the approximately 2200 edges are affected. This finding supports the hypothesis that FFLs may be critical for systems performing information-processing tasks.

Fig. 7.9 shows according results for the transcriptional network of the yeast S. cerevisiae, another biological network in which the FFL pattern is a motif. Since there are no other triadic motifs expressed in the graph structure (compare Fig. 6.4(a) on page 116), we also investigate a targeted removal in terms of the vertices' PageRank and betweenness, respectively. Again the FFL has the highest impact on the spectral gap, serving as further evidence for the relevance of the pattern in biological contexts.

# 8. Conclusions

Modeling complex systems as networks is a powerful approach to investigate their structure. With more and more data becoming available, over the last 15 years, network science has established itself as an own scientific field. Using methodologies from statistical physics, computer science and mathematics, with applications ranging from neuroscience and biology to economy, sociology, and politics, network science serves as a paradigm for interdisciplinary research. In the center of attention is the ambition to reveal the relationship between graph structure and system function.

## 8.1. Summary

The focus of this dissertation has been on the role of three-node subgraph structures. After a general introduction to the theory of network science, in Chapter 3, we discussed existing work on the abundance of non-random triadic structure. Examples are the high clustering coefficients that have been observed in many real-world networks, or the discovery of overrepresented patterns (motifs) in systems of various fields. There has been evidence for the functional importance of those motifs. Investigating isolated subgraphs, it has been reported that some patterns enhance the reliability of modeled information-processing dynamics in the presence of noise [81], and that there are patterns which stabilize the steady states of certain simulated dynamical processes [120]. Moreover, it was found that the reliability and stability associated with triadic subgraphs coincided with their overrepresentation in real-world systems in which the respective measures are assumed to be essential.

*8. Conclusions*

Furthermore, synthetic networks have been optimized, e.g. in order to make them robust against link failure. In the resulting graphs, motifs have been detected that are also abundant in real-world networks corresponding to the modeled processes [25, 71, 73].

All of the aforementioned investigations suggest the functional importance of triadic subgraph patterns. However, they either focussed on isolated subgraphs or they provide for evidence of the type *form follows function*, i.e. systems with a certain function all have a certain form. To test, on the other hand, whether *form implies function*, appropriate models to generate synthetic networks with predefined triadic subgraph structure are necessary. Nevertheless, there has been a lack of such triadic network models. A major challenge when attempting to induce triadic graph structure is that each dyad is part of multiple triads. Hence, not all triads can be specified independently. However, there are partitions of a graph's vertices into triples which actually *can* be specified at the same time, so called Steiner triple systems (STSs). Since two nodes appear together in exactly one such Steiner triple (ST), the system is not overdetermined by defining the configurations on all STs. In Chapter 4 we suggested a rather general class of generative network models based on Steiner triple systems.

Subsequently, in Chapter 5, we studied the most basic of such models – which we termed the triadic random graph model (TRGM) – in great detail. It assumes a probability distribution over all possible triad patterns. In order to generate a network instantiation of the model, for all Steiner triples in the system we drew a pattern from the distribution and adjusted it randomly on the ST. The TRGM can be considered the triadic analgon to Erdös-Rényi (ER) graphs. In the latter, the probabilities for all dyad configurations are the same and independent of each other. In the former, the probability distributions for the configurations on all Steiner triples are the same and independent of each other. In Section 5.3, we calculated the degree distributions of the triadic random graphs analytically and found it to be similar, yet not identical, to a Poissonian distribution which is typical for ER graphs. Depending on the input distribution, the degree distribution of the TRGM is broader than a Poissonian.

By means of extensive samplings we proved that TRGMs are capable of inducing non-vanishing $Z$ scores. We could demonstrate the strong impact of the probability distributions on the Steiner triples, $\vec{\mathcal{P}}$, on the observed $Z$-score profiles over the whole network. These allow us to design ensembles of networks with predefined triadic $Z$-score profiles. Hence, our suggested triadic random graphs help to overcome the lack of generative models needed for modeling triadic structure. They facilitate a systematic study of the effect of motifs on network dynamics. Furthermore, we discovered inevitable correlations between the $Z$ scores of certain triad patterns. These occur solely for statistical reasons and therefore should be taken into account when attributing functional relevance to particular motifs in real systems.

In order to investigate the functional importance of triad motifs in real networks, it is further necessary to assess whether motifs appear homogeneously or heterogeneously distributed over a graph. Therefore, in Chapter 6, we studied triadic subgraph structures in each node's neighborhood individually. We introduced the algorithm of **No**de-**S**pecific **Pa**ttern **M**ining (NoSPaM) for both directed unsigned (Section 6.2), and undirected signed networks (Section 6.4). Analyzing gene-transcription networks – in which the feed-forward loop (FFL) pattern is conjectured to be functionally critical – we found that the FFL is distributed highly heterogeneously, concentrated around only very few vertices. Evidence for the potential vulnerability of systems with respect to the failure of these vertices was found in Chapter 7.2, in which we studied the evolution of a graph's spectral gap under node removal. Furthermore, analyzing networks in terms of the homogeneity and homophily of their node-specific triadic structure, we found that these features differ strongly between systems of different origin. Moreover, clustering the vertices of graphs with respect to their node-specific triadic structure, we analyzed structural groups in the neural network of C. elegans, the international airport-connection network, and the global network of diplomatic sentiments between countries. For the latter we found indications for the instability of unbalanced triangles,

*8. Conclusions*

as suggested by Heider's theory [60, 61].

Finally, having our triadic random graph model available, in Chapter 7, we investigated ensembles of networks with similar triadic substructure in terms of the evolution of dynamical processes acting on their vertices. Considering oscillators, coupled along the graphs' edges, we found that triad motifs impose a clear signature on the systems even when embedded in a larger network structure. Moreover, using our newly suggested node-specific triadic $Z$ scores, we studied the effect of targeted node-removal, with respect to the nodes' contributions to certain subgraph patterns, on the spectral gap of the system. The spectral gap is the difference of the magnitude between the two strongest eigenvalues. For many dynamical processes, a non-vanishing spectral gap is critical for the systems to be able to synchronize. For the two analyzed systems – the neural network of C. elegans, and the transcriptional network of S. cerevisiae – we found the FFL to have the strongest impact on the spectral gap, compared to other motifs and graph measures. This observation serves as further evidence of its importance for the systems' function.

## 8.2. Outlook

The research presented in this dissertation offers various starting points for further investigations. In many real-world systems, individual node properties play a crucial role. Future models based on Steiner triple systems may include those characteristics in Eq. (4.9), in order to model, e.g., desired degree distributions. Furthermore, triadic generative models may be utilized to predict hitherto undiscovered links.

In this work we have studied the influence of triad motifs on the dynamics of coupled oscillators. However, there are various kinds of dynamics left to investigate, e.g. epidemic spreading, neuron dynamics, chaotic systems and many others. All those types of dynamics may potentially be analyzed with respect to the influence of certain triadic subgraph patterns.

In addition, over the last years, a multitude of networks has been analyzed in terms of their triadic subgraph structure. Using our suggested methodology for node-specific pattern mining, all of these systems can now be investigated with respect to their node-specific triadic substructure. The introduced measures for homogeneity and homophily will further allow to compare the detailed structure of different systems with each other and to extend the illustration in Fig. 6.8 on page 123. The analysis may unveil structural differences, even for those networks with similar ordinary triadic $Z$-score profiles.

Moreover, there is an enormous amount of data describing the political relationships beetween countries available. In Chapter 6.4.2 we already received an impression of the potential predictive power of signed, node-specific triadic $Z$-score profiles for the evolution of sentiments between countries. Studying the available data extensively, one might gain further insight into the underlying social processes.

# A. Appendix

## A.1. Bayes' Theorem

*Bayes' theorem* relates the conditional probabilities of events, $X$ and $Y$, to their prior probabilites,

$$\mathcal{P}(Y|X)\,\mathcal{P}(X) = \mathcal{P}(X|Y)\,\mathcal{P}(Y). \tag{A.1}$$

$\mathcal{P}(Y|X)$ is the conditional probability distribution for event $Y$, given the fact that $X$ has been observed.

Consider, e.g. a probability distribution, $\mathcal{P}\left(\bm{D}|\vec{\theta}\right)$, over networks conditioned on the parameters $\vec{\theta}$ of a model. If we want to fit the parameters to an observed adjacency matrix, $\bm{A}$, we can use Eq. (A.1) to find a probability distribution for the $\vec{\theta}$,

$$\mathcal{P}\left(\vec{\theta}|\bm{A}\right) = \frac{\mathcal{P}\left(\bm{A}|\vec{\theta}\right)\,\mathcal{P}\left(\vec{\theta}\right)}{\mathcal{P}(\bm{A})}. \tag{A.2}$$

Knowing the numerator of Eq. (A.2), the denominator can be inferred straightforwardly,

$$\mathcal{P}(\bm{A}) = \int\int ... \mathcal{P}\left(\bm{A}|\vec{\theta}\right)\,\mathcal{P}\left(\vec{\theta}\right)\,d\theta_1\,d\theta_2\,...\ . \tag{A.3}$$

The four probability distributions in Eq. (A.2) are referred to as the *posterior* $\mathcal{P}\left(\vec{\theta}|\bm{A}\right)$, the *likelihood* $\mathcal{P}\left(\bm{A}|\vec{\theta}\right)$, the *prior* $\mathcal{P}\left(\vec{\theta}\right)$, and the *marginal* $\mathcal{P}(\bm{A})$. It is thus

$$\text{posterior} \propto \text{likelihood} \times \text{prior}. \tag{A.4}$$

*A. Appendix*

As priors are often unknown when trying to learn the parameters of a model, so called *maximum-likelihood (ML)* techniques are useful. For an observed (i.e. fixed) matrix, $\boldsymbol{A}$, the latter aim to find the $\vec{\theta}$ that maximizes $\mathcal{P}\left(\boldsymbol{A}|\vec{\theta}\right)$.

## A.2. Datasets

The following datasets have been used in this dissertation. Here we give a brief description of the data. For more details, see the cited references.

### A.2.1. Directed Networks

**E. coli transcriptional** [7, 93]**:** 424 nodes, 519 edges. Nodes are operons, each edge is directed from an operon that encodes a transcription factor to an operon that it directly regulates (an operon is one or more genes transcribed on the same mRNA).

**Yeast transcriptional** [7, 32]**:** 688 nodes, 1,079 edges. Transcriptional network of the yeast S. cerevisiae. Nodes are genes, edges point from regulating genes to regulated genes. It is not distinguished between activation and repression.

**Neural network of C. elegans** [8, 142]**:** 279 nodes, 2,194 edges. Nodes are the neurons in the largest connected component of the somatic nervous system of the nematode C. elegans. Edges describe the chemical synapses between the neurons.

**Scientific citations** [49, 87, 89]**:** 27,700 nodes, 352,807 edges. Nodes are high-energy physics papers on the arXiv, submitted between January 1993 and April 2003. Edges from node A to B indicate that paper A cites paper B. Although it may seem unintuitive, there are papers citing each other. This may happen as papers can be updated continuously in time.

**Political blogs** [1, 109]**:** 1,224 nodes, 19,025 edges. Largest connected component of a network where the nodes are political blogs. Edges represent links between the blogs recorded over a period of two months preceding the 2004 US Presidential election.

**Enron-email network** [83, 89, 90]**:** 36,692 nodes, 183,831 edges. Email communication network from Enron. Nodes of the network are email addresses and if an address A sent at least one email to address B, the graph contains an undirected edge between A and B.

*A. Appendix*

**Leadership social network** [7, 100]: 32 nodes, 96 edges. Social network of college students in a course about leadership.

**Prisoners social network** [7, 100]: 67 nodes, 182 edges. Social network of inmates in prison.

**English book** [7, 100]: 7,381 nodes, 46,281 edges. Word-adjacency network of an English book. Nodes are words; an edge from node A to node B indicates that word B directly follows word A at least once in the text.

**French book** [7, 100]: 8,325 nodes, 24,295 edges. Word-adjacency network of a French book.

**Japanese book** [7, 100]: 2,704 nodes, 8,300 edges. Word-adjacency network of a Japanese book.

**Spanish book** [7, 100]: 11,586 nodes, 45,129 edges. Word-adjacency network of a Spanish book.

**Airport-connections network:** 3,438 nodes, 34,775 edges. Nodes are airports, an edge from airport A to airport B indicates a direct flight connection from A to B. Data processed from `http://openflights.org/`.

**Electronic Circuit s208** [7, 102]: 122 nodes, 189 edges. Network of electronic circuits. The nodes in these circuits represent electronic components e.g., logic gates in digital circuits and resistors, capacitors, or diodes in analogic circuits. Edges are directed connections between the elements. Parsed by Milo et al. from the ISCAS89 benchmark set of sequential logic electronic circuits [24, 26].

**Electronic Circuit s420** [7, 24, 26, 102]: 252 nodes, 399 edges. Network of electronic circuits.

**Electronic Circuit s838** [7, 24, 26, 102]: 512 nodes, 819 edges. Network of electronic circuits.

## A.2.2. Signed Undirected Networks of Political Sentiment Between Countries

Various networks composed of 223 nodes. The nodes are countries and edges represent the sentiment of the relationship between the countries.

Networks are processed from data retrieved from `http://gdeltproject.org/` [84].

The data provides for a description of the relationships of countries between the years 1979 and 2013 on a daily basis. The relationships are estimated by mining international sentiments from news sources. In the original form of the dataset, for each day the sentiment of country $a$ towards country $b$ is described by a four-dimensional vector,

$$\vec{QC}_{a \to b} = \begin{pmatrix} \text{verbal cooperation} \\ \text{verbal conflict} \end{pmatrix}. \quad \text{(A.5)}$$

The dataset is incomplete, i.e., for every day not all pairwise relationships are specified.

**Preprocessing**

In our analysis we want to focus on triadic substructures in the signed network of relationships. In particular, we want to focus on the *node-specific signed triad patterns* shown in Fig. 6.16. In order to keep the space of non-isomorphic, node-specific patterns managable we will focus on *unweighted, undirected* graphs. These graphs are generated from the data source by the following preprocessing.

To obtain the sentiment of a country $a$ towards a country $b$ at a given time, we subtract the respective conflict value from the corresponding cooperation value. The sign of the result shall indicate the sentiment.

As mentioned above, we will treat the graphs as undirected. In the case of sentiment in a social context we believe this to be a reasonable assumption since inconsistent relationships should appear rather infrequently. The mapping from the directed relationships $a \overset{sign}{\to} b$ and $b \overset{sign}{\to} a$ to the undirected one $(a, b, sign)$ is done as shown in Table A.1.

Moreover, we aggregate the data to a monthly resolution which is advantageous in many aspects. First, we average out noisy fluctuations in the monitored sentiments as the typical time scale for changes in

## A. Appendix

| $a \to b$ | $b \to a$ | $(a, b)$ |
|---|---|---|
| + | + | + |
| + | 0 | + |
| 0 | + | + |
| + | - | 0 |
| - | + | 0 |
| 0 | 0 | 0 |
| - | - | - |
| - | 0 | - |
| 0 | - | - |

**Table A.1.:** Mapping of directed sentiments to their undirected representations. There is no distinction between neutral and ascent / unknown edges.

the relationships between countries are much longer than single days. Second, we further complete the network, since not all pairwise relations are reported every day.

From the daily, signed coupling matrix, $\boldsymbol{G}^t \in \{-1, 0, 1\}^{N \times N}$, the monthly coupling matrix, $\tilde{\boldsymbol{G}}^t \in \{-1, 0, 1\}^{N \times N}$, is obtained via

$$\tilde{G}_{ij}^t = signum\left( \sum_{t'=1+30(t-1)}^{30t} G_{ij}^{t'} \right). \tag{A.6}$$

*A. Appendix*

# A.3. Network Structure of C. Elegans

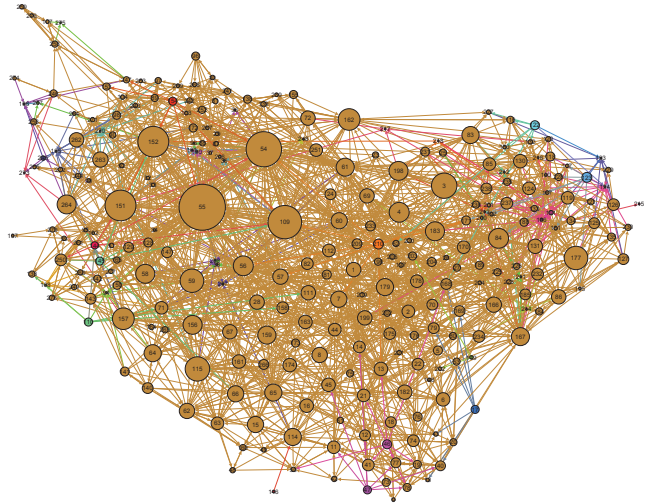

**(a)** 3 nodes removed

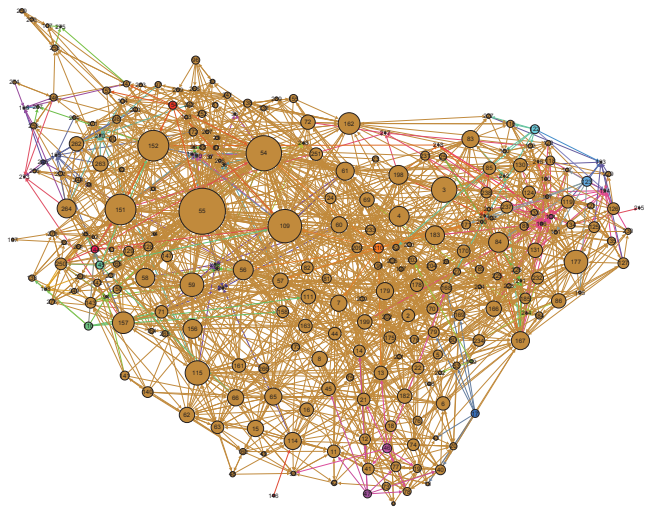

**(b)** 16 nodes removed

**Figure A.1.:** Graph structure of the neural network of C. elegans after consecutive removal of vertices with the strongest contribution to the feed-forward loop pattern. After a removal of 16 nodes, the spectral gap of the network drops to zero (see Fig. 7.8 on page 154). Vertex sizes reflect their out degree, colors indicate strongly connected components of the original graph.

# Index

*Index*

*Index*